\documentstyle[aps,psfig,epsf,bbox]{revtex}
\begin{document}
\newcommand{\vAi}{{\cal A}_{i_1\cdots i_n}}
\newcommand{\vAim}{{\cal A}_{i_1\cdots i_{n-1}}}
\newcommand{\vAbi}{\bar{\cal A}^{i_1\cdots i_n}}
\newcommand{\vAbim}{\bar{\cal A}^{i_1\cdots i_{n-1}}}
\newcommand{\htS}{\hat{S}}
\newcommand{\htR}{\hat{R}}
\newcommand{\htB}{\hat{B}}
\newcommand{\htD}{\hat{D}}
\newcommand{\htV}{\hat{V}}
\newcommand{\cT}{{\cal T}}
\newcommand{\cM}{{\cal M}}
\newcommand{\cMs}{{\cal M}^*}
\newcommand{\vk}{{\vec k}}
\newcommand{\vK}{{\vec K}}
\newcommand{\vb}{{\vec b}}
\newcommand{{\vp}}{{\vec p}}
\newcommand{{\vq}}{{\vec q}}
\newcommand{\vQ}{{\vec Q}}
\newcommand{\vx}{{\vec x}}
\newcommand{\tr}{{{\rm Tr}}}
\newcommand{\beq}{\begin{equation}}
\newcommand{\eeq}[1]{\label{#1} \end{equation}}
\newcommand{\half}{{\textstyle \frac{1}{2}}}
\newcommand{\gton}{\stackrel{>}{\sim}}
\newcommand{\lton}{\mathrel{\lower.9ex
                  \hbox{$\stackrel{\displaystyle <}{\sim}$}}}
\newcommand{\ee}{\end{equation}} \newcommand{\ben}{\begin{enumerate}}
\newcommand{\een}{\end{enumerate}} \newcommand{\bit}{\begin{itemize}}
\newcommand{\eit}{\end{itemize}} \newcommand{\bc}{\begin{center}}
\newcommand{\ec}{\end{center}} \newcommand{\bea}{\begin{eqnarray}}
\newcommand{\eea}{\end{eqnarray}}
\newcommand{\beqar}{\begin{eqnarray}}
\newcommand{\eeqar}[1]{\label{#1} \end{eqnarray}}
\newcommand{\bra}[1]{\langle {#1}|}
\newcommand{\ket}[1]{|{#1}\rangle}
\newcommand{\norm}[2]{\langle{#1}|{#2}\rangle}
\newcommand{\brac}[3]{\langle{#1}|{#2}|{#3}\rangle}
\newcommand{\hilb}{{\cal H}}
\newcommand{\pleft}{\stackrel{\leftarrow}{\partial}}
\newcommand{\pright}{\stackrel{\rightarrow}{\partial}}

\begin{flushright}
CU-TP-979
\end{flushright}
\vspace{1cm}

\begin{center}
{\Large {REACTION OPERATOR APPROACH  TO NON-ABELIAN \\
  \vskip .5cm 
ENERGY LOSS}}

\vspace{1cm}

{ M. Gyulassy$^{1}$, P. Levai$^{1,2}$ and I. Vitev$^{1}$ }

\vspace{.8cm}

{\em {$^1$  Dept. Physics, Columbia University, 
       538 W 120-th Street,\\ New York, NY 10027, USA\\
$^2$ KFKI Research Institute for Particle and Nuclear Physics, \\
P.O. Box 49, Budapest, 1525, Hungary} }
\vspace{.5cm}

{June 8, 2000}

\end{center}

\vspace{.5cm}

\begin{abstract}
  A systematic expansion of the induced inclusive gluon radiation associated
  with jet production in a dense QCD plasma is derived using a reaction
  operator formalism. Analytic expressions for the transverse momentum and
  light-cone momentum distributions are derived to all orders in powers of the
  gluon opacity of the medium, $N\sigma_g/A=L/\lambda_g$.  The reaction
  operator approach also leads to a simple algebraic proof of the ``color
  triviality'' of single inclusive distributions and to a solvable set of
  recursion relations. The analytic solution generalizes previous continuum
  solutions (BDMPS) for applications to mesoscopic QCD plasmas.  The solution
  is furthermore not restricted to uncorrelated geometries and allows for
  evolving screening scales as well as the inclusion of finite kinematic
  constraints. The later is particularly important because below LHC energies
  the kinematic constraints significantly decrease the non-abelian energy loss.
  Our solution for the inclusive distribution also generalizes the finite order
  exclusive (tagged) distribution case studied previously (GLV1). The form of
  the analytic solution is well suited for numerical implementation in Monte
  Carlo event generators to enable more accurate calculations of jet quenching
  in ultra-relativistic nuclear collisions.  Numerical results illustrating the
  constributions of the first three orders in opacity are compared to the
  ``self-quenching'' hard radiation intensity.  A surprising result is that the
  induced gluon radiation intensity is dominated by the (quadratic in $L$)
  first order opacity contribution for realistic geometries and jet energies in
  nuclear collisions.
\end{abstract}

\vspace{.5cm}


\vspace{.5cm}

\section{Introduction and Summary}

One of the expected signatures of the quark-gluon plasma created at
RHIC and LHC energies of $\sqrt{s} \simeq 200$~AGeV and $\sqrt{s}
\simeq 1500$~AGeV is jet
quenching~\cite{Bjorken:1982tu}-\cite{GLVPRL}. At the SPS energies of
$\sqrt{s} \simeq 20$~AGeV, on the other hand, no quenching of moderate
$p_\perp < 4$~GeV hadrons~\cite{XWWA98} was observed even in $Pb+Pb$..
This may be due to the break-down of pQCD at such low momentum scales
and the difficulty of disentangling non-perturbative multiparticle
production effects such as the Cronin and intrinsic $k_T$ effects
~\cite{MGPL}. Also finite kinematic constraints may limit the energy
loss at low jet energies. In addition, at SPS energies 
jet propagation in high density matter is limited to at most a few
fm/c due to rapid longitudinal expansion. This motivated our previous work
(GLV1)~\cite{GLV1A,GLV1B} to study the problem of energy loss in
``thin'' plasmas.  In that work we considered the exclusive (tagged)
case of energy loss associated with a fixed number of interactions.
Rough numerical estimates for nominal 5~GeV jets suggested that at
those relative small energies and small opacities, the radiative
energy loss may indeed be much smaller than predicted by Baier,
Dokshitzer, Mueller, Peign\'e and Schiff
(BDMPS)~\cite{BDMPS1,BDMPS2,BDMPS3,Baier:1999ds} and
Zhakharov~\cite{ZAH,ZAH2} for asymptotic jet energies.
  
Several approaches have been advanced to compute non-abelian energy
 loss.  One approach is aimed at treating relatively ``thick'' targets,
 which while small compared to the jet coherence length are large
 enough so that many collisions occur in the medium. This leads to  a
 continuum formulation of the problem.
 In~\cite{BDMPS1,BDMPS2,BDMPS3,Baier:1999ds} an effective 2D
 Schroedinger equation formulation was introduced, and
 in~\cite{ZAH,ZAH2,Kopeliovich:1999nw,URS,Wiedemann:2000ez} a path
 integral formulation of the problem was developed.  Another
 approach~\cite{MGXW,Wang:1995fx,GLV1A,GLV1B,GLVPRL}, which we extend
 here, aims to address the problem of radiative energy loss in
 ``thin'' plasmas, a few mean free paths thick, by computing directly
 the radiation pattern from the finite number of Feynman diagrams for
 the case of a few collisions.

The advantage of the former is that it can make direct contact with
the conventional Landau-Pomeranchuk-Migdal effect~\cite{LAND,MIGD} in
QED. It has the disadvantage that the continuum solutions obtained for
the case of many collisions cannot be directly applied to the case of
a few collisions. Also the effects of finite kinematical constraints
are difficult to include in such approaches. The advantage of the
second approach is that the few collision case can be computed
directly from the finite number of amplitudes involved.  The
disadvantage is that the LPM limit is out of reach, and the work
involved in summing diagrams increases exponentially with the number
of collisions.  In the present paper we introduce a new approach to
bridge the gap between these approaches.  Our new approach
is based on the construction of a suitable reaction operator,
$\htR_n$, from which recursion relations for the inclusive gluon
distribution can be derived and solved analytically at arbitrary
order $n$.

For ``thin'' QCD plasmas that can be formed in nuclear collisions, 
the gluon radiation intensity can be studied
systematically through an expansion in powers of opacity defined by
the mean number of collisions in the medium 
\beq
\bar{n}=\frac{L}{\lambda}= \frac{N\sigma_{el}}{A_\perp} =\int dz \int
d^2{\bf q}\, \frac{d\sigma_{el}(z)}{d^2{\bf q}}
\rho\left(z,\tau=\frac{z}{c} \right) \approx \, \frac{dN}{dy} \,
\frac{\sigma_{el}}{2 \pi R_G^2} \, \log \frac{R_G}{\tau_0} \;,
\eeq{opac} 
where $N$ is the number of targets in the medium of
transverse area $A_\perp$. An opacity expansion in terms of the path
integral formulation was introduced
in~\cite{URS,Wiedemann:2000ez,UOPAC}.  In~\cite{GLV1A,GLV1B} we
considered the exclusive tagged case where all $N$ target partons
interacting with the jet. Here (and \cite{GLVPRL}) we extend that
calculation to the inclusive case where fluctuations of the number of
collisions are allowed and only the geometry is fixed.

For a homogeneous rectangular target of thickness, $L$,
the density is $\rho=N/(L A_\perp)$, and the 
mean free path  is  $\lambda=1/(\rho \sigma_{el})$.
The opacity
is then simply $L/\lambda=N\sigma_{el}/A_\perp$.
For application to high transverse momentum
jets propagating through  cylindrical nuclear reaction geometries, we 
can interpret $L=1.2\, A^{1/3} \equiv R_s$.
For  more realistic  3+1D Bjorken and
transverse expanding Gaussian cylindrical  geometry such as
\beq
\rho(\vec{\bf x},\tau)=\rho_0\, \left(\frac{\tau_0}{\tau}\right) 
\exp\left({ -\frac{\vec{\bf x}^2+\Delta \tau^2}{R_G^2} }\right)
I_0 \left( \frac{2|\vec{\bf x}| \Delta \tau}{R_G^2} \right) 
\; \; ,\eeq{gaus}
 $L$ is replaced by the equivalent rms Gaussian transverse radius
$R_G=0.75\, A^{1/3}$~fm.  The rightmost expression in (\ref{opac})
is obtained by averaging the number of collisions of a transverse jet
in this expanding Gaussian cylinder with
the initial jet production coordinate, $\vec{\bf x}_0$, averaged over
$\rho(\vec{\bf x}_0,\tau_0)$. Here 
$\tau_0$  is the formation time of the plasma. 
At RHIC energies ($\sqrt{s}\simeq 200$~AGeV) the  expected rapidity
density of the gluons
is $dN/dy \simeq 1000$ for  $A=200$~\cite{KARI} . For a typical
 elastic gluon-gluon cross section
$\sigma_{el}\sim 2$~mb and a plasma formation time $\sim 0.5$~fm/c,
the opacity is moderately small ${\bar{n}} < 10$. 
This suggests that neither the thin nor thick plasma approximations
may apply for applications to nuclear collisions.
Therefore, an alternate method is needed
to handle the intermediate (mesoscopic) case.

An important simplification may be anticipated
due the  non-abelian analog~\cite{MGXW,BDMPS1,GLVPRL} 
of the Landau-Pomeranchuk-Migdal (LPM) effect~\cite{LAND,MIGD}. 
While the number of scatterings is moderately large,
the radiation intensity angular distribution and total energy loss are 
controlled by the combined effect of the number of scatterings
$\bar{n}=L/\lambda$ and  the formation probability
$p_f\sim  L/l_f = L |{\bf k}|^2/(2xE)$ of the gluon in the medium.
Here  $x$ is  the light-cone momentum fraction and
$|{\bf k}|\ll x E$ is the transverse momentum  of the radiated gluon. 
Therefore, the opacity expansion may 
 converge  more rapidly than one would first expect.
One of the surprising  results derived here and summarized in
\cite{GLVPRL} is that the {\em inclusive}
induced gluon radiation intensity is 
dominated by the (already quadratic in $L$) 
 first order opacity contribution 
for realistic geometries and jet energies in nuclear collisions.

In GLV1~\cite{GLV1B}, we showed in contrast
that in the exclusive tagged case, i.e.
with all recoiled partons measured 
in coincidence with the jet and gluon,
a much more complex and nonlinear non-abelian radiation pattern
emerges. The tagged recoil
partons act as additional color dipole antennas that
interfere in a nontrivial way with the jet's color  dipole antenna.
The resulting color algebra cannot be reduced to simple powers 
of the color Casimirs. We found that 
 the gluon number  distribution
for the case of $n_s=N$ tagged target partons has the form
\beq
{dN_g^{(n_s)}}
= {dN_g^{(0)}}
\left[ 1 + ((1+R)^{n_s}-1) f^{(n_s)}(\kappa,\xi) \right]/Z_{n_s}\;\;,
\eeq{glv1scal} 
where $f^{(n_s)}$ depends on two dimensionless ratios
$\kappa = {\bf k}^2/ \mu^2 $ and 
$\xi = \lambda/l_f= \lambda \mu^2 / 2 x E$,
and depends only weakly on $n_s$. The color dependence
enters through a rapidly increasing function
of $R=C_A/C_R$, the ratio of the 
Casimirs in the adjoint ($d_A=N^2_c-1$ dimensional)
gluon representation and  the Casimir, $C_R$, of the
$d_R$ dimensional jet representation.  
In GLV1 unitarity was imposed by
a wave function renormalization factor, $Z_{n_s}$,
 that was computed perturbatively.
For  the inclusive case studied here, unitarity is assured
 by the inclusion of certain contact double Born (Virtual) terms
as emphasized in BDMS~\cite{BDMPS3}.

To address energy loss of hard probes produced  in
nucleus-nucleus collisions we concentrate as in~\cite{GLV1B,GLVPRL} on
the case of hard jet produced inside the plasma at a finite point
$(t_0,z_0,{\bf x}_0)$ rather than on the  Gunion-Bertsch
problem~\cite{BG}  in which the jet is replaced by
a high energy beam of quarks or gluons prepared in the remote
past. We employ as in GW~\cite{MGXW,GLV1B} static color-screened
Yukawa potentials to model interactions in a deconfined quark gluon
plasma.  The  Fourier and color structure of those potentials
are assumed to have the form
\beq V_n = V(q_n)e^{iq_n x_n} = 2\pi \delta(q^0) v(\vec{\bf
q}_n) e^{-i \vec{\bf q}_n\cdot\vec{\bf x}_n} \; T_{a_n}(R)\otimes
T_{a_n}(n) \;\; ,
\eeq{gwmod}
where $\vec{\bf x}_n$ is the location  of the $n^{\rm th}$ (heavy)
target parton and 
\beq
v(\vec{\bf q}_n)\equiv
\frac{4\pi \alpha_s}{\vec{\bf q}_n^{\;2}+\mu^2}=
\frac{4\pi \alpha_s}{(q_{nz}+i\mu_{n})(q_{nz}-i\mu_{n})}
\;\; ,
\eeq{vq}
where $\mu_{n}^2=\mu_{n\perp}^2 = \mu^2+{\bf q}_{n}^{\;2}$.
The small transverse momentum transfer elastic cross section 
between the jet and target partons in this
model is
\beq
\frac{d\sigma_{el}(R,T)}{d^2{\bf q}}=
\frac{C_RC_2(T)}{d_A}\frac{|v({\bf q})|^2}{(2\pi)^2}
\;\; . \eeq{sigel}
In our notation
transverse 2D vectors are denoted as ${\bf p}$, 3D vectors 
as $\vec{\bf p}=(p_z,{\bf p})$, and four vectors by $p=(p^0,\vec{\bf p})=
[p^0+p^z,p^0-p^z,{\bf p}]$.

The color exchange bookkeeping with the target parton $n$ is handled
by an appropriate $SU(N_c)$ generator, $T_a(n)$, in the $d_n$
dimensional representation of the target.  ($\tr\, T_a(n)=0$ and
$\tr\,(T_a(i)T_b(j))=\delta_{ij} \delta_{ab} C_2(i)d_i/d_A$. We will
assume that all target partons are in the same $d_T$ dimensional
representation with Casimir $C_2(T)$.) We denote the generators in the
$d_R$ dimensional representation of the  jet by $a\equiv t_a$
with $aa=C_R{\bf 1}$.  The elastic cross section of the jet 
with any target parton
 is therefore proportional to the product of Casimirs,
$C_R C_2(T)$.

The analytic results derived below with the reaction operator approach
do not depend on the actual form of $v$, but the Yukawa form will be
used for numerical estimates.  Recall that in a thermally equilibrated
medium at temperature $T$, the color screening mass in pQCD is given
by $\mu=4 \pi \alpha_s T^2$.  Also there is a cut-off frequency for
soft gluon modes $\omega_{pl} \sim \mu$.  We take here $\mu\simeq
\omega_{pl} \simeq 0.5$ GeV for numerical estimates.  In perturbation
theory, $\mu^2/\lambda \approx 4\pi \alpha_s^3 \rho$ provides a
measure of the $\rho$, the density of plasma partons weighed by
appropriate color factors.

This paper is organized as follows: in Sec.~II we review the effect of
unitarity corrections in the simple elastic jet scattering case. We
show how the jet probability is conserved up to first order in opacity
through the inclusion of contact double Born graphs.  This is the
simplest example illustrating how a factor $-\half$ arises from
longitudinal momentum contour integrations in the contact limit.  In
Sec.~III we proceed to the case of gluon bremsstrahlung associated
hard probes produced in a dense medium.  We discuss the basic
diagrammatic rules and specify the assumptions and approximations used
in this work. We extend the algebraic classification of diagrams given
in Ref.~\cite{GLV1B} to include virtual double Born amplitudes needed
in the inclusive case.  Sec.~III~C summarizes the rules of
diagrammatic calculus that emerge from  detailed analysis of 
diagrams in Appendix~A through Appendix~E.  

In Sec.~IV the new reaction operator formalism is developed.  First,
operators $\htD_n,\htV_n$ in Eqs.~(\ref{didamit},\ref{dvid}) are
constructed from the diagrammatic rules.  Products of these operators
create partial sums of direct and virtual amplitudes from the initial
hard vacuum amplitude. Those partial sums, Eq.~(\ref{atens}), form
$3^n$ classes of diagrams that can be conveniently enumerated via a
tensor notation and used to construct recursion relations.  In
Sec.~IV~B, the reaction operator, $\htR_n=\hat{D}_n^\dagger
\hat{D}_n+\hat{V}_n+\hat{V}_n^\dagger$, is constructed
to relate the $n^{\rm th}$
order in opacity inclusive radiation probability distribution to
classes of diagrams of order $n-1$.  The resulting simple recursion
relation, Eq.~(\ref{rp0}), can be solved in closed form.  The general
solution, Eq.~(\ref{iter2}), is suitable for implementation in Monte
Carlo event generators to study observable consequences of jet
quenching in nuclear collisions.

Color triviality of the inclusive distribution is proven to all orders
algebraically with  Eq.~(\ref{iter2}). 
 The proof is much simpler and  more transparent
 than in the path integral formulations~\cite{ZAH,Wiedemann:2000ez,UOPAC}
and is not limited to quark jets.
  
In Sec~IV~C a compact general expression for the momentum transfer
averaged inclusive distributions, Eq.~(\ref{ndifdis}) is derived.
Appendix F provides an independent check of this solution through
second order starting from the amplitude iteration
technique. Numerical results comparing angular distributions of gluons
up to the first three orders in opacity are presented in Sec.~V~A.
Analytic and numerical results for the angular integrated intensity
distributions are compared in Sec.~V~B.  It is shown that the induced
intensity is dominated by the first order in opacity result that is
already quadratic in $L$.

A brief summary of these results up to second order in opacity was
 reported in Ref.~\cite{GLVPRL}. The main result of this paper is the
 new reaction operator derivation  of the
 solutions, Eq.~(\ref{iter2},\ref{ndifdis}), that specify the inclusive
 non-abelian radiation distribution to any order in opacity.
 
\section{Elastic scattering and unitarity }

To illustrate how the double Born graphs cancel direct contributions
to preserve unitarity we review here the simplest case of elastic
scattering.  Consider a wave packet $j(p)$ of a parton prepared at
time $t_0$ and localized at $\vec{\bf x}_0=(z_0,{\bf x}_0)$ in color
representation $R$.  The (color matrix) amplitude to measure its
momentum as $\vec{\bf p}$ in the absence of final state interactions
is
\beq
M_0\equiv i  e^{ipx_0} j(p) \,\times  {\bf 1}\;\;.
\eeq{p0}
Multiplying $|M_0|^2$ by the invariant one particle phase space
element $d^3\vec{\bf p}/((2\pi)^3 2|\vec{\bf p}|)$ and taking the color trace gives the unperturbed 
inclusive distribution of jets in the wave packet:
\beq
d^3N_0= {\rm Tr}\;
|M_0|^2 \frac{d^3 \vec{\bf p} }{2|\vec{\bf p} |(2\pi)^3}
 =  |j(p)|^2   \frac{d_R \; d^3 \vec{\bf p}}{2|\vec{\bf  
 p} |(2\pi)^3}
\;\; . 
\eeq{n0} 
Consider next the effect of final state elastic interactions
with an array of 
static potentials localized  at $\vec{\bf x}_i=(z_i,{\bf b}_i)$ using 
\beqar
H_I(t)&=&\int d^3 \vec{\bf x} \, \sum_{i=1}^N
v(\vec{\bf x}-\vec{\bf x}_1) T_a(i)
\phi^\dagger(\vec{\bf x},t)T_a(R) \hat{D}(t)\phi(\vec{\bf x},t)
\;\;,
\eeqar{hj}
where $\hat{D}(t)=i\stackrel{\leftrightarrow}{\partial_t}$
and $\tr T_a(i)T_b(j)=\delta_{ij}\delta_{ab} C_2(T) d_T/d_A$.
We will compute  the three graphs in Fig.~1.
The first order, direct amplitude to scatter with one of 
the (static) target partons is
\beqar
M_1&=& 
i  e^{ipx_0}
\int \frac{d^4 q}{(2\pi)^4} j(p-q)\Delta(p-q)v(q)D(2p-q)
\sum_{j=1}^N e^{iq(x_j-x_0)} T_a(j)T_a(R)\;\;,
\eeqar{p1}
where $\Delta(p)\equiv (p^2+i\epsilon)^{-1}$ and $D(p)=p^0$.
The sum of double Born  amplitudes in the {\em same} 
external potential is
\beqar
M_2&=&
i  e^{ipx_0}T_a(R)T_b(R)
\int \frac{d^4 q_1}{(2\pi)^4}\frac{d^4 q_2}{(2\pi)^4} j(p-q_1-q_2)
\Delta(p-q_1-q_2) D(2p-2q_2-q_1) v(q_1) \, \times \nonumber \\[1ex]
&\;& \hspace{1.5in}  \times \, \Delta(p-q_2)D(2p-q_2) v(q_2)
\sum_{j=1}^N e^{i(q_1+q_2)(x_j-x_0)}T_a(j)T_b(j)
\;\; . 
\eeqar{p2}
\begin{center}
\vspace*{8.7cm}
\includegraphics{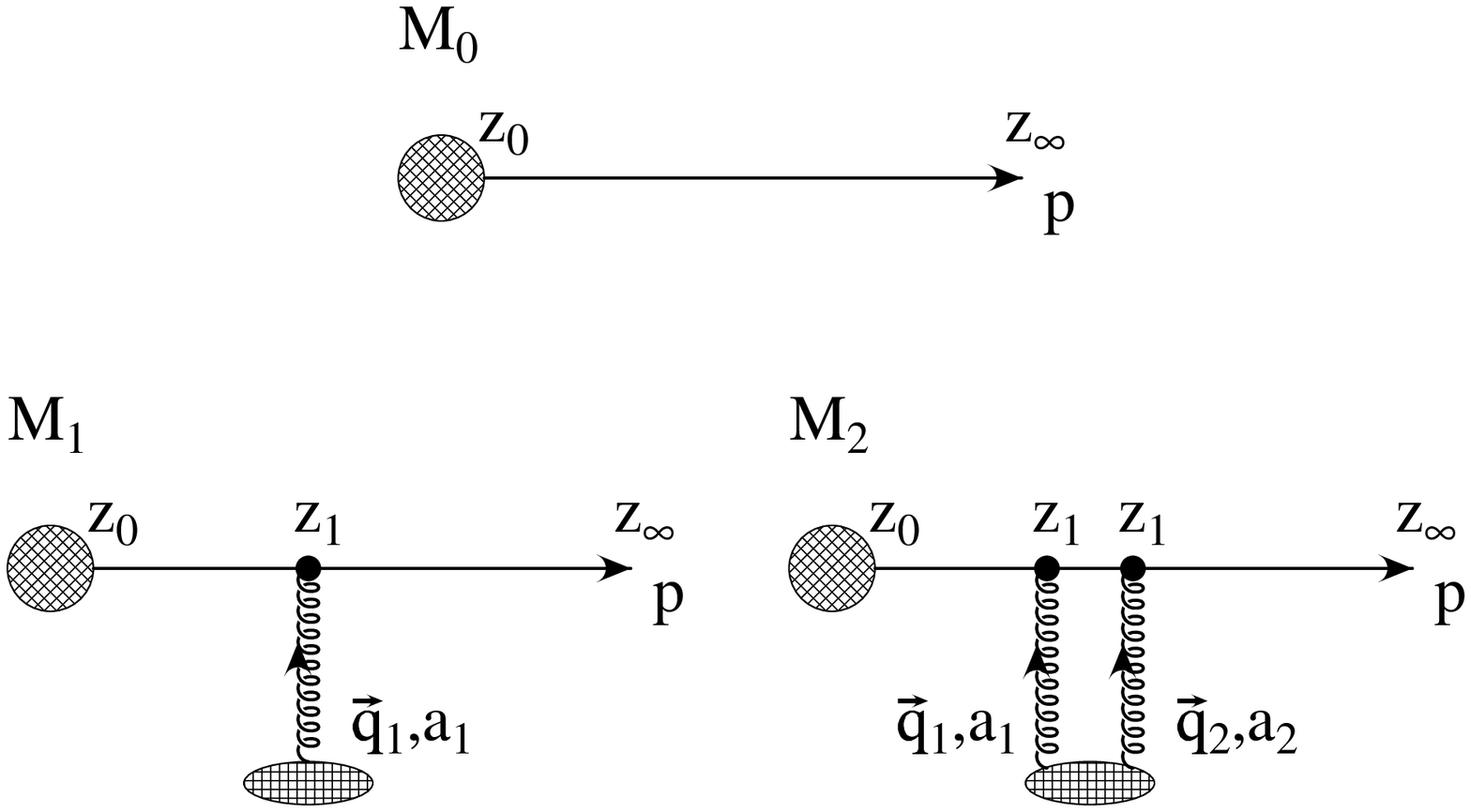}
\vskip -30pt
\begin{minipage}[t]{15.0cm}
{\small {FIG~1.}  Graphs that produce
a jet with $\vec{\bf p}$ from
an initial wave packet formed at $z_0$ followed by single and double
Born scattering center located at $z_1$.
}
\end{minipage}
\end{center}
\vskip 4truemm 
Double Born with two different centers will not
contribute because ${\rm Tr} T_a(j)=0$. To first order in opacity, the
probability distribution of jets is given by 
\beqar
d^3N&=&\frac{1}{d_T}{\rm Tr}\;|M_0+M_1+M_2 + \cdots |^2
\frac{d^3\vec{\bf p}}{2|\vec{\bf p}|(2\pi)^3} \nonumber \\[1ex]
&=& d^3N_0 + d^3N_1+ \frac{1}{d_T}{\rm Tr}\;
\left[2\, {\rm Re} (M_1 M_0^*) + 
2\, {\rm Re} (M_2 M_0^*)\right]
\frac{d^3\vec{\bf p}}{2|\vec{\bf p}|(2\pi)^3} \; +  \; \cdots \;\; ,
\eeqar{dn12th}
where  we separated
 the direct $(|M_1|^2)$  and unitary correction
contributions with  
\beqar
 2|\vec{\bf p}|(2\pi)^3 \,\frac{d^3N_1}{d^3\vec{\bf p}}
&=& 
\int\frac{d^4 q}{(2\pi)^4} \frac{d^4 q^\prime}{(2\pi)^4} 
\; d_R\;j(p-q)j^*(p-q^\prime) D(2p-q) D(2p-q^\prime) \, \times \nonumber \\[1ex]
&\;& \times \; \frac{1}{(p-q)^2+i\epsilon} \; 
\frac{1}{(p-q^\prime)^2-i\epsilon} \;
 v(q)v^*(q^\prime) \frac{C_RC_2(T)}{d_A} \; \sum_{j=1}^Ne^{i(q-q^\prime)(x_j-x_0)}\;\; .
\eeqar{dn1}
To proceed further, we consider a Yukawa potential 
as in Eqs.~(\ref{gwmod},\ref{vq}) and assume that all
the $x_j$ are distributed with the same 
density 
\beq
\rho(\vec{\bf x})=\frac{N}{A_\perp} \bar{\rho}(z)
\;\; , \eeq{densit}
where $\int dz\bar{\rho}(z)=1$.
Second,  we assume that the observed
$p=(E,E,0)=[E^+,0,0]$  is 
high as compared to the potential screening scale, i.e.
\beq
E^+ \simeq 2E \gg \mu \;\; .
\eeq{assume2}
We also assume that the distance between the source and scattering
centers are  large compared to the interaction range:
\beq
z_i-z_0 \gg 1/\mu \;\; .
\eeq{assume3}
Finally, we assume that the source current or packet
$j(p)$ varies slowly over the range of 
momentum transfers supplied by the potential. We can then approximate 
the $q_z$ contour integrals ignoring the $q_z$ dependence of the 
source $j(p \pm q)$.  With (\ref{assume3}) we can also neglect the 
contributions to the $q_z$  contour integrals due to singularities of 
the potentials at $\pm i \mu_\perp, 
\;\mu^2_\perp=\mu^2+{\bf q}^2 $. With these simplifying assumptions, 
only the residues with $q_{z}\approx -i\epsilon +{\bf q}^2/E^+$ and  
$q_{z}^\prime \approx +i\epsilon + {\bf q}^{\prime 2}/E^+$ 
contribute:
\beqar
 2|\vec{\bf p}|(2\pi)^3 \,\frac{d^3N_1}{d^3\vec{\bf p}} &=&  
Nd_R
\int\frac{d q_z d^2{\bf q}}{(2\pi)^3} 
\frac{dq_z^\prime d^2{\bf q}^{\prime}}{(2\pi)^3} 
\; j(p-q)j^*(p-q^\prime)\; v({\bf q} \,)v^*({\bf q}^{\prime}) 
\frac{C_RC_2(T)}{d_A} \; \times 
\nonumber \\[1ex]
&\;& \times \frac{E^+}{E^+q_z-q^2_z-{\bf q}^2+i\epsilon} \;
\frac{E^+}{E^+q^\prime_z-q^{\prime 2}_z-{\bf q}^{\prime 2}-i\epsilon}
\; \langle e^{-i( \vec{\bf q}-\vec{\bf q}^{\; \prime})
\cdot(\vec{\bf x}_1-\vec{\bf x}_0)}\rangle
\nonumber \\ [1.ex] 
&\approx&N d_R
\int \frac{d^2{\bf q}}{(2\pi)^2} \;\frac{d^2{\bf q}^{\; \prime}}
{(2\pi)^2} j(p-{\bf q}) j^*(p-{\bf q}^{\; \prime})
 \langle
e^{-i({\bf q}-{\bf q}^{\; \prime})\cdot
({\bf x}_{ 1}-{\bf x}_{ 0})}\rangle
 v({\bf q})v^*({\bf q}^{\; \prime})\frac{C_RC_2(T)}{d_A} \;\; .
\eeqar{dn1a}

A major simplification (that we will
also use in the radiation case)  occurs if the 
relative transverse coordinate (impact parameter)
${\bf b}={\bf x}_{i }-
{\bf x}_{0 }$  varies over a large 
transverse area, $A_\perp$, relative to the interaction area $1/\mu^2$. 
In this case,  the ensemble average over the scattering center
location reduces to an impact parameter average
 as follows:
\beq
\langle \, \cdots \, \rangle = \int \frac{d^2{\bf b}}{A_\perp} \cdots
\eeq{impact}
The ensemble average over the phase factor in Eq.~(\ref{dn1a})
then yields
\beq
\langle \, e^{-i({\bf q}-{\bf q}^{\prime})\cdot
{\bf b}} \, \rangle= \frac{(2\pi)^2}{A_\perp} 
\delta^2({\bf q}-{\bf q}^{\prime})
\eeq{bave}
Note that we ignored a small phase shift in Eq.~(\ref{dn1a})
$\propto \exp[-{\bf q}^2 (z_1-z_0)/ E^+]\approx 1$ on account of
our  high energy (eikonal) assumption. 
 The ensemble averaged first order in opacity direct 
contribution to the jet distribution  therefore reduces to the familiar 
form noting Eq.~(\ref{sigel})
\beqar
 \frac{2|\vec{\bf p}|(2\pi)^3}{d_R}
 \,\frac{d^3N_1}{d^3\vec{\bf p}} &\approx& 
\frac{N}{A_\perp} \int\frac{d^2{\bf q}}{(2\pi)^2}|j(p-{\bf q})|^2 
\frac{d\sigma_{el}(R,T)}{d^2{\bf q}}
\nonumber \\
&\approx&|j(p)|^2 \int d^2{\bf q}  
\frac{N}{A_\perp}\frac{d\sigma_{el}(R,T)}{d^2{\bf q}} 
= \frac{L}{\lambda} \; |j(p)|^2 \;\; ,
\eeqar{dn1f}
where the last line only holds if the initial
packet is very wide in momentum space compared to the momentum
transfer scale $\mu$. 

Next we turn next to the unitary corrections. 
The first order term $\langle M_1 M_0^*\rangle=0$ on account
of ${\rm Tr}\; T_a(j)=0$.
A non-vanishing unitarity correction arises however from
\beqar
\frac{1}{d_T}{\rm Tr}\;
\langle M_2 M_0^*\rangle&=&
\int \frac{d^4 q_1}{(2\pi)^4}\frac{d^4 q_2}{(2\pi)^4} \; j(p-q_1-q_2)j^*(p)
D(2p-q_2)D(2p-q_1-q_2)\; \times
\nonumber \\[1ex]
&\;& \times \; \frac{V(q_2)}{(p-q_1-q_2)^2+i\epsilon}\;
\frac{V(q_1)}{(p-q_2)^2+i\epsilon}\;
{\rm Tr}(T_a(R)T_b(R)) \;\nonumber \\
&\;& \times
\langle \; \sum_{j=1}^N e^{i(q_1+q_2)(x_j-x_0)} \frac{1}{d_T}
{\rm Tr}(T_a(j)T_b(j))\ \; \rangle\; \; . 
\eeqar{p2p0}
Note that the impact parameter average constrains
 ${\bf q}_{ 1}+{\bf q}_{ 2}=0$ in this case
($q_1^0=q_2^0=0$ for static potentials). The 
phase factor requires  that we  close
the $q_{1z}$ contour in the lower half-plane. We pick up the 
$-2\pi i\, {\rm Res}(q_{1z}=-i \epsilon-q_{2z})$. This results in setting 
$\vec{\bf q}_1+\vec{\bf q}_2=0$ throughout. Again the residue from 
the second pole in the  lower half-plane $q_z=-i\mu_\perp$ is 
suppressed by the phase factor. We are left with
\beqar
\frac{1}{d_T}\tr\;
\langle M_2 M_0^*\rangle &\approx& -i\,N d_R |j(p)|^2
\int \frac{d^2 {\bf q}}{(2\pi)^2}\frac{d q_{2z}}{(2\pi)} 
\frac{(4 \pi \alpha_s)^2}{A_\perp} \;  \frac{C_R C_2(T)}{d_A}
\frac{1}{(q_{2z}^2 +  \mu^2_\perp)^2}\;
\frac{E^+}{E^+q_{2z}-q_{2z}^2-{\bf q}^{2}+i\epsilon}   \; \; , 
\eeqar{p2p0pr}
where $\mu^2_\perp=\mu^2+{\bf q}^{2}$.
The $q_{2z}$ can be performed yielding  
\beqar
\int \frac{dq_{2z}}{2\pi} \frac{1}{(q_{2z}^2 + \mu^2_\perp)^2}\;
\frac{E^+}{E^+q_{2z}-q_{2z}^2-{\bf q}^{\;2}+i\epsilon} 
 \approx \frac{-i}{2} \frac{1}{({\bf q}^{2}+\mu^2)^2} \;\;.
\eeqar{sct}
The double Born ``contact'' contribution to the differential yield at
first order in opacity is therefore 
\beq 
\frac{1}{d_T}\tr\;2\, {\rm Re} \langle M_2
M_0^*\rangle \approx -\frac{L}{\lambda}\; |j(p)|^2 d_R\;\; .
\eeq{final20} The inclusive first order in opacity elastic
distribution is therefore
 \beqar 
\frac{2|\vec{\bf p}|(2\pi)^3}{d_R}
\frac{d^3N}{d^3\vec{\bf p}} &=& |j(p)|^2(1-\frac{L}{\lambda}) +
\frac{N}{A_\perp} \int\frac{d^2{\bf q}}{(2\pi)^2}|j(p-{\bf q})|^2
\frac{d\sigma_{el}(R,T)}{d^2{\bf q}} + O\left(\frac{L}{\lambda}\right)^2
\nonumber \\
&=& |j(p)|^2+\int dz \rho(z,0)
 \int\frac{d^2{\bf q}}{(2\pi)^2}\left\{
\frac{d\sigma_{el}(R,T)}{d^2{\bf q}}-\sigma_{el}\delta^2(
{\bf q})\right\} |j(p-{\bf q})|^2
 + O\left(\frac{L}{\lambda}\right)^2
\; \;\; . 
\eeqar{glaub1}
This is the first order term in the
 Glauber multiple scattering series. Note
that when integrated over momentum,
the double Born term cancels  exactly 
direct contribution and therefore  enforces
 probability conservation. However, 
for very narrow coordinate space packets, i.e. wide
in momentum space so that $|j(p-{\bf q})|^2\approx
|j(p)|^2$,
we see that the double Born  contribution actually cancels 
approximately the first order direct term  differentially.

 We derived this cancelation assuming ${\bf p}=0$ throughout, but it
holds generally since all we needed was that the phase shift
$(z_i-z_0)({\bf q}-{\bf p})^2/E^+\ll 1$.  The point of this brief
review was to emphasize that the unitarity cancelation arose due to
the factor of $\half$ that appeared in Eq.~(\ref{sct}) from the
longitudinal momentum contour integration and from the factor $-1$
that appeared due to moving the potential from one side of a cut to
another. This is a generic property of contact interactions that also
holds in the radiation case also as pointed out in \cite{BDMPS3}
and derived here in the Appendix.

\section{Soft gluon radiation}
\subsection{Kinematics, triple gluon vertices, and color algebra}

Consider a source $J$ that produces a jet
that subsequently radiate a gluon with
four momentum $k$, polarization $\epsilon(k)$,
and emerges  with momentum $p$. In light-cone components
\beq
k=[xE^+,k^-\equiv \omega_0,{\bf k}] \;\;, 
\epsilon(k)=[0,2\frac{\bbox{\epsilon} \cdot
{\bf k}}{xE^+},\bbox{\epsilon}]\;\;, 
\;\;  p=[(1-x)E^+,p^-,{\bf p} ] \;\;.
\eeq{kinem}
Soft radiation is defined as $x\ll 1$ so that, for
example, $p^+\gg k^+$ and
 $p^-={\bf p}^2/(1-x)E^+\ll k^-=\omega_0= {\bf k}^2 /x E^+$.

We adopt the shorthand notation of Refs.~\cite{GLV1A,GLV1B}
\beq
\omega_0=\frac{{\bf k}^{2}}{2\omega}\; ,\;
\omega_i=\frac{({\bf k}-{\bf q}_{i})^2}{2\omega}\; ,\;
\omega_{(ij)}=\frac{({\bf k}-
{\bf q}_{i}-{\bf q}_{j})^2}{2\omega}
\;,\; \omega_{(i\cdots j)}=\frac{({\bf k}-
\sum_{m=i}^j {\bf q}_{m})^2}{2\omega}
\; \; .
\eeq{shorth}
In the soft eikonal kinematics, that we consider
\beq
E^+\gg k^+\gg \omega_{(i\cdots j)}\gg 
\frac{({\bf p}+ {\bf k})^2}{E^+} \;\;.
\eeq{eorder}

The  soft gluon and soft rescattering approximations
will allow us to  simplify analytic results using:
\beqar
J(p-Q+k)&\approx& J(p+k)\approx J(p) \;\;, \nonumber \\
\epsilon_\mu(k) (p-Q+k)^\mu &\approx& \frac{\bbox{\epsilon}
\cdot{\bf k}}{x}
 \;\;, \nonumber \\
k(p-Q)&\approx& kp \approx \frac{{\bf k}^2}{2x}
 \;\;, \nonumber \\
\Delta(p-Q)\Delta(p-Q+k) &\approx&
\frac{1}{kp} \left( \Delta(p-Q)
- \Delta(p-Q+k) \right) \;\;. 
\eeqar{softness}

Scattering of the jet or radiated gluon with the potential 
centered at position $\vec{\bf x}_n$ introduces an integration 
$\int d^4q_n/(2\pi)^4$ in the amplitude over the potential 
$V(q_n)$. We use the GW model potential Eqs.~(\ref{gwmod},\ref{vq}).   
If the scattering involves a momentum exchange
with the high energy jet, the vertex factor is simply
$-iE^+$ in the eikonal limit. Neglecting the spin
of the jet parton, each intermediate jet propagator
brings in a factor $+i\Delta(p-Q)$, where $\Delta(p)=1/(p^2+i\epsilon)$. 
The gluon  emission vertex then gives in this approximation
a factor $ +ig_s(2p+k-2Q)^\mu$, where $Q$ is the sum of the
subsequent momentum transfers. 

For scattering of the radiated gluon,
consider the triple gluon tensor, $\Gamma^{\alpha 0 \gamma}(k,q)$
for a $g(k-q,\alpha,a) + V(q,\beta=0,b) \rightarrow g(k,\gamma,c)$.
In our model, $q_n^0=0$, $|q_{nz}|=|q_n^+|=|q_n^- |\ll |{\bf q}_{n}|$,
and thus the $q_n$  correspond to  small (almost) transverse momentum transfers. 
If the last interaction of the gluon is with center $m$,
then that last vertex is
\beq
\Gamma^{\alpha0\gamma}(k,q_m,k-q_m)= 2\omega g^{\alpha\gamma}-
(k+q_m)^\alpha g^{0\gamma} + (2q_m-k)^\gamma g^{0\alpha} \;\;.
\eeq{gam1}
The color factor for the above ``clockwise'' kinematic convention
is $-f^{cba}$.

The contraction of Eq.~(\ref{gam1})  with the final polarization gives
\beq
\Gamma^{\alpha}(k;q_m)\equiv \Gamma^{\alpha\gamma}(k,q_m)\epsilon_\gamma(k)=
2\omega \epsilon^\alpha -(k+q_m)^\alpha \epsilon^{0} + 
2g^{0\alpha} (q_m\epsilon) \;\;.
\eeq{g1}
For the case of two gluon rescatterings the two vertices combine
\beqar  
\Gamma^{\alpha}(k;q_n,q_m)&\equiv &\Gamma^{\alpha\mu}(k-q_m,q_n)g_{\mu\nu}
\Gamma^{\nu}(k;q_m) \nonumber \\[1ex]
\Gamma^{\alpha}(k;q_n,q_m)&=& 4\omega^2 \epsilon^\alpha - 
\omega (3k+q_n+q_m)^\alpha \epsilon^0 
+  g^{0\alpha}(4 \omega(q_n+q_m)\epsilon) \nonumber \\[1ex]
&\;& \;  - g^{0\alpha} (2q_n(k-q_m)+q_m^2)\epsilon^0
-2(k-q_m+q_n)^\alpha (q_m\epsilon) \;\;.
\eeqar{g2}
Note that using $+g^{\mu\nu}$ to combine the
vertices above requires us to insert $-i\Delta(k-Q)$ for each gluon propagator, where use
$\Delta(p)=1/(p^2+i\epsilon)$. 

The gluon vertex factors  contracted by  emission vertex
factors then give
\beqar
\Gamma_m&\equiv&(2p+k-q_m)_\alpha\Gamma^\alpha(k;q_m)
\approx 2E^+ ( \bbox{\epsilon}
\cdot ({\bf k} - {\bf q}_{m})) \;\;,\nonumber \\[1ex]
\Gamma_{mn}&\equiv& (2p+k-q_m-q_n)_\alpha \Gamma^\alpha(k;q_n,q_m)\approx 
2E^+ k^+( \bbox{\epsilon}
\cdot ({\bf k}-{\bf q}_{m}-{\bf q}_{n})) \;\;.
\eeqar{g12}
We neglected above  corrections
of ${\cal O}(x)$. Eq.(\ref{g12}) also applies approximately if
the emission occurs while the jet is off shell with $p-Q$,
with $Q^\mu=(0,\delta{\bf q})$ total momentum transfer
still to be acquired from subsequent jet re-interactions. Replacing
$p$ by $p-Q$ adds for example $2(k^+)^2\bbox{\epsilon}\cdot{\bf q}$
which is again ${\cal{O}}(x=k^+/E^+)$ smaller than the terms retained above. 

The full triple glue vertices including coupling and color algebra 
for producing a final color $c$ and initial color $a$
followed by color potential interactions $a_m$ and $a_m$
are then given by
\beqar
\Lambda_m &\equiv& \Gamma_m (-f^{ca_m a})(ig_st_a)(T_{a_m}(m))
\nonumber \\[1ex] 
&\approx&   \; -2g_s E^+{ \bbox{\epsilon}}
\cdot({\bf k}-{\bf q}_{m}) [c,a_m] T_{a_m}(m) \;\;,  \nonumber \\[1ex]
\Lambda_{mn}&\equiv&  \Gamma_{mn}
(-f^{ca_n e})(-f^{ea_m a})(ig_st_a) (T_{a_n}(n))(T_{a_m}(m))
\nonumber \\[1ex]
&\approx& 
-2ig_s E^+ k^+{  \bbox{\epsilon}}
\cdot({\bf k}-{\bf q}_{1}-{\bf q}_{2}) [[c,a_n],a_m] 
(T_{a_n}(n)T_{a_m}(m)) \;\;.
\eeqar{lam2}

To complete the Feynman rules for our problem,
we note that there is a factor $$iJ(p+k-Q)e^{i(p+k-Q)x_0}$$ 
for the jet production vertex
at $x_0^\mu$ if the net subsequent target momentum transfer
to the jet plus gluon system is $Q$. 
The hard jet radiation amplitude to emit a gluon with
momentum, polarization, and color $(k,\epsilon,c)$ without
final state interactions is then 
\beqar
M_0&=&iJ(p+k)e^{i(p+k)x_0}(ig_s)(2p+k)_\mu\epsilon^\mu(k) 
i\Delta(p+k)c \nonumber \\[1ex]
&\approx& J(p+k)e^{i(p+k)x_0}(-2ig_s)
\frac{\bbox{\epsilon}\cdot{\bf k}}{k^2}c
\approx J(p)e^{ipx_0}(-2ig_s)
\frac{\bbox{\epsilon}\cdot{\bf k}}{k^2} 
\; e^{i\omega_0z_0}\; c \;\;,
\eeqar{m0}
corresponding to Eqs.~(9,10) in Ref.~\cite{GLV1B}.

\subsection{Graphical shorthand}

In Ref.~\cite{GLV1A} we introduced the following notation, $M_{n_s,m,l}$
for the direct (tagged) $n_s$ scattering center amplitude for emitting a gluon
between points $z_m$ and $z_{m+1}$ with a final state interaction
pattern encoded by $l=\sum_1^{n_s} \sigma_i 2^{i-m-1}$,
with $\sigma_i=0,1$ depending on whether the interaction was with the jet 
of the gluon. To include virtual double Born corrections,
it convenient to associate a more specific graphical coding
to keep track of the many different possibilities..

A graph consists of a gluon emission vertex, $G_m$, for emitting
a gluon between centers $z_m$ and $z_{m+1}$ as before, but
also a specific set of direct interactions, $X_{i,\sigma_i}$ with
$\sigma_i=0,1$ for center $i$, and double Born interactions denoted,
 $O_{j,a_j}$, for center $j$. The index  $a_j =0,1,2$ denotes a contact  
interaction at center $j$ with the jet, gluon, and both 
jet+gluon respectively. In this notation
\beq
M_{n,m,l}=  \left[ \, \prod_{i=0}^m X_{i,0} \right]  \, G_m
 \left[ \, \prod_{j=m+1}^n X_{j,\sigma_j} \right]
\label{mnml} \;\; , 
\eeq
where $\sigma_j=0,1$ is the j$^{\rm th}$ binary bit in the expansion
of $l\times 2^{m+1}$. For example, the graph corresponding to gluon 
emission before the first scattering center followed by a jet interaction 
with center ``1'' and the gluon rescattering at center ``2'' is  given by
$M_{2,0,2}=G_0 X_{1,0} X_{2,1}$. This is simply an algebraic way of 
representing the corresponding graph. Similarly, $M_{0,0,0}=G_0$ and
$M_{1,0,1}=G_0 X_{1,1}$. This notation is particularly useful 
to specify  more complicated diagrams that arise from multiple
direct and double Born interactions.  It extends the algebraic approach
to the topologically distinct graphs of Ref.~\cite{GLV1B}  by the inclusion of
virtual corrections. Any diagram can be written in the form 
\beq
M = \left[ \, \prod_{i=0}^m T_{i,\alpha_i} \right] \, G_m  
\left[  \, \prod_{i=m+1}^n T_{j,\beta_j} \right] \, , 
\quad  T_{i,\alpha_i}, T_{i,\beta_i} \in 
\left(  X_{i,\sigma_i}, O_{i,a_i}  \right) 
\eeq{graphgen}

While Eq.~(\ref{graphgen}) can be used to enumerate all ``single gluon
emission with rescatterings'' diagrams arising from a target with $n$
aligned centers, it will prove convenient to group diagrams into
classes of graphs that can be iteratively built from a diagrammatic
kernel. The two important kernels correspond to the two distinct
possibilities for preparing an off-shell parton.  In the first case
the jet originates asymptotically {\em at infinity} and then enters
the medium.  It needs at least one inelastic interaction to get
off-shell. This is the more extensively studied Gunion-Bertsch (GB)
limit~\cite{BDMPS1,BDMPS2,BG}.  Depending on whether the first
interaction in the plasma is {\em real} or {\em virtual} 
\beq 
{\rm Ker}^{(GB)} = \left\{ \begin{array}{ll} B_1^{(R)} = - 2 i g_s \,
\bbox{\epsilon} \cdot \left( \frac{ {\bf k}}{{\bf k}^2} - \frac{{\bf
k} - {\bf q}_{1} } { ({\bf k} -{\bf q}_{1})^2 } \right) \, e^{i
\omega_0 z_1 } \, [c,a_1] \; , & \qquad {\rm Real\;\; initiator}
\\[2ex] B_1^{(V)} = - 2 i g_s \, \frac{C_A}{2} \, \bbox{\epsilon}
\cdot \left( \frac{ {\bf k}}{{\bf k}^2} - \frac{{\bf k} - {\bf q}_{1}
} { ({\bf k} -{\bf q}_{1})^2 } \right) \, e^{i \omega_0 z_1 } \, c \;
, & \qquad {\rm Virtual \;\;initiator} 
\\ \end{array} \right.
\eeq{bgcur}
 is the effective color current~\cite{MGXW,BDMPS1} (here
given in the eikonal approximation in the small $x$ limit). In
Eq.~(\ref{bgcur}) $c$ and $a_1$ are the color matrices of the radiated
gluon and the soft momentum transfer.

In the second and more 
experimentally relevant case,
 the parton is produced {\em inside} 
the medium (e.g., through  $A+A \rightarrow q +\bar{q} +X$ with high
$Q^2 \equiv
E^{+2}$) with accompanying gluon  radiation. 
The kernel or initiator for the hard production vertex 
is
\beq
{\rm Ker}^{(H)} = G_0 =  - 2 i g_s \, { \bbox{\epsilon} \cdot
{\bf k} \over  {\bf k}^2 } \, 
e^{i \omega_0 z_0 } \, c \;, 
\eeq{hvert}
where $c$ the color (generator) 
of the radiated the gluon.
 Classes
for both {\em Hard} and {\em Gunion-Bertsch} cases can 
be constructed starting from the appropriate kernel.
We focus here the hard jet case
relevant to nuclear collisions 
with the hard production
vertex localized at $z_0$, as in Eq.~(\ref{hvert}).
We consider the effect of final state interactions
 at positions $z_i > z_0$ along the direction of the jet.
 
Let ${\cal A}$ denote a class of graphs with $N_{\cal A}$ members in
which the last interaction has occured at position $z_j < z_i$.  Each
class includes one special graph, denoted ${\cal A}_{0}$, where the
gluon is emitted after all real and virtual interactions in that
class.  We can enlarge this class of graphs to include a {\em direct}
hit at $z_i$ and label it by ${\cal A}D_i$, where $D_i$ specifies the
direct insertion iteration. This new class contains $2N_{\cal A}$
diagrams ${\cal A} X_{i,0}$ and ${\cal A} X_{i,1}$ obtained by a
direct interaction of either the jet or the previously emitted
gluon. In addition, there is new special diagram, ${\cal A}G^{-1}
X_{i,0}G_i=({\cal A}D_i)_0$, where all interactions (direct or
virtual) including the one at $z_i$ are with the jet and the gluon is
emitted after all interactions at $z_i < z < \infty$.  ($G^{-1}$
amputates the gluon emission vertex of ${\cal A}_{0}$.) The new class
with $2N_{\cal A}+1$ graphs is constructed operationally as 
\beq {\cal
A}\;\; \Rightarrow \;\; {\cal A}D_i = {\cal A}X_{i,0} + {\cal
A}X_{i,1} + {\cal A} G^{-1} X_{i,0}G_i 
\;\;, \eeq{dirit} 
Starting with
${\cal A}=G_0$, ${\cal A}D_1\cdots D_{n_s}$ is then the set of
$2^{n_s+1}-1$ direct interactions with centers ``1'' through ``$n_s$''
are constructed in~\cite{GLV1B}.

Similarly we can consider the possibilities that arise from inserting
a {\em double Born} hit at location $z_i$. This case differs from the
above only in that now a new subclass appears because one of the legs
can be attached to the jet line and the other one to the gluon line,
i.e.  ${\cal A}O_{i,2}$. The 
new class ${\cal A}V_i$ with $3N_{\cal A} +1$ graphs
including these virtual interactions at center $i$
are constructed  operationally as 
 \beq {\cal A}\;\; \Rightarrow
\;\; {\cal A}V_i = {\cal A}O_{i,0} + {\cal A}O_{i,1} + {\cal A}O_{i,2}
+ {\cal A}G^{-1}O_{i,0}G_i \;\;.  \eeq{virit} 

The classes and conjugate classes
contributing to the gluon radiation distribution
in the opacity expansion at order 
$(L/\lambda)^n \propto (\sigma_{el}/A_\perp)^n$  for
$n=0,1,2$ are listed in the table below.
Each class is obtained through 
ordered insertions
from the hard kernel ${\rm Ker}^{(H)}$ indicated 
by $G_0$.  
\begin{enumerate}
 \item $(L/\lambda)^0 \propto (\sigma_{el}/A_\perp)^0$ contributions
   \begin{itemize}
     \item 1 $(n_s=0)\times(n_s=0)^\dagger$    
        \beqar
           G_0 G_0^\dagger \;&\;&\; 1\;{\rm graph} 
        \label{h0} \eea
  \end{itemize} 
 \item $(L/\lambda)^1 \propto (\sigma_{el}/A_\perp)^1$ contributions
   \begin{itemize}
       \item 4 $(n_s=2)\times(n_s=0)^\dagger$
         \beqar
           G_0 V_1 G_0^\dagger \;&\;&\; 4\;{\rm graphs}
         \label{h1_1} \eea
       \item 9 $(n_s=1)\times(n_s=1)^\dagger$
         \beqar
          G_0 D_1 D_1^\dagger G_0^\dagger \;&\;&\; 9\;{\rm graphs}
          \label{h1_2} \eea
    \end{itemize} 
 \item $(L/\lambda)^2 \propto (\sigma_{el}/A_\perp)^2$ contributions
    \begin{itemize}
       \item 13 $(n_s=4)\times(n_s=0)^\dagger$
         \beqar
           G_0 V_1 V_2 G_0^\dagger \;&\;&\; 13\;{\rm graphs}
           \label{h2_1} \eea
       \item 57 $(n_s=3)\times(n_s=1)^\dagger$
         \beqar
          G_0 D_1 V_2 D_1^\dagger G_0^\dagger \;&\;&\; 30\;{\rm graphs} 
          \label{h2_2}     \\
          G_0 V_1 D_2 D_2^\dagger G_0^\dagger \;&\;&\; 27\;{\rm graphs}
        \label{h2_3} \eea 
       \item 65 $(n_s=2)\times(n_s=2)^\dagger$
         \beqar
       G_0 D_1 D_2 D_2^\dagger D_1^\dagger G_0^\dagger 
 \;&\;&\; 49\;{\rm graphs} 
        \label{h2_4}   \\
       G_0 V_1 V_2^\dagger G_0^\dagger \;&\;&\; 16\;{\rm graphs}
        \label{h2_5} \eea 
   \end{itemize} 
\end{enumerate}
The left column sorts contributions in terms of
diagrams with fixed number of interactions $n_s, n_s^\prime$
in the amplitude and complex conjugate amplitude. For example,
there are 57 contributions involving amplitudes with 3 interactions
times complex conjugate aplitudes with 1 interaction
that contribute to second order in opacity. Since two of the three
interactions in the amplitude must involve the same center,
these 57 diagrams subdivide naturally into the two class structures
(\ref{h2_2},\ref{h2_3}) indicated on the right column.
We note that (\ref{h0}), (\ref{h1_2}) and (\ref{h2_4}) correspond
to the direct diagrams 
computed in~\cite{GLV1B} for the exclusive tagged spectrum.

\subsection{Diagrammatic calculus}

Whereas the time-ordered perturbation
 techniques~\cite{BDMPS1}
used in~\cite{GLV1B} greatly facilitated the 
listing of higher order 
contributions, it is useful to
 show how those shortcuts emerged from Feynman diagrams
especially to check the shortcut rules that apply
to the {\em double Born} contributions. Detailed diagramatic
evaluation from which those rules emerge
are presented
in Appendix~A through Appendix~E.
They exhaust all possibilities
that we can encounter for single and double Born interactions in a 
potential model with multiple centers. It is important  to note that
a double Born interaction at position $z_i$ can be thought of as
the ``contact'' limit of two direct hits at positions 
$z_i$ and $z_{i+1}$ respectively   
\beq
{\cal A} O_{i,a_i} \;\;  \propto \;\; \lim_{z_{i+1} \rightarrow z_i} 
  {\cal A} X_{i,b_i} X_{i+1,b_{i+1} } \;\;.
\eeq{contlim} 
We summarize here the kinematical
and color parts of this diagrammatic calculus:
\begin{itemize}
  \item In all diagrams the jet production factors out in the
     form $$J(p)\, e^{i
px_0}$$ under the soft-gluon  soft-interaction 
     assumption Eq.~(\ref{softness}). Therefore the jet 
     distribution can be factored  out under the assumption
     that $J$ is broad.  
   \item For graphs involving direct interactions only, i.e. of class 
      $G_0D_1D_2 \cdots D_i \cdots $, the Feynman diagrams
      reproduce the time ordered perturbation results listed
       in Ref.~\cite{GLV1B}. The simplest one interaction example is
       discussed in Appendix~A. We note that although Appendix~B 
       through Appendix~E are aimed at clarifying the contact limits,
        their intermediate results    in the well-separated 
       $z_{i+1}-z_i=\lambda \gg 1/\mu$ case 
        are directly comparable to~\cite{GLV1B}.
   \item A diagram which involves a contact contribution of the form
         ${\cal A} O_{i,0}$ or ${\cal A} O_{i,1}$, i.e. both interactions 
         are on a single line, 
       is kinematically identical to a diagram in the well-separated
         case , $\propto X_{i,0}X_{i+1,0}$ and  
          $\propto X_{i,1}X_{i+1,1}$ respectively, but with
         $z_i=z_{i+1}$~(\ref{contlim}).  This can be seen in 
   from Appendix~B for the final state virtual 
        interaction of the radiated gluon and in Appendix~C for a virtual 
         final state scattering of the jet.     
           There is, however,  a {\em multiplicative} factor of 
         $\half$ per contact interaction that emerges from 
         contour integrals. Eq.~(\ref{i3a2})
 shows  how the strength interpolates 
between 1 and $\half$ as the separation varies for a  Yukawa
potential. Eq.~(\ref{lims}) shows 
the generality of the factor $\half$ in the
contact limit.

   \item For the case  when one of the legs in the  contact contribution 
         is attached to the jet line and the other to the gluon line, 
         i.e. ${\cal A} O_{i,2}$, the amplitude 
is  kinematically identical 
         to a diagram 
         ${\cal A} X_{i,0}X_{i+1,1}$  in the well-separated case 
         with  $z_i=z_{i+1}$~(\ref{contlim}). 
         There are {\em no} additional numerical factors for such mixed
       contact term   as shown in Appendix~D. We also note that the 
         contact limit of the diagram 
         $\propto X_{i,1}X_{i+1,0}$ should not be added to 
          avoid over counting 
         (i.e,  they are topologically identical in the contact limit). 
    \item We show in Appendix~E that the class of diagrams 
          that have the gluon emission vertex 
          {\em between} the two momentum  transfers of a double Born 
          interaction are suppressed in the ``contact'' limit by  factors
          ${\cal O}(k^+/E^+)$ and  ${\cal O}(\mu/k^+)$ and thus 
          can be neglected in the framework of our  
           approximations~(\ref{eorder}). This verifies
 the naive argument from time-ordered perturbation theory  that 
        such diagrams are of zero-measure and vanish in the contact
limit.
    \item Each power of the potential comes with a weight 
       (elastic scattering amplitude)   
       $$ (-i)\int\frac{d^2 {\bf q}_{i}}{(2\pi)^2} 
        \,v(0,{\bf q}_{i}) \,
           e^{-i{\bf q}_{i}\cdot{\bf b}_{i}} T_{a_i}(i) \;\;, $$    
        that factors out as well under the assumptions 
         (\ref{softness},\ref{eorder}) and is still to be integrated over.
        The amplitudes above eventually build into the elastic scattering 
        cross section $\propto \sigma_{el}$.  
    \item The  appropriate color factor for the jet+gluon part can be 
          constructed following Ref.~\cite{GLV1B}. In the case of 
          a virtual interaction certain simplifications, namely
\beq      
 a_i a_i = C_R{\bf 1}, \quad a_i [c, a_i] = 
       - \frac{C_A}{2} \,c \;\;,
\eeq{color}
 can occur even at the amplitude 
       level.         
     \item The last critical rule that emerges is
that  {\em direct} terms come with a {\em plus} sign while
each  {\em virtual} interaction contributes 
     a {\em minus}.
\end{itemize}

These rules are illustrated to compute the classes of amplitudes
needed up to second order in opacity  in Appendix F.

\subsection{Impact parameter averages}

Consider next the averaging  over the transverse
impact parameters ${\bf b}_i={\bf x}_i-{\bf x}_0$. 
A typical term that appears to second order in opacity 
is $G_0X_{1,0} O_{2,2} X_{1,0}^\dagger G_0^\dagger$.
This involves one direct interaction with center ``1'' and
a double Born with center ``2''. We consider separately the average over
${\bf b}_i$ involving
direct and virtual interaction centers.

For a scattering center, $i$, that appears in a direct interaction
the  impact parameter average takes the form 
\beqar
\langle \cdots \rangle_{A_\perp} &=& \langle \cdots
\int \frac{d^2{\bf b}_{i}}{A_\perp} \,
 (-i)\int\frac{d^2 {\bf q}_{i}}{(2\pi)^2} \,v(0,{\bf q}_{i}) \,
 e^{-i{\bf q}_{i}\cdot{\bf b}_{i}} \, 
(+i)\int\frac{d^2 {\bf q}_{i}^{\prime}}{(2\pi)^2} 
\,v^*(0,{\bf q}_{i}^{\prime}) \, e^{+i{\bf q}_{i}^{\; \prime}
\cdot{\bf b}_{i}}  \; \cdots \rangle \; \nonumber \\
&=& \cdots \int \frac{d^2 {\bf q}_{i}}{(2\pi)^2} \,
\frac{d^2 {\bf q}_{i}^{\prime}}{(2\pi)^2} 
\frac{(2\pi)^2 \delta^2 ({\bf q}_{i} - 
{\bf q}_{i}^{\prime})}{A_\perp} 
v(0,{\bf q}_{i})v(0,{\bf q}_{i}^{\prime}) \; \cdots 
\;\; \nonumber \\
&=& 
\cdots \frac{\sigma}{A_\perp} 
\int \frac{d^2 {\bf q}_{i}}{(2\pi)^2} |\bar{v}({\bf q}_i)|^2 \,
\; \int d^2 {\bf q}_{i}^{\prime} 
\delta^2 ({\bf q}_{i} - 
{\bf q}_{i}^{\prime}) 
\; \cdots \, \;\; ,
\label{travd}
\eea
where $|\bar{v}|^2$ is defined as the normalized
distribution of momentum transfers from the scattering centers.
In terms of Eq.~(\ref{sigel}), it is given by
\beq
|\bar{v}({\bf q})|^2\equiv \frac{1}{\sigma_{el}}
\frac{d^2 \sigma_{el}}{d^2{\bf q}}=\frac{1}{\pi}
\frac{\mu_{eff}^2}{({\bf q}^{2}+\mu^2)^2}
\;\; ,
\eeq{vbar}
where in the Yukawa example,
 the  normalization depends on the kinematic bounds through
\beq
\frac{1}{\mu_{eff}^2}=\frac{1}{\mu^2}-\frac{1}{{\bf q}_{\max}^2+\mu^2}
\;\; ,\eeq{mueff}
and insures that $\int^{{\bf q}_{\max}} d^2{\bf q} |\bar{v}({\bf q})|^2=1.$
In numerical estimates we take ${\bf q}_{max}^2=s/4\approx 3E\mu$.
For asymptotic energies $\mu_{eff}\approx \mu$.
Note also that unlike $v$, the barred 
$\bar{v}$ is independent of color factors
and thus applies to both quark and gluon rescattering.

If the interaction with center $i$ is
 a double Born rather than a direct one,
then the only change relative to (\ref{travd}) is that
both ${\bf q}_i$ and ${\bf q}_i^\prime$
come with the same $(\pm i)^2 =-1$ rather
than the $(-i)(+i)=1$ factor and the phase shifts
change relative sign. If the double Born interaction is in the amplitude,
as in the example noted above, then the impact parameter average
leads to 
\beqar
\langle \cdots \rangle_{A_\perp} &=&  \langle \cdots
\int \frac{d^2{\bf b}_{i}}{A_\perp} \,
 (-i)\int\frac{d^2 {\bf q}_{i}}{(2\pi)^2} \,v(0,{\bf q}_{i}) \,
 e^{-i{\bf q}_{i}\cdot{\bf b}_{i}} \, 
(-i)\int\frac{d^2 {\bf q}_{i}^{\prime}}{(2\pi)^2} 
\,v(0,{\bf q}_{i}^{\prime}) \, e^{-i{\bf q}_{i}^{\; \prime}
\cdot{\bf b}_{i}}  \; \cdots \rangle\; \nonumber \\
&=& \cdots (-1) \int \frac{d^2 {\bf q}_{i}}{(2\pi)^2} \,
\frac{d^2 {\bf q}_{i}^{\prime}}{(2\pi)^2} 
\frac{(2\pi)^2 \delta^2 ({\bf q}_{i} + 
{\bf q}_{i}^{\prime})}{A_\perp} 
v(0,{\bf q}_{i})v(0,{\bf q}_{i}^{\prime}) \; \cdots 
\;\; \nonumber \\
&=& 
\cdots (-1) \frac{\sigma}{A_\perp} 
\int \frac{d^2 {\bf q}_{i}}{(2\pi)^2} |\bar{v}({\bf q}_i)|^2 \,
\; \int d^2 {\bf q}_{i}
 \delta^2 ({\bf q}_{i} + 
{\bf q}_{i}^{\prime}) 
\; \cdots \, \;\; .
\label{travv}
\eea
The same holds true if the double Born is in the conjugate amplitude.
The results  above  is  the basis for the last graphical
rule of the last section. A contribution to $n^{\rm th}$ opacity
involving $m$ ``virtual''
hits produces a factor
$(-1)^m (\sigma/A_\perp)^n$. 
 
A key to the analytic derivation in the next section 
 is that the transverse impact parameter
averages in Eqs.~({\ref{travd},\ref{travv})
{\em  diagonalize} the products of amplitudes and complex
conjugate amplitudes in the ${\bf q}_i, {\bf q}_i^\prime$
variables. 

The question of {\em which} cross section $\sigma$ enters above
depends on the details of the color algebra. At this point it is not
at all clear since any particular contribution to a given order in
opacity has an extremely complicated color structure in
general~\cite{GLV1B}.  However, as we prove in the next section the
color algebra simplifies drastically in the sum over all possible real
and virtual contributions~\cite{BDMPS2,ZAH}.  The color ``triviality''
of single inclusive distributions proved in the next section via the
reaction operator approach, states that after summing all
contributions, each integration over ${\bf q}_i$ is accompanied simply
by a factor $(C_AC_2(T)/d_A)$. This implies that $\sigma=\sigma_g$ is
the {\em gluon} elastic jet cross section that appears in both
Eqs.~(\ref{travd},\ref{travv}), rather than the jet one
$(\propto C_R C_2(T))$ .  Recall that 
\beq
\sigma_g=\frac{C_A}{C_R} \sigma_{el}(R,T) \eeq{sigmag}
 and hence 
\beq
C_A \lambda_g = C_R \lambda
\;\; . \eeq{mfprelat}
Color triviality implies that
the opacity expansion is in  terms of powers
of the gluon rather than the jet elastic cross section,
i.e. $\propto (L/\lambda_g)^n$.

\section{Recursive method for  opacity expansion}

The inclusive gluon distribution to 
${\cal O} ((L/\lambda)^1 \propto (\sigma_{el}/A_\perp)^1)$ 
is a sum of $3^2$  direct and  $2\times 4$  virtual cut diagrams. 
If we want to proceed to second order in opacity
$ {\cal O} ( (L/\lambda)^2 \propto (\sigma_{el}/A_\perp)^2)$ 
there are  $7^2$ direct plus $2\times 86$ contributions. It is 
therefore useful to develop an recursive  procedure for writing the sum of
amplitudes in a certain class of diagrams. We will ignore the jet 
and elastic scattering  parts of the diagrams  which leaves
us with simpler amplitudes similar to the ones discussed in 
Refs.~\cite{GLV1A,GLV1B}. We recall the definitions of the 
Hard, Gunion-Bertsch and Cascade terms
\beqar
{\bf H}&=&{{\bf k} \over {\bf k}^2 }\; , \qquad \qquad
{\bf C}_{(i_1i_2 \cdots i_m)}={({\bf k} - {\bf q}_{i_1} - 
{\bf q}_{i_2}- 
\cdots -{\bf q}_{i_m} ) 
\over ({\bf k} - {\bf q}_{i_1} - 
{\bf q}_{i_2}- 
\cdots -{\bf q}_{i_m}   )^2 } \;, 
\nonumber \\[1.ex]
{\bf B}_i &= &{\bf H} - {\bf C}_i \; , \qquad
{\bf B}_{(i_1 i_2 \cdots i_m )(j_1j_2 \cdots i_n)} = 
{\bf C}_{(i_1 i_2 \cdots j_m)} - {\bf C}_{(j_1 j_2 \cdots j_n)}\;\; .
\eeqar{hbgcdef}

\subsection{Amplitude iteration}

There are two basic iteration steps that one has to consider
to construct the inclusive distribution of gluon.  
The first one represents  the addition of
a ``direct'' 
interaction  $D_n$ that
changes the color or momentum of the target parton
located at $z_n$ as  illustrated by Fig.~2.  
\begin{center}
\vspace*{7.9cm}
\includegraphics{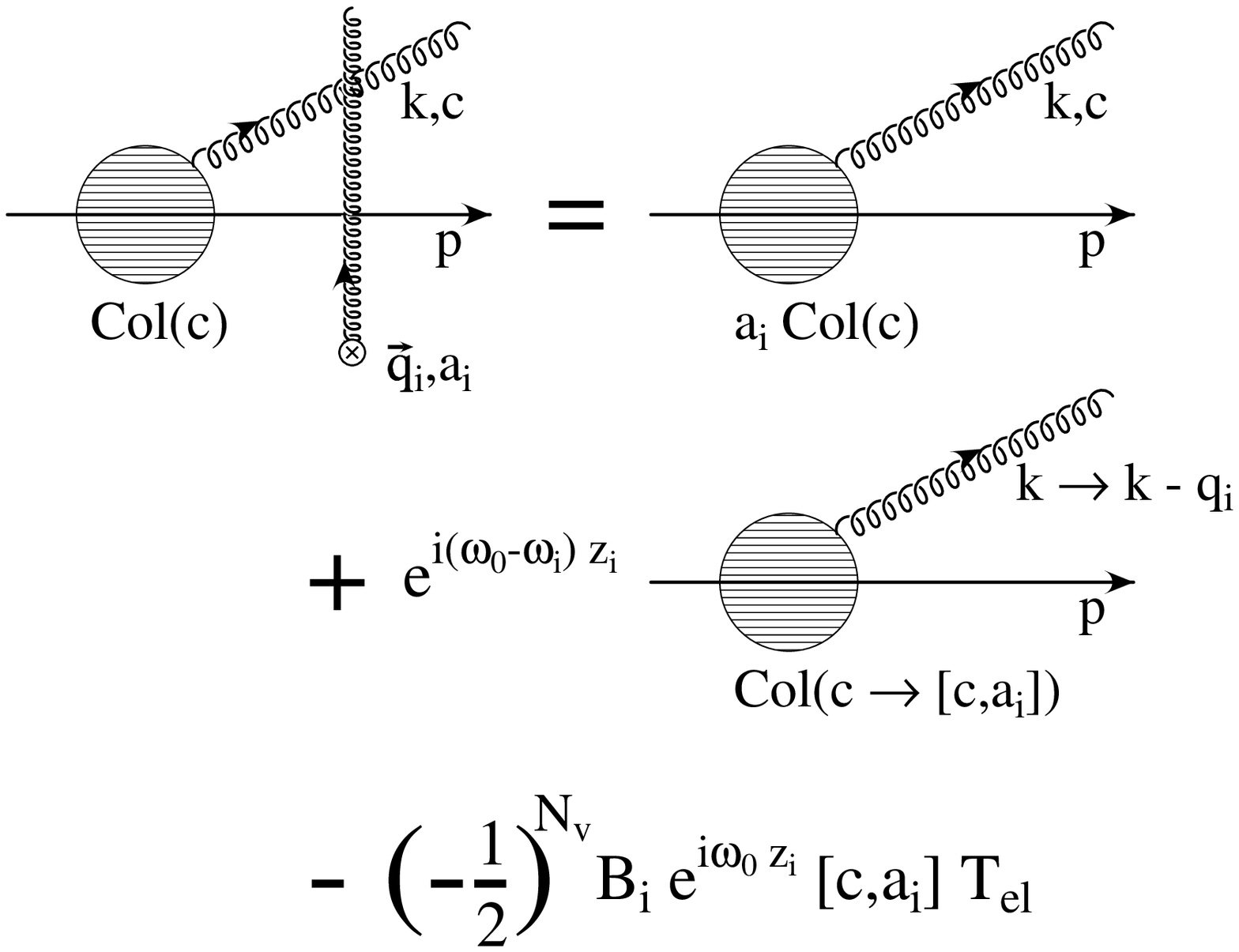}  
\vskip -0pt
\begin{minipage}[t]{15.0cm}
{\small {FIG~2.}
Diagrammatic representation of  the sum of amplitudes  generated by 
${\cal A}D_i = {\cal A}X_{i,0} + {\cal A}X_{i,1}  + 
{\cal A}G^{-1}X_{i,0}G_i$.} 
\end{minipage}
\end{center}
\vskip 4truemm
The second iteration step  corresponds to a double Born ``virtual''
interaction $V_n$ that leaves both the color and momentum of
the target parton unchanged and  is illustrated in Fig.~3.

Our goal here is to construct operators $\htD_n, \htV_n$
 that can be used
to construct recursively the new amplitudes.

It is important to note that each new diagram produces 
a  difference of {\em two} phase factors with the exception of 
one  special diagram, 
$({\cal A}D_n)_0\equiv {\cal A}G^{-1}X_{n,0}G_n$,  corresponding
to the emission of the gluon after, $z_n$. The
upper limits of emission is $z=\infty$ in that case,
and only the lower bound phase factor, $-e^{i\omega_0z_n}$
survives on account 
of the adiabatic damping factor   $\exp(-\epsilon |z|)$.
Therefore all but one  diagram  can be divided into two 
parts which can then be recombined
into more easily interpretable 
Hard, Gunion-Bertsch and Cascade amplitudes~(\ref{hbgcdef}) 
as in ~\cite{GLV1A,GLV1B} . 

The sum of amplitudes in  class ${\cal A}$ can be denoted by
\begin{equation}   
{\cal A}(x,{\bf k},c)\equiv \sum_\alpha 
{\cal A}_\alpha(x,{\bf k}) Col(c)_\alpha
\;\; ,
\end {equation}
where ${\cal A}_\alpha(x,{\bf k})$ 
represents the kinematical part, $Col(c)_\alpha$ stands for the color
matrix for the distinct graphs in this class
enumerated by $\alpha$. There is of course considerable
freedom  in rearranging the terms of this sum.
Since by definition classes are constructed
by repeated operations of either of one of three
operations, $\hat{\bf 1},\htD_i,\htV_i$
it is convenient to enumerate the $3^n$ different classes of diagrams
via a tensor
notation, ${\cal A}_{i_1\cdots i_n}$, where
the indices $i_j=0,1,2$ specify whether there was no, a  direct,
or a virtual interaction with  the target parton at $z_j$.
The sum of amplitudes for class, ${\cal A}_{i_1\cdots i_n}$,
can be written explicitly in the form
\beq
\vAi(x,{\bf k},c)=\prod_{m=1}^n
\left( \delta_{0,i_m} + \delta_{1,i_m} \htD_m + \delta_{2,i_m} 
\htV_m
\right) G_0(x,{\bf k},c)
\; \; . \eeq{atens}
Here $G_0$ is the color matrix amplitude that corresponds
to the initial source or kernel of the jet and gluon in the limit
of no final state interactions.

The inclusive induced ``probability'' distribution
at order $n$ in the opacity expansion
can then be computed
from the following sum of products over the $3^n$ classes that
contribute at that order:
\beq
P_n(x,{\bf k})=\vAbi(c)\vAi(c)\equiv
\tr \sum_{i_1=0}^2\cdots \sum_{i_n=0}^2
\bar{\cal A}^\dagger_{i_1\cdots i_n}(x,{\bf k},c)    
A_{i_1\cdots i_n}(x,{\bf k},c) 
\;\; ,\eeq{pnten}
where  the unique {\em complementary} class that contracts
with  $\vAi$ is defined by
\beqar
\bar{\cal A}_{i_1\cdots i_n}(x,{\bf k},c)&\equiv& \prod_{m=1}^n
\left( \delta_{0,i_m} \hat{V}_m + \delta_{1,i_m} \hat{D}_m 
+ \delta_{2,i_m} 
\right) G_0(x,{\bf k},c) \;\;,  \nonumber \\
\vAbi(x,{\bf k},c)&\equiv& G_0^\dagger
(x,{\bf k},c) \prod_{m=1}^n
\left( \delta_{0,i_m} \hat{V}_m^\dagger + 
\delta_{1,i_m} \hat{D}_m^\dagger 
+ \delta_{2,i_m} 
\right) 
\; \; .\eeqar{atens2}
For example, $\bar{\cal A}_{0210}=\htD_3\htV_2 G_0$ for
${\cal A}_{2012}=\htV_4\htD_3\htV_1G_0$.
Note that $\bar{\cal A}{\cal A}$ includes products of classes and complementary
classes with different powers of the external coupling, $g_s$, 
but every product in the sum is the same order, $\alpha_s^{2n+1}$.
These mixed terms are  the 
unitarity corrections 
to the direct $\bar{\cal A}^{1\cdots 1}{\cal A}_{1\cdots 1}$ term
that contribute to inclusive processes when the target recoils 
are not tagged, i.e. measured exclusively. 
Note that $P_n$ is not positive definite except for $n=0$, and
we refer to it as a ``probability'' only figuratively.

With this tensor classification and construction, it
becomes possible to construct
$P_n$ recursively  from lower rank (opacity) classes 
through the insertion of a  ``reaction'' operator as follows:
\beq
P_n=\bar{\cal A}^{i_1\cdots i_{n-1}}\htR_n {\cal A}_{i_1\cdots i_{n-1}}
\;, \qquad \htR_n = \hat{D}_n^\dagger
\hat{D}_n+\hat{V}_n+\hat{V}_n^\dagger
\; \; . \eeq{rop}
We emphasize that it is possible to write the inclusive
probability in this simple form only because
the transverse impact parameter averages in
Eqs.~(\ref{travd},\ref{travv}) {\em diagonalize} the products
of amplitudes and complementary conjugate 
amplitudes in the ${\bf q}_i,{\bf q}_i^\prime$
variables. If the transverse
area of the target were not large in comparison to the
cross section, then off diagonal components in those variables
would appear that would have to folded with a
suitable transverse Glauber profile functions
 as discussed in~\cite{MGXW}.
The simple algebraic structure of the problem in Eq.~(\ref{rop})
therefore hinges on the validity of the transverse
averaging  via Eqs.~(\ref{travd},\ref{travv}). 

To construct $\hat{D}_n$, consider first the contribution 
${\cal A}X_{n,0}$ from Eq.~(\ref{dirit}).  In this case, a  new direct
 interaction  is attached to the jet line at $z_n$.  For the partial 
sum of  amplitudes in ${\cal A}_{i_1\cdots i_{n-1}}$
corresponding to  emitting the gluon before the last real or virtual
 interaction  at $z_{f}\le z_{n-1}$, 
the  direct interaction of the jet at $z_n$ 
 simply multiplies that partial sum  by $a_n$. No extra phases are
 introduced since the energy of the jet is assumed to be very
 high. 
As noted before,
for every class, there is a term in ${\cal A}_{i_1\cdots i_{n-1}}$
 corresponding  to a graph where the gluon 
is emitted after the last interaction point $z_{f}$. That term 
 has the explicit  form 
\beq 
\left(-\half\right)^{N_v(\vAim)}  {\bf H}
(-e^{i\omega_0 z_{f}} ) \, cT_{el}(\vAim)
\;\; , 
\eeq{lastt}
where the elastic color  matrices for that class  and its complementary 
are denoted by
\beq
T_{el}({\cal A}_{i_1\cdots i_{n-1}})\equiv (a_{n-1})^{i_{n-1}}\cdots (a_1)^{i_1}
\;, \qquad  T_{el}^\dagger(\bar{\cal A}^{i_1\cdots i_{n-1}})
\equiv (a_1)^{2-i_1} \cdots (a_{n-1})^{2-i_{n-1}}
\; \;  . 
\eeq{tel}
For example, $T_{el}({\cal A}_{1021}=
\htD_4\htV_3\htD_1G_0)=a_4 (a_3  a_3)  a_1= C_R a_4 a_1$.
Eq.~(\ref{tel}) implies the following  identity holds for elastic color factors:
\beq
T^\dagger_{el}(\bar{\cal A}^{i_1\cdots i_{n-1}})
T_{el}({\cal A}_{i_1\cdots i_{n-1}}) = C_R^{n-1}
\eeq{elid}
valid for all $3^{n-1}$ components of rank $n-1$ classes.

The factor $(-\half)^{N_v}$ in Eq.~(\ref{lastt}) arises 
because every virtual contact interaction in the class acquires
a factor $-\half$  from the the contact limit
of the contour integration over longitudinal momentum
(see Sec.~II and the Appendix). 
The numbers, $N_v,\bar{N}_v$, of such contact
interactions in class, $\vAim$, and its complementary class, $\vAbim$, are 
\beqar
N_v=N_v(\vAim)=\sum_{m=1}^{n-1} \delta_{2,i_m}\;,  \qquad
\bar{N}_v=N_v(\vAbim)=\sum_{m=1}^{n-1} \delta_{0,i_m} 
\;\; .
\eeqar{kv}
For example $N_v=2,\bar{N}_v=2$ for ${\cal A}_{01202}=\htV_5 \htV_3 \htD_2G_0$.
A useful bookkeeping  identity  that we will need  is
\beq
\sum_{i_1,\cdots,i_m} 
\left(-\half\right)^{N_v(\bar{\cal A}_{i_1\cdots i_m})}
\left(-\half\right)^{N_v({\cal A}_{i_1\cdots i_m})}
=\left(-\half -\half +1\right)^m=0
\; \; .
\eeq{kvid}

Now consider how a jet interaction at $z_n$ changes
the  graph Eq.~(\ref{lastt}). Besides being multiplied 
by $a_n$, the only other change  is that the phase factor changes 
into a difference of phase factors as
\beq
 -e^{i\omega_0 z_{f}}  \stackrel{X_{n,0}}{
\longrightarrow}
 e^{i\omega_0 z_n} -e^{i\omega_0 z_{f}}   \;\; . 
\eeq{step1} 
The part, $\propto e^{i \omega_0  z_{f}}$, on the right hand side
can be added back to the sum of all the other graphs with
 emission before $z_{f}$. The modified
sum $a_n \vAim$ therefore includes this part of Eq.~(\ref{step1}).
In summary, the component 
of the $\htD_n$ operator that takes into
account the jet rescattering at $z_n$ is simply multiplication
by the color matrix,
$a_n$, in the jet representation.
This then identifies the first part of the $\htD_n$ operator as
\beq
\htD_n= \hat{\bf 1} a_n + \cdots
\;\; .
\eeq{dnpart1}
The left over part  $\propto e^{i\omega_0 z_n}$ of 
 Eq.~(\ref{step1}) can be combined with the  graph $({\cal A}
 G^{-1}X_{n,0}G_n)$, e.g., $M_{1,1,0}, M_{2,2,0}$ in  Appendix A and C,
that accounts for  gluon emission after $z_{n}$.
This  combination leads to the following
extra term in $\vAi$:
\beq 
- \left(-\half\right)^{N_v(\vAim)}\; 
{\bf H} e^{i\omega_0
 z_n} \; [c,a_n] \, T_{el}(\vAim) \;\; . 
\eeq{step2} 
The sum of graphs ${\cal A}(X_{n,0}+G^{-1}X_{n,0}G_n)$ 
is therefore $a_n \vAi$ plus  Eq.~(\ref{step2}). 

To finish 
constructing $\htD_n$, we need to consider 
the effect of an elastic scattering of the gluon produced
at an earlier point in the medium.
The gluon ``cascade'' interaction  graph ${\cal A}X_{n,1}$
(see e.g., $M_{2,0,3}$ in Appendix B)
introduces the following changes in the amplitude:
First an extra  phase shift 
$\exp[i(\omega_0-\omega_n)z_n ]$ arises  due to the difference
of gluon energies before and after the transverse 
momentum exchange, ${\bf q}_{n}$, at $z_n$.
This is in contrast 
to the jet rescattering where that energy change was negligible.
Second, the gluon 
color is rotated by $c \rightarrow [c,a_n]$. 
Finally, the transverse momentum  in ${\cal A}(x,{\bf k}, c)$ is shifted
backward by ${\bf k}\rightarrow {\bf k}-{\bf q}_{n }$
in order that  same final state ${\bf k}$ is reached.
This part of the $\htD_n$ operator therefore simply  transforms
\beq
\vAim({\bf k},c)\; \stackrel{X_{n,1}}{\longrightarrow}
\htS_n \vAim =
 e^{i(\omega_0-\omega_n)z_n } 
 \;  \vAim({\bf k}- {\bf q}_{n},[c,a_n]) \;\;.
\eeq{di2}
  The ``shift'' or gluon scattering operator
that implements  gluon rescattering at $z_n$
is thus defined by 
\beq
\htS_n =if^{ca_n d}\times e^{i(\omega_0-\omega_n)z_n } 
e^{i{\bf q}_{n} \cdot \hat{\bf b}} 
\;\; .
\eeq{ops}
Here, $\hat{\bf b}\equiv i 
\stackrel{\longrightarrow}{\bbox{\nabla}_{\bf k}}$ is  the 
impact parameter operator conjugate to ${\bf k}$ such that
$$ e^{i{\bf q}\cdot\hat{\bf b}} f({\bf k})=f({\bf k}-{\bf q})\; \; .$$
This transverse momentum shift together with a
class independent phase shift   
and a  color rotation (via the structure constants $f^{abc}$)
 enlarges $\vAim$
to include  all but one of the gluon final state
interaction  diagrams with an interaction at $z_n$.

The left over  graph  not
included in~(\ref{di2}) is the one where the jet  scatters elastically
up to previous last interaction point $z_f$, 
emits a gluon with transverse momentum ${\bf k}-{\bf q}_{n}$
between $z_{f}$ and $z_n$, and the gluon scatters at $z_n$
(see e.g., $M_{1,0,1}$ in Appendix A). 
The integral over the emission
interval  leaves a characteristic  differences of phases
$e^{i\omega_n z_n}-e^{i\omega_n z_f}$, while the elastic scattering
introduces  a   phase shift factor $e^{i(\omega_0-\omega_n)z_n}$ as well.
The kinematic part of this cascade amplitude is ${\bf C}_n$. 
 The part proportional to $e^{i\omega_n z_n}$ contributes 
therefore the following extra term
\beq
+ \left(-\half\right)^{N_v}
{\bf C}_n e^{i(\omega_0 -\omega_1)z_n}e^{i\omega_1 z_n} 
[c,a_n]\, T_{el}(\vAim)
\;\; , \eeq{step3}
where again $N_v$ is the number of contact interactions 
before $z_n$.
Recalling that ${\bf B}_n={\bf H}-{\bf C}_n$
has the physical interpretation of a Gunion-Bertsch gluon source 
term at $z_n$, the two
extra terms  from Eq.~(\ref{step3})
and Eq.~(\ref{step2}) can be conveniently combined  to form
\beqar
\htB_n \vAim(c) &\equiv&  - \left(-\half\right)^{N_v(\vAim)}
if^{ca_n d}\times {\bf B}_n\, 
e^{i \omega_0 z_n}\; d\; T_{el}(\vAim) 
\nonumber \\[1ex]
&=&-  \prod_{m=1}^{n-1} \left(-\frac{1}{2} 
\,\right )^{\delta_{2,i_m}}   {\bf B}_n\, 
e^{i \omega_0 z_n} [c,a_n] \; a_{n-1}^{i_{n-1}}\cdots
a_1^{i_1} \;\;.  
\eeqar{di3}  
 
Combination of three steps $X_{n,0}+X_{n,1}+
G^{-1}X_{n,0}G$ can therefore be summarized by the following
operation on rank $n-1$ class elements:
\beqar
 \htD_n \vAim(x,{\bf k},c) &\equiv&   (a_n + \htS_n + \htB_n) 
\vAim(x,{\bf k},c)
 \nonumber \\[1ex]
&=& a_n \vAim(x,{\bf k},c) + e^{i(\omega_0-\omega_n)z_n } 
\vAim(x,{\bf k}- {\bf q}_{n},[c,a_n]) \nonumber \\[1ex]
&\;&  - \left(-\half \,\right )^{N_v(\vAim)} {\bf B}_n \, 
e^{i \omega_0 z_n} [c,a_n] T_{el}(\vAim) \;\;.
\eeqar{didamit}  
We note that the common factor to all diagrams 
$2ig_s \bbox{\epsilon} \cdot (\cdots )$ 
 is suppressed for clarity.

We now turn to the second class of diagrams, $\htV_n\vAim$
created by inserting contact interaction at $z_n$.
This new  class is constructed via
Eq.~(\ref{virit}) and illustrated 
in  Fig.~3.
\begin{center}
\vspace*{8.9cm}
\includegraphics{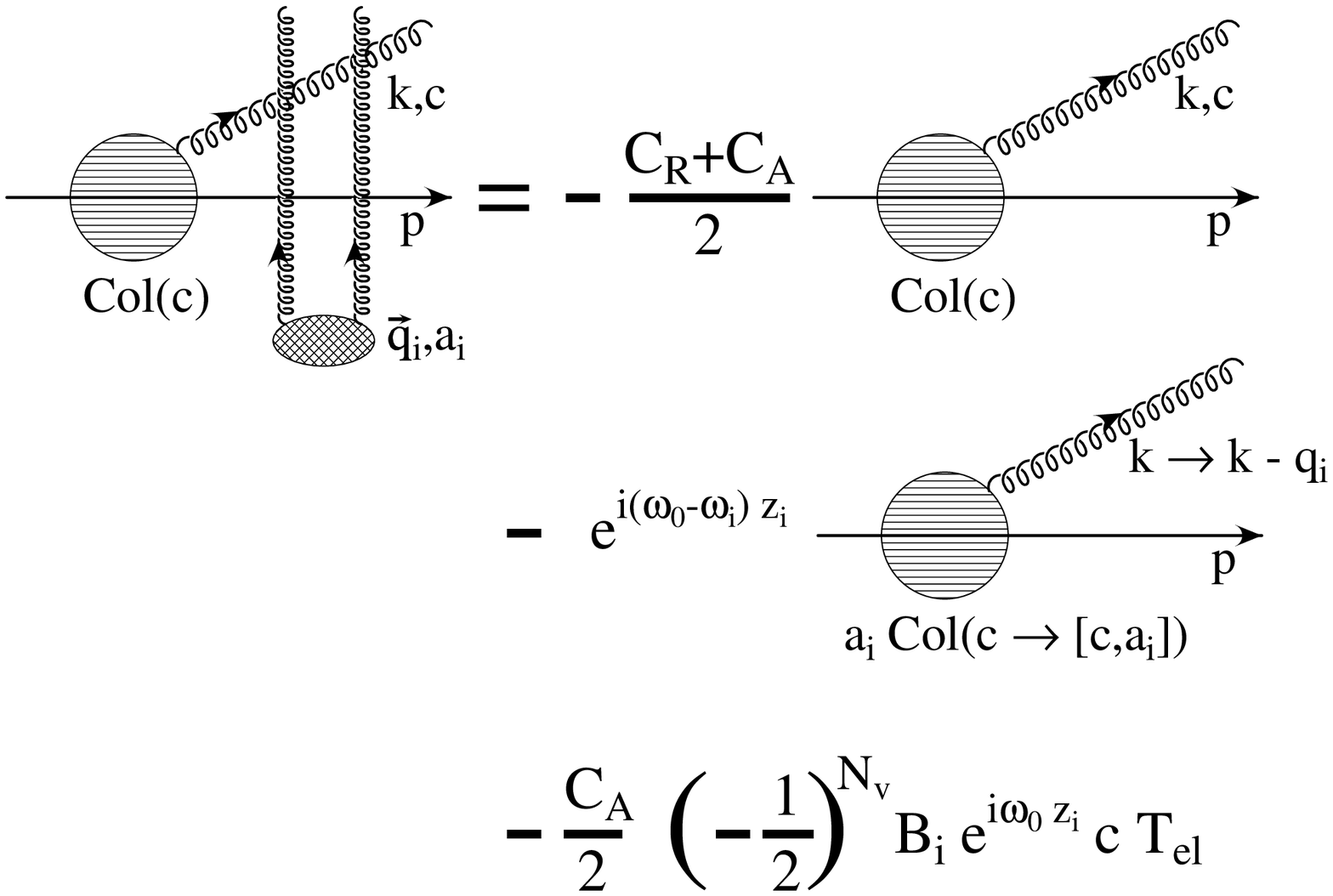}  
\vskip -0pt
\begin{minipage}[t]{15.0cm}
{\small {FIG~3.}
Diagrammatic representation of  the sum of amplitudes, generated by 
${\cal A}V_i = {\cal A}O_{i,0} + {\cal A}O_{i,1}  +  
{\cal A}O_{i,2}  +    {\cal A}G^{-1}O_{i,0}G_i$.} 
\end{minipage}
\end{center}
\vskip 4truemm
For a double Born interaction at $z_n$, 
 the transverse coordinate
integration can be performed 
before multiplying by a complementary
hermitian conjugate amplitudes since the coordinates of a virtual
interaction cannot occur in the complementary class.
The transverse coordinate integration on the center of the
$i^{\rm th}$ interaction enforces (see Eq.~(\ref{bave}))
that the net momentum transfered in the two legs
vanishes  via a factor 
$\delta^2({\bf q}_{i} + {\bf q}^{\prime}_{i})$.

Consider first the sum of graphs  where
 the two legs of the contact  interaction are attached to the 
jet line (${\cal A}O_{n,0}$) (see,e.g. $M^c_{2,0,0}$ in Appendix C). 
The situation is similar 
to Eq.~(\ref{dnpart1}) except that two color factors $a_na_n=C_R$
multiply the rank $n-1$ amplitudes instead of one. In addition
a  factor of $-\half$ appears due to the contact
limit of the contour integration.
Thus, this part of the virtual interaction simply transforms
\beq
\vAim \stackrel{O_{n,0}}{\longrightarrow} 
- \frac{C_R}{2} \vAim
 \;\;.
\eeq{vi1}  
The second term, ${\cal A}O_{n,1}$ in Eq.~(\ref{virit})
is the contact with the gluon at $z_n$ (see $M^c_{2,0,3}$ in
Appendix B).
Since $if^{ca_nd}if^{d a_n e}=C_A\delta_{c,e}$, the color
structure of this term is also trivial. In addition,
since  the total  momentum transfer is now zero,
no  extra phase factor arises and  the gluon momentum 
in $\vAim$ remains unchanged.
The contact gluon interaction at $z_n$ therefore contributes
\beq
\vAim \stackrel{O_{n,1}}{\longrightarrow} 
- \frac{C_A}{2} \vAim
 \;\;.
\eeq{vi2}  
The mixed contact interaction  ${\cal A}O_{i,2}$ (see, e.g. $M^c_{2,0,1}$ 
in Appendix D) is the most subtle.
Even though the net momentum transfer vanishes, in this case
both the jet and gluons receive (opposite) non-vanishing
transverse momentum transfers, $\pm{\bf q}_{n}$. This leads
to a highly non-local  modification of $\vAim$.
The contact limit of the longitudinal momentum transfer
contour integration produces in this case a factor of $-1$
as shown in the Appendix. 
The elastic scattering of the gluon
produces  multiplicative phase factor, $e^{i(\omega_0-\omega_n)z_n } $.
The jet elastic scattering
introduces no new phase in the small $x$ limit as before.
The transverse momentum exchange,  to the gluon requires
that ${\bf k}$ is again shifted back by 
${\bf q}_{n}$ as in Eq.~(\ref{di2}).
The gluon color is also rotated 
as in (\ref{di2}). Furthermore, the scattering of the jet
introduces another color matrix $a_n$.
This mixed contact interaction therefore simply
produces the following variant of~(\ref{di2}):
\beq
\vAim({\bf k},c)\; \stackrel{O_{n,2}}{\longrightarrow}
a_n\htS_n \vAim = e^{i(\omega_0-\omega_n)z_n } 
 \;  a_n \vAim({\bf k}- {\bf q}_{n},[c,a_n]) \;\;,
\eeq{dv3}
with again one extra term that can be
recombined with the  amplitude of  gluon radiation
after the contact interaction at $z_n$ to form a
virtual Gunion-Bertsch source
\beq
-a_n \htB_n\vAim= - \left(- \half\,\right)^{N_v(\vAim)} \frac{C_A}{2} 
 \, {\bf B}_n \, e^{i \omega_0 z_n} c T_{el}(\vAim) \;\;,
\eeq{vi3}  
where we used $a_n[c,a_n]=-\half C_A c$  to simplify  the color algebra.

The double Born interaction at $z_n$  can therefore
be implemented by the following operator
\beqar
\htV_n  &=&  -\half(C_A+C_R) - a_n \htS_n- a_n\htB_n
= -a_n \htD_n - \half(C_A-C_R)  \; \; .
\eeqar{dvid} 
Specifically, the virtual iteration of a rank $n-1$ class gives
\beqar
\htV_n \vAim(x,{\bf k},c)
&= & - \frac{C_R+C_A}{2} \vAim(x,{\bf k},c)
-  e^{i(\omega_0-\omega_n)z_n } a_n
\vAim(x,{\bf k}- {\bf q}_{n}, [c,a_n])
\nonumber \\[1ex]
&\;& - \left(-\half\,\right)^{N_v}
\frac{C_A}{2} \, {\bf B}_n \,
e^{i \omega_0 z_n} c a_{n-1}^{i_{n-1}}\cdots a_1^{i_1} \;\;.
\eeqar{vidamit} 

\subsection{Reaction operator recursion to all orders in opacity}

We emphasize that this recursive process is a completely general
and  applies  both the hard emission kernel ${\rm Ker^{(H)}}$ 
Eq.~(\ref{hvert}) and the asymptotic Gunion-Bertsch ${\rm Ker^{(GB)}}$
Eq.~(\ref{bgcur}).  

The key to unraveling the interplay between direct and virtual
interaction is the operator identity  Eq.~(\ref{dvid})
between $\htD_n$ and $\htV_n$. 
This  makes it possible
to relate the  $n^{\rm th}$
order in opacity  gluon emission ``probability'' distribution
$P_n$ to the probability at $(n-1)^{\rm th}$ order
by expressing the reaction operator $\htR_n$ in Eq.~(\ref{rop}) as
\beqar
\htR_n=  (\htD_{n}-a_{n})^\dagger  (\htD_{n}-a_{n}) - C_A =
(\htS_{n}+\htB_{n})^\dagger  (\htS_{n}+\htB_{n}) -C_A \;\; .
\eeqar{np1}
A further major simplification
occurs because both $\htS$ and $\htB$ involve the same
gluon color rotation through  $if^{ca_{n} d}$.
We show next how this unravels the color algebra
and reduces it to simply powers of $C_A$ and $C_R$.
We therefore prove  color ``triviality'' of the reaction operator,
a property that is implicit in the work of Refs.~\cite{BDMPS2,ZAH}
and which has been  proved in a somewhat more
involved way via path integral techniques in~\cite{UOPAC} for quark
jets only.

First, we note that the  shift operator has the property  that
\beq
\htS_{n}^\dagger \htS_{n} = C_A 
e^{-{\bf q}_{n} \stackrel{\leftarrow}{{\nabla}_{{\bf k}}}}
e^{-{\bf q}_{n} \stackrel{\rightarrow}{{\nabla}_{{\bf k}}}}
\;\; .
\eeq{ss}
Therefore  $\htS^\dagger \htS$ is {\em diagonal} in color space.
This allows us to compute the effect of an extra gluon cascade
 interactions  through simple transverse momentum shifts
\beqar
\vAbim(\htS_{n}^\dagger \htS_{n}-C_A)\vAim 
&=& C_A \left(\vAbim({\bf k}- {\bf q}_{n},c)
\vAim({\bf k}-{\bf q}_{n},c)  
- \vAbim({\bf k},c)\vAim({\bf k},c) \right) \nonumber \\[1.ex]
&=& C_A\left( P_{n-1}({\bf k}-{\bf q}_{n})
            - P_{n-1}({\bf k})\right)
= C_A\left( e^{i{\bf q}_{n}\cdot\hat{\bf b}}-1 \right) P_{n-1}({\bf k})
\;\; .
\eeqar{assa}
Next, consider  the modulus square  of the extra Gunion-Bertsch source
for two or more scatterings of the jet+gluon system ($n > 1$):
\beqar
\vAbim\htB_{n}^\dagger \htB_{n}\vAim
&=& C_A \, |{\bf B}_n|^2 \sum_{i_1,\cdots,i_{n-1}} 
\left(-\half\right)^{\bar{N}_v}
\left(-\half\right)^{N_v}
 T_{el}^\dagger(\vAbim) \, c c \, T_{el}(\vAim) \nonumber \\[1ex]
&=& C_A C_R^n\, |{\bf B}_n|^2 \sum_{i_1,\cdots,i_{n-1}} 
\prod_{m=1}^{n-1}
\left(-\half\right)^{ \delta_{2,i_m}}
\left(-\half\right)^{\delta_{0,i_m}}\left(1\right)^{\delta_{1,i_m}}
= 0 \; \; , 
\eeqar{bb}
where we used Eq.~(\ref{elid}), and
the bookkeeping identity~(\ref{kvid}).
 Therefore, the  virtual corrections 
cancel the direct  extra Gunion-Bertsch source term for $n>1$.
For the special case with $n=1$, however, this diagonal term survives
\beq
G_0^\dagger\htB_{1}^\dagger \htB_{1} G_0= C_A C_R |{\bf B}_1|^2  
\; \; .
\eeq{bb1}
Finally, we have to  calculate the interference  between the cascade
and Gunion-Bertsch source terms in~(\ref{rop}). 
The two gluon color factor $f^{ca_n d}$ in Eqs.~(\ref{ops},\ref{di3})
again contract to form a color diagonal factor
$C_A$. This leaves
\beqar
2 {\bf Re}\, \vAbim\htB_{n}^\dagger \htS_{n}\vAim
&=& -2 C_A\, {\bf B}_n\cdot \left({\rm Re}\; e^{-i\omega_nz_n} 
e^{i{\bf q_n}\cdot\hat{\bf b}} {\bf I}_{n-1}\right)
\;\; , 
\eeqar{bcn}
where the ${\bf I}_{n-1}$  term in the iteration  step is 
easily seen to be
\beqar
{\bf I}_{n-1}&\equiv&\sum_{i_1\cdots i_{n-1}}
\left(-\half\right)^{N_v(\bar{\cal A}^{i_1\cdots i_{n-1}})}
 T_{el}^\dagger(\bar{\cal A}^{i_1\cdots i_{n-1}}) c
 {\cal A}_{i_1\cdots i_{n-1}} \nonumber \\[.5ex]
&=&\sum_{i_1\cdots i_{n-2}}
\left(-\half\right)^{N_v(\bar{\cal A}^{i_1\cdots i_{n-2}})}
 T_{el}^\dagger(\bar{\cal A}^{i_1\cdots i_{n-2}}) \nonumber \\[.5ex]
&\;& \hspace{0.5in} 
\times \sum_{i_{n-1}=0}^2\left\{(-\half)^{\delta_{0,i_{n-1}}}
a_{n-1}^{2-i_{n-1}}\right\}c
 (\delta_{0,i_{n-1}} + \delta_{1,i_{n-1}} \htD_{n-1}+
\delta_{2,i_{n-1}} \htV_{n-1}){\cal A}_{i_1\cdots i_{n-2}} \nonumber \\[.5ex]
&=&\sum_{i_1\cdots i_{n-2}}
\left(-\half\right)^{N_v(\bar{\cal A}_{i_1\cdots i_{n-2}})}
 T_{el}^\dagger(\bar{\cal A}_{i_1\cdots i_{n-2}}) 
 (-\half a_{n-1}a_{n-1} c + a_{n-1}c \htD_{n-1}+
 c \htV_{n-1}){\cal A}_{i_1\cdots i_{n-2}} 
\;\; .
\eeqar{bs2}
This contribution builds up on the class of diagrams 
${\cal A}_{i_1 \cdots i_{n-2}}$ under the action of an iteration
operator which 
can be reduced using Eqs.~(\ref{ops},\ref{dvid}) into the form
\beqar
-\half a_{n}a_{n} c + a_n c \htD_n+c\htV_n&=&
 [a_n,c]\htD_n -\frac{C_A}{2} c = [a_n,c](\htS_n+\htB_n) -C_A c
\nonumber \\[1ex]
&=&C_A(e^{i(\omega_0-\omega_n)z_n}
 e^{i{\bf q}_{n}\cdot\hat{\bf b}} -1) c
+[a_n,c]\htB_n
\; \; .  
\eeqar{bs3}
Next, we note that the contributions from $\htB$ actually
cancel exactly for $n> 2$
in the sum over all classes due to the bookkeeping
identity:
\beqar
&\;&\sum_{i_1\cdots i_{n-2}}
\left(-\half\right)^{N_v(\bar{\cal A}^{i_1\cdots i_{n-2}})}
 T_{el}^\dagger(\bar{\cal A}^{i_1\cdots i_{n-2}}) 
 [a_{n-1},c]\htB_{n-1}{\cal A}_{i_1\cdots i_{n-2}} \nonumber \\[1ex]
&=&-C_A C_R {\bf B}_{n-1} e^{i\omega_0   z_{n-1}}
\sum_{i_1\cdots i_{n-2}}
\left(-\half\right)^{N_v(\bar{\cal A}_{i_1\cdots i_{n-2}})}
\left(-\half\right)^{N_v({\cal A}_{i_1\cdots i_{n-2}})}
 T_{el}^\dagger(\bar{\cal A}_{i_1\cdots i_{n-2}}) 
 T_{el}({\cal A}_{i_1\cdots i_{n-2}})  = 0
\; \; .
\eeqar{bnm1} 
For the special case with $n=2$ we can evaluate ${\bf I}_1$  directly
from the definition Eq.~(\ref{bs2}) and the initial 
hard amplitude to obtain
\beqar
{\bf I }_1 &=&C_A(e^{i(\omega_0-\omega_1)z_1}
 e^{i{\bf q}_{1}\cdot\hat{\bf b}} -1) cG_0
+[a_1,c]\htB_1 G_0 
= -C_A C_R\left( (e^{i(\omega_0-\omega_1)z_1}
 e^{i{\bf q}_{1}\cdot \hat{\bf b}} -1) {\bf H}e^{i\omega_0 z_0}
+{\bf B}_{1} e^{i\omega_0 z_{1}}\right)
\nonumber \\[1ex]
&=& -C_A C_R\left(e^{i(\omega_0-\omega_1) z_1} 
                  e^{i{\bf q}_{1}\cdot \hat{\bf b}} -1
\right) {\bf H}(e^{i\omega_0 z_0}-e^{i\omega_0 z_1})
\;\; .
\eeqar{bfi1}
In the last line the Gunion-Bertsch amplitude was rewritten 
with the help of the shift operator Eq.~(\ref{ops}) as
being derived from the hard vertex kernel
\beq
{\bf B}_{1} e^{i\omega_0 z_{1}}=-\left(e^{i(\omega_0-\omega_1) z_1} 
                  e^{i{\bf q}_{1}\cdot \hat{\bf b}} -1\right) 
                 {\bf H}e^{i\omega_0 z_1}
\;\; . 
\eeq{bg1r}
Eqs.~(\ref{bs2},\ref{bs3},\ref{bfi1})
therefore imply that ${\bf I}_n$ obeys
 the recursion relation
\beqar
{\bf I }_n &=&C_A(e^{i(\omega_0-\omega_n)z_n}
 e^{i{\bf q}_{n}\cdot\hat{\bf b}} -1) {\bf I}_{n-1}
-\delta_{n,1} C_A C_R {\bf B}_{1} e^{i\omega_0 z_{1}}
\;\; . 
\eeqar{bfin}
With ${\bf I}_0= -C_R {\bf H}e^{i\omega_0 z_0}$  and Eq.~(\ref{bg1r}) 
for $n\ge 1$ we solve~(\ref{bfin}) in a closed form  
\beq
{\bf I}_n= C_R C_A^n\left\{\prod_{m=1}^n(e^{i(\omega_0-\omega_m)z_m}
 e^{i{\bf q}_{m}\cdot\hat{\bf b}} -1)\right\} 
\; {\bf H}(e^{i\omega_0 z_1}-e^{i\omega_0 z_0})
\;\;, 
\eeq{reci}
where the product is understood as  ordered for left to right
in decreasing order in  operators labeled by $m$.

Combining Eqs.~(\ref{assa},\ref{bb},\ref{bb1},\ref{bcn},\ref{reci})
that specify how the reaction operator acts between the contraction
of rank $n-1$ classes, we find that the inclusive radiation
probability, Eq.~(\ref{rop}), obeys a simple 
recursion  relation:
\beqar
P_{n}({\bf k})&=& C_A
(P_{n-1}({\bf k}-{\bf q}_{n}) - P_{n-1}
({\bf k)})
-2 C_A\, {\bf B}_n\cdot \left( {\bf Re}\; e^{-i\omega_nz_n} 
e^{i{\bf q_n}\cdot\hat{\bf b}} {\bf I}_{n-1} \right)
 + \delta_{n,1} C_A C_R |{\bf B}_1|^2  
\;\; .  \eeqar{iter}
We can solve Eq.~(\ref{iter}) with the initial condition
\beq
P_0=C_R \, {\bf H}^2
\;\; . \eeq{rp0}
We introduce the following notation for the 
separation of the scattering centers $\Delta z_n \equiv z_n -  z_{n-1}$. 
For $n=1$ the solution is
\beqar
P_1&=&C_AC_R\left( {\bf C}_1^2 - {\bf H}^2 + {\bf B}_1^2 + 2{\bf B}_1\cdot
{\bf C}_1 \cos(\omega_1 \Delta z_1) \right)
= -2C_A C_R {\bf B}_1\cdot
{\bf C}_1(1-\cos(\omega_1 \Delta z_1)) 
\;\; . \eeqar{rp1}
For $n=2$ we need
\beqar
{\bf Re}\; e^{-i\omega_2 z_2}I_1({\bf k}-{\bf q}_2)
&=& -C_A C_R\, {\rm Re}\;e^{-i\omega_2 z_2}
\left( {\bf C}_{(12)}(e^{i(\omega_2 z_1-\omega_{(12)} \Delta z_{1})}-
e^{i\omega_2 z_1})
- {\bf C}_2(e^{i\omega_2 z_0}-e^{i\omega_{2} z_1}) \right)
\nonumber \\[1.5ex]
&=& C_A C_R \, \left[
- {\bf C}_{(12)}(\cos(\omega_2 \Delta z_{2}+\omega_{(12)} \Delta z_{1})-
\cos( \omega_2 \Delta z_{2})) \right.
\nonumber \\[1.5ex]
&& \left. \qquad   \quad \;
  +{\bf C}_2(\cos(\omega_2 ( \Delta z_{1} +  \Delta z_{2})
-\cos(\omega_{2} \Delta z_{2})) \; \right]
\; \; .
\eeqar{bi2}
Therefore, the $n=2$ distribution is given by
\beqar
P_2&=& \; 2C^2_A C_R \left[\;  {\bf B}_1\cdot
{\bf C}_1(1-\cos(\omega_1 \Delta z_{1}) )
- {\bf B}_{2(12)}\cdot
{\bf C}_{(12)}(1-\cos(\omega_{(12)} \Delta z_{1})) 
\right. \nonumber \\[1.5ex]
&\;&   \qquad \qquad + {\bf B}_2\cdot{\bf C}_{(12)}
(\cos(\omega_2 \Delta z_{2}+\omega_{(12)} \Delta z_{1})-  
\cos( \omega_2 \Delta z_{2}))
\nonumber \\[1.5ex]
 &\;& \left.   \qquad  \qquad - {\bf B}_2\cdot{\bf C}_2
(\cos(\omega_2 (\Delta z_{1} + \Delta z_{2})
-\cos(\omega_{2} \Delta z_{2})) \; \right]
\;\; . 
\eeqar{rp2}

For $n>1$, we can use~(\ref{reci}) to obtain the general solution
for the gluon probability at n$^{\rm th}$ order in opacity
\beqar
P_{n}({\bf k})&=& 
 C_A(e^{i{\bf q}_{n}\cdot \hat{\bf b}}-1)  
P_{n-1}({\bf k})\nonumber  \\
&\;& \qquad  - 2 C_R C_A^n \, {\bf B}_n\cdot {\bf Re}\; e^{-i\omega_n z_n} 
e^{i{\bf q_n}\cdot \hat{\bf b}}
\left\{\prod_{m=1}^{n-1}(e^{i(\omega_0-\omega_m)z_m}
 e^{i{\bf q}_{m}\cdot\hat{\bf b}} -1)\right\} 
\; {\bf H}(e^{i\omega_0 z_1}-e^{i\omega_0 z_0}) 
\nonumber \\
&=& -2C_R C_A^n \, {\bf Re}
\sum_{i=1}^n 
\left\{\prod_{j=i+1}^n( e^{i{\bf q}_{j}\cdot \hat{\bf b}} - 1)  \right\}
{\bf B}_{i} \cdot\; 
e^{i{\bf q}_{i}\cdot \hat{\bf b}}   e^{-i\omega_0 z_i} \times
\nonumber \\ 
&\;& \hspace{1.5in}  \left\{
    \prod_{m=1}^{i-1}(e^{i(\omega_0-\omega_m)z_m}
     e^{i{\bf q}_{m}\cdot \hat{\bf b}} -1)
  \right\}
\; {\bf H}(e^{i\omega_0 z_1}-e^{i\omega_0 z_0}) 
\; \; ,  \eeqar{iter2}
where the last applied to $n=1$ as well.
This is therefore the complete solution to the problem
and is the central result of this paper.
Eq.~(\ref{iter2})  provides
not only an algebraic proof of color triviality of the
inclusive gluon distribution ($P_n\propto C_R C_A^n$),
but in fact gives the complete angular dependence
for arbitrary $z_i$ and ${\bf q}_{n}$. It also applies to both
quark and gluon jets.
The derivation through the reaction operator and recursion relations
is furthermore much simpler and transparent than 
previous derivations~\cite{BDMPS2,ZAH,UOPAC}.
The result is also more general and versatile for applications
to nuclear collisions. 

In this form, it is for example possible 
to implement $P_n$ directly as an ``after-burner'' in a Monte
Carlo event generator, e.g. ZPC~\cite{ZHANG,MGYPBZ} or 
MPC~\cite{MOLNAR,MPC}.
Such programs solve  relativistic 3+1D transport theory
and provide much more realistic  distributions of   target
parton coordinates  $z_i$ along the path of a high
energy jet in nuclear collisions. 
Eq.~(\ref{iter2})  also makes it possible
to consider evolving
potentials since the distribution  of the ${\bf q}_n$ remain
unspecified. For example, the screening scale $\mu$, that
characterizes the mean ${\bf q}_{n}$ transfered at $z_n$,
 could itself depend on $n$ because of the variation of
the local density at that point.
In this way, jet quenching in realistic, evolving nuclear geometries
can  be calculated more accurately.

\subsection{Ensemble average over momentum transfers}

We now consider the integrated gluon distribution
over the ${\bf q}_{m}$ via 
the normalized squared scattering potential Eq.~(\ref{vbar})
\beq
\left \langle \;  P_n({\bf k}) \; \right \rangle_v = 
 \int \prod_{m=1}^n \left\{d{\bf q}_{m}
\, \bar{v}^2_m({\bf q}_{m})\right\}  
 \;   P_n({\bf k}; {\bf q}_{1}
\cdots {\bf q}_{n})  \;\; ,
\eeq{vnormp}
where we allow for the possibility that the effective potential
could change along the path of the jet. We hold the $z_m$ fixed as yet.
>From Eq.~(\ref{iter}) we need to consider for example
\beqar
&\;& \int d{\bf q}_{n}\, \bar{v}^2_n({\bf q}_{n})  
 \;    (P_{n-1}({\bf k}-{\bf q}_{n})-
P_{n-1}({\bf k}))
= \int d{\bf q}_{n}\, \left(\bar{v}^2_n({\bf q}_{n})-
\delta^2({\bf q}_{n})\right)  
 \;    P_{n-1}({\bf k}-{\bf q}_{n}) \;\;.
\eeqar{intit1}
Since $\delta^2({\bf q}_{n}){\bf B}_n\equiv 0$, we can also replace
$\bar{v}^2_n({\bf q}_{n})  $ by $\bar{v}^2_n({\bf q}_{n})  -
\delta^2({\bf q}_{n})$ in the average over the inhomogeneous
term. However, due to the special form of the
operators defining ${\bf I}_n$ in Eq.~(\ref{reci})
we can also rewrite
the integral over  ${\bf q}_{i}$ for $i=1,\cdots,n-1$
in analogous form
\beqar
&\;&\int \prod_{j=1}^n\left\{
d{\bf q}_{j}\, \bar{v}^2_j({\bf q}_{j})
\right\}  
 \;   {\bf B}_n\; e^{-i\omega_n z_n} 
e^{i{\bf q_n}\cdot\hat{\bf b}}
\prod_{m=1}^{n-1}(e^{i(\omega_0-\omega_m)z_m}
 e^{i{\bf q}_{m}\cdot\hat{\bf b}} -1)
\nonumber \\[.5ex]
 \qquad &\qquad& 
\qquad=\; \int \prod_{j=1}^n\left\{
d{\bf q}_{j}\, \left[
\bar{v}^2_j({\bf q}_{j})-\delta^2({\bf q}_{j})\right]\right\}  
 \;    {\bf B}_n e^{-i\omega_0 z_n} 
 \prod_{m=1}^{n}\; e^{i(\omega_0-\omega_m) z_m} 
e^{i{\bf q_m}\cdot\hat{\bf b}}
\; \; ,\eeqar{intit2}
where the product of operators is again path ordered.
This product of operators can be simplified  as follows
\beqar
e^{-i\omega_0 z_n}
\prod_{j=1}^{n}\; e^{i(\omega_0-\omega_j) z_j}e^{i{\bf q_j}\cdot
\hat{\bf b}} &=&
e^{-i\omega_{n} z_n} 
e^{i(\omega_{n}-\omega_{(n-1,n)}) z_{n-1}}  
\cdots \; e^{i(\omega_{(2\cdots n)} -\omega_{(1\cdots n)}) z_1}
e^{i {\bf Q}\cdot\hat{\bf b}}
\nonumber \\[1.ex]
&=& \exp\left(-i\sum_{j=1}^n 
\omega_{(j\cdots n)}(z_j-z_{j-1})  
\right) e^{-i\omega_{(1\cdots n)} z_0}
e^{i {\bf Q}\cdot\hat{\bf b}}
\;\; ,
\eeqar{prd}
where ${\bf Q}=\sum {\bf q}_m$ is the total momentum transfer,
and 
\beq
\omega_{(m,\cdots , n)}=
\frac{({\bf k}-{\bf q}_m-\cdots-{\bf q}_n)^2}{2 x E}
\;\; . \eeq{omegm}
Therefore, we can evaluate 
\beqar
&\;& e^{-i\omega_0 z_n}\prod_{j=1}^n\; e^{i(\omega_0-\omega_j) z_j} 
e^{i{\bf q_j}\cdot\hat{\bf b}}\; {\bf H}(e^{i\omega_0 z_1}-e^{i\omega_0 z_0}) 
= {\bf C}_{(1\cdots n)} e^{i\Phi_{n,n}}
( e^{i\omega_{(1\cdots n)} (z_1-z_0)}-1)
\; \; ,\eeqar{intit30}
where $\Phi_{n,n}$ is the  gluon elastic scattering
phase shift from $z_0$ to $z_n$. 
The partial (eikonal) phase shift 
($\int_{z_{m-1}}^{z_n} dz\;(k_z(z)-xE)$) due to gluon rescattering
from $z_{m-1}$ to $z_n$ is given by
\beqar
\Phi_{n,m} &=& -\, \sum_{k=1}^m \omega_{(k\cdots n)} (z_k-z_{k-1})
= -\,  \sum_{k=1}^m \omega_{(k\cdots n)} \Delta z_k
\;\; . 
\eeqar{phin}
Eq.~(\ref{intit30}) makes it possible
to write the recursion 
relation for the  momentum transfer averaged probability for $n>1$
in the form
\beqar
\left \langle \;  P_n  \; \right \rangle_v &=& C_A \int 
d{\bf q}_{n}\, \left(
\bar{v}^2_n({\bf q}_{n})-\delta^2({\bf q}_{n})\right)
\left \langle \;  P_{n-1}({\bf k}-{\bf q}_{n})
  \; \right \rangle_v
\nonumber \\
&\;& - \,  2 C_R C_A^n\int \prod_{i=1}^n\left\{
d{\bf q}_{i}\, \left(
\bar{v}^2_i({\bf q}_{i})-\delta^2({\bf q}_{i})\right)\right\}  
 \;    {\bf B}_n\cdot {\bf C}_{(1\cdots n)} \;
{\rm Re}\; e^{i\Phi_{n,n}}
(e^{i\omega_{(1\cdots n)} (z_1-z_0)} - 1)
\; \; .
\eeqar{intit3}
With $P_1$ given by Eq.~(\ref{rp1}), the solution for $n>0$ is
\beqar
\left \langle \;  P_n  \; \right \rangle_v &=& 
 - \,  2 C_R C_A^n\int \prod_{i=1}^n\left\{
d{\bf q}_{i}\, \left(
\bar{v}^2_i({\bf q}_{i})-\delta^2({\bf q}_{i})\right)\right\}  
\nonumber \\
&& \qquad \qquad \quad \qquad \times \;
\sum_{m=1}^n {\bf B}_{(m+1,\cdots,n)(m \cdots n)} 
\cdot {\bf C}_{(1 \cdots n) } \, {\bf Re} \;\; 
\left( e^{i\Phi_{n,m} }
(e^{i\omega_{(1\cdots n)}(z_1-z_0) }-1 ) \, \right) \;\;.
\eeqar{intit3cl}
where for $m=n$, 
${\bf B}_{(n+1\cdots n)(n)}\equiv
{\bf B}_{(0)(n)}\equiv {\bf B}_{(n)}$.

Recall that we cannot take the contact limit $z_1=z_0$, 
which apparently vanishes, because our derivation
assumed that $z_m-z_{m-1}$ were larger than the range
of the force $1/\mu$. However, it is clear from
Eq.~(\ref{intit3cl}), that it is the
first step phase difference 
\beq
e^{i\omega_{(1\cdots n)} (z_1-z_0)}-1 
\;\;  
\eeq{dominant}
that controls the
bulk of the LPM destructive interference effect that
suppresses radiation when the formation length 
$1/\omega_{(1\cdots n)}$ is long compared to $(z_1-z_0)$. 
The subsequent interactions  merely modulate this effect
with an elastic scattering phase shift.

Finally, we can restore the proportionality constants by
multiplying with $\alpha_s/\pi$ for the production vertex,
$1/\pi$ for the $d^2{\bf k}$ measure, and
$\prod_j (\sigma_g(j)/A_\perp) $ 
along the path to convert the $\bar{v}_j$ back
into $v_j$ from (\ref{vbar}).
In addition, we must multiply by the combinatorial
factor,
\beq
\frac{N!}{n!(N-n)!}\approx \frac{N^n}{n!} \;\;\;\;\;, 
\eeq{combi}
that counts the number of ways $n$ target partons out of $N$
can be within the interaction range of the jet+gluon system.
Including these factors,
the general formula for the induced gluon
number distribution can finally be written as
\beqar
x\frac{dN^{(n)}}{dx\, d^2 {\bf k}} &=&
\frac{C_R \alpha_s}{\pi^2} \frac{1}{n!} 
\left( \frac{L}{\lambda_g(1)} \right)^n
 \int \prod_{i=1}^n \left\{d{\bf q}_{i}\, 
\left(\frac{\lambda_g(1)}{\lambda_g(i)}\right)
(\bar{v}_i^2({\bf q}_{i}) - \delta^2({\bf q}_{i}) )\right\}  
\nonumber \\[1.ex]
&\;& \times \left[ -2\,{\bf C}_{(1, \cdots ,n)} \cdot 
\sum_{m=1}^n {\bf B}_{(m+1, \cdots ,n)(m, \cdots, n)} 
\left( \cos \left (
\, \sum_{k=2}^m \omega_{(k,\cdots,n)} \Delta z_k \right)
-   \cos \left (\, \sum_{k=1}^m \omega_{(k,\cdots,n)} \Delta z_k \right)
\right)\; \right]    
\nonumber \\[1.5ex]
\eeqar{ndifdis} 
where $\sum_2^1 \equiv 0$ is understood.
  
We emphasize that  Eq.~(\ref{ndifdis}) is not restricted
to uncorrelated geometries as in \cite{BDMPS2,UOPAC}.
Also it allows the inclusion of finite kinematic boundaries
on the ${\bf q}_i$ as well as different functional forms and
elastic cross sections $\sigma_g(i)$ along the eikonal path.

For particular models of the target geometry
one can proceed further analytically.
For a sharp rectangular 
geometry, the average over the longitudinal target profile with
\beq
\bar{\rho}
(z_1,\cdots,z_n)= \frac{n!}{L^n} \;\theta(L-z_n)\theta(z_n-z_{n-1})
\cdots \theta(z_2-z_{1})
\eeq{boxgeom}
leads to an  oscillatory pattern~\cite{UOPAC} that  
is an artifact of the assumed sharp edges.

A somewhat more realistic model~\cite{GLV1B,GLVPRL}
utilizes normalized exponential
longitudinal distributions of scattering center separations:
\beq
\bar{\rho}(z_1,\cdots,z_n)=  \prod_{j=1}^n
\frac{\theta(\Delta z_j)}{L_e(n)}e^{-\Delta z_j/L_e(n)} \;\;,
\eeq{expgeom}
This converts the oscillating formation physics factors
in Eq.~(\ref{ndifdis}) into simple 
 Lorentzian factors
\beq
\int d\bar{\rho}
\cos \left (\, \sum_{k=j}^m \omega_{(k,\cdots,n)} \Delta z_k \right)
={\rm Re}\;\prod_{k=j}^m \frac{1}{1+i\omega_{(k,\cdots,n)}L_e(n)}
\; \; , \eeq{loren} 
In order to fix $L_e(n)$, we require
that $\langle z_k-z_0 \rangle
=k L/(n+1)$ 
for both geometries~(\ref{boxgeom},\ref{expgeom}). This constrains
$L_e(n)=L/(n+1)$ \cite{GLV1B,GLVPRL}.

Note finally that the ratio
of the medium induced to the medium independent factorization
 gluon distributions vanish at both small and large
$|{\bf k}|$
\beqar
&& \lim_{|{\bf k}|=0 \;{\rm and}\; \infty}
 \frac{dN^{(n)}}{dN^{(0)}} = 0
\eeqar{limn}
As  $|{\bf k}|\rightarrow 0$, $dN^{(0)}\rightarrow\infty$
while all but one term in $dN^{(n)}$ is finite.
The potentially singular term $\propto (1/{\bf k}^2) \int \sum_j 
{\bf k}  \cdot  {\bf q}_{j }$ however 
 vanishes due to  azimuthal integrations. 
In the $|{\bf k}|\rightarrow \infty$ limit, on the other hand,
all Gunion-Bertsch currents ${\bf B}_{(..),(..)}\rightarrow 
O(1/{\bf k}^4)$ 
and hence vanish  faster than $1/{\bf k}^2$ from $dN^{(0)}$.
This observation partly justifies neglecting the kinematical 
$|{\bf k}|$ boundaries ~(\ref{klimits}) in analytical approximations.

\section{Numerical Estimates}

\subsection{Angular distributions}

Figs.~4a-c illustrate  Eq.~(\ref{ndifdis}) 
for a exponential geometry~(\ref{expgeom}) equivalent
to a box of thickness $L=5$ fm with  $\lambda_g=1$ fm
including kinematical constraints appropriate for a 50 GeV jet.
We used adaptive Monte-Carlo integration~\cite{DADMUL1,DADMUL2} 
to integrate  over the
momentum transfers ${\bf q}_1 \cdots {\bf q}_n$.  

The medium induced gluon differential 
distributions up to third order in opacity are  plotted divided
by the zeroth order in opacity, hard distribution  Eq.~(\ref{hdist})
for radiated 
gluons with light-cone momentum fractions $x=0.05,\, 0.2,\, 0.5$. 
Note  that $dN^{(ind)}/dN^{(0)}$ tends to vanish at both small and large
$k$  in accordance with  Eq.~(\ref{limn}). 
As $x$ increases the relative magnitude of the medium
induced contribution decrease and the overall magnitude is set
already by the  first order in opacity result.
Higher orders redistribute the moderate $k^2/\mu^2 \sim 10$
toward higher values due to elastic rescattering of the gluon,
but they also tend to fill in the low $k$ region with additional soft
radiation. The overall angular pattern appears somewhat complex, 
but one must keep in mind that the variation is rather
slow since the logarithmic scale varies over several orders of magnitude.

Fig.~4d shows the actual rather  modest size of the
induced radiation contribution on top of the
hard $1/{\bf k}^2$ distribution for the case of $x=0.05$.

\subsection{Intensity distribution and energy loss to first order 
in opacity. Analytical approach}

\subsubsection{Zero Opacity Limit}

The jet distribution in the absence of final state interactions
is given by~\cite{GLV1B}
\beq
d^3N_J = \rho^{(0)}(\vec{\bf p})\, \frac{d^3\vec{\bf p}}{2|\vec{\bf p}| (2\pi)^3 } =
d_R |J(|\vec{\bf p}|,\vec{\bf p})|^2\, \frac{d^3\vec{\bf p}}{2|\vec{\bf p}| (2\pi)^3 }
\;\; ,
\eeq{rho0}
In the Leading Pole Approximation (LPA) 
approximation~\cite{FIELD}, the radiation distribution
accompanying the such hard processes for a spin $\half$ jet is given by  
\begin{equation}  
x\frac{dN^{(0)}}{dx\, d {\bf k}^2}\approx 
\frac{ C_R \alpha_s}{\pi}\, \left( 1-x + \frac{x^2}{2} \right)\, 
\frac{1}{{\bf k}^2} \;\;, 
\label{hdist}
\end{equation}
where in the eikonal approximation the light-cone momentum fraction 
$x=k^+/E^+ \approx \omega/E$ and ${\bf k}$ is assumed to be small
compared to $\omega$. For other spin jets, another
 suitable splitting function replaces the $x$ dependence above.
\begin{center}
\vspace*{16.5cm}
\includegraphics{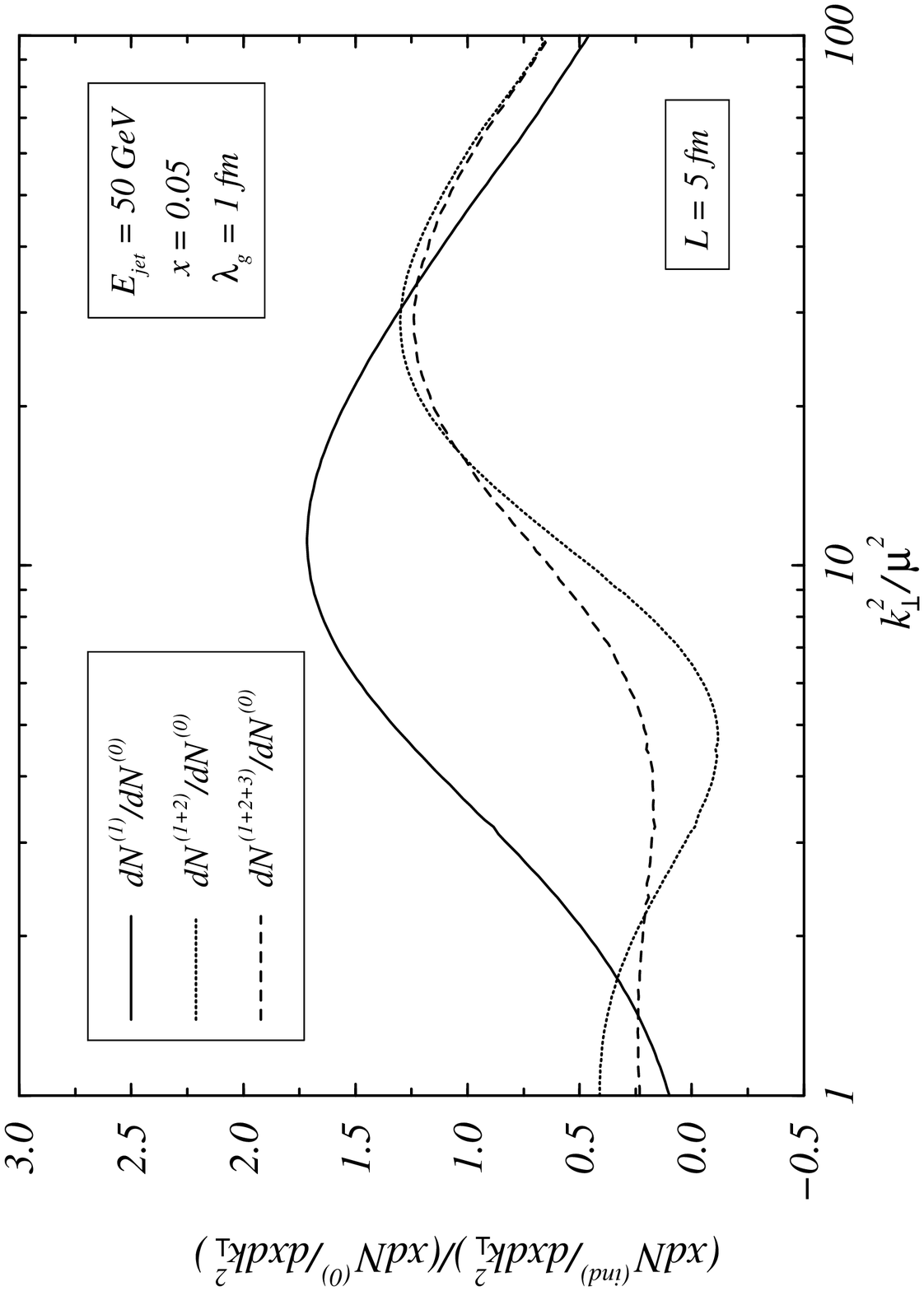}
\includegraphics{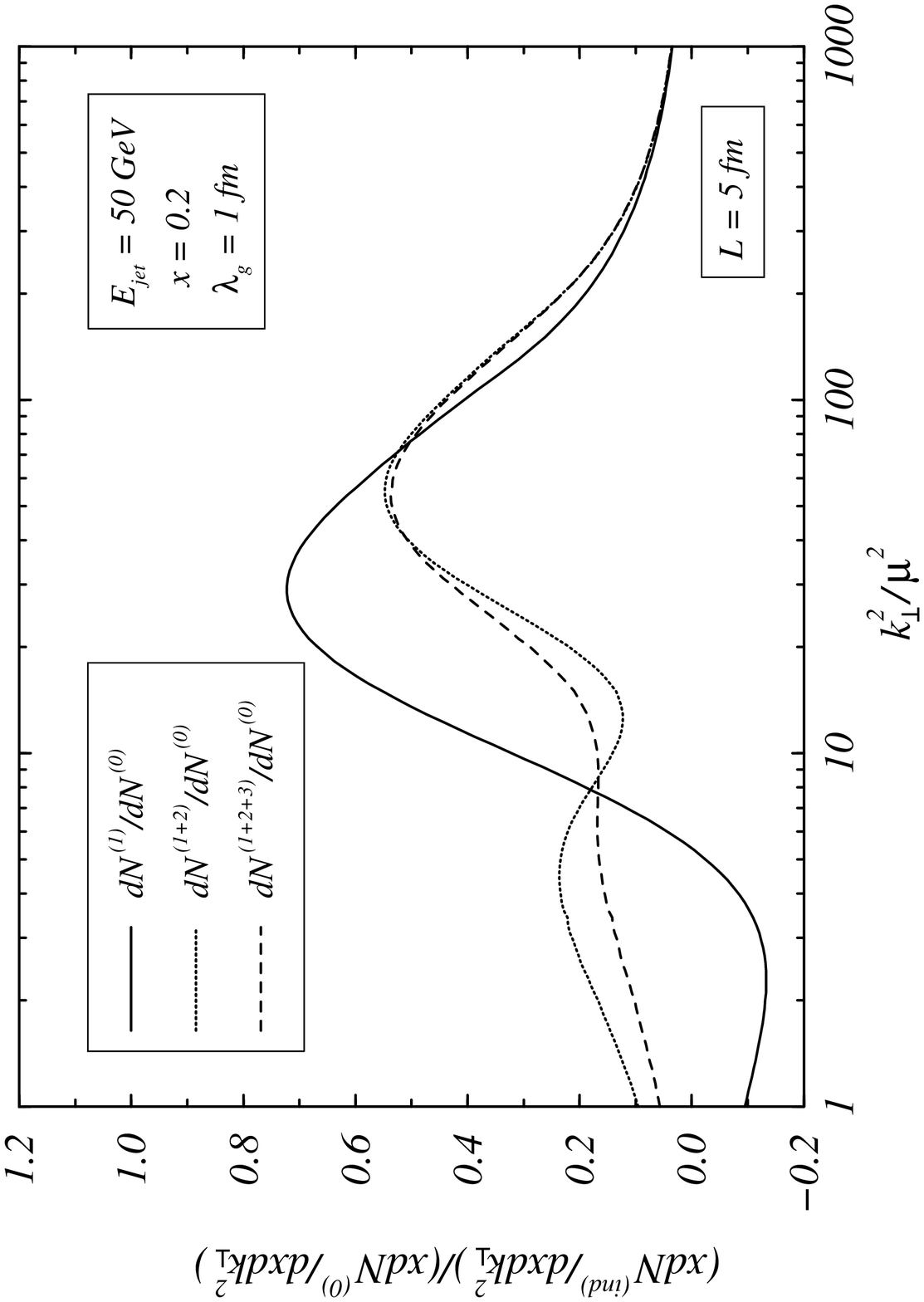}
\includegraphics{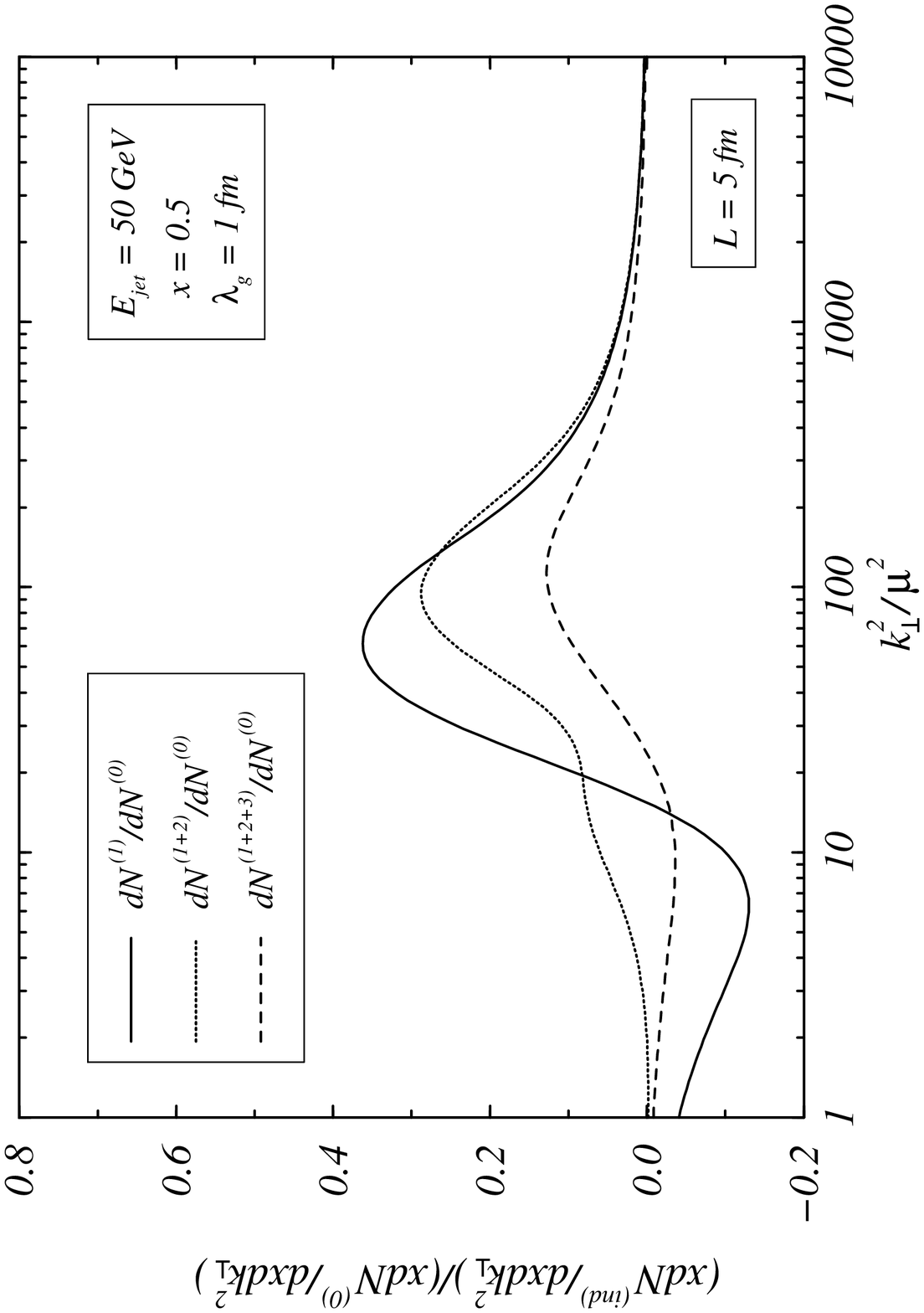}
\includegraphics{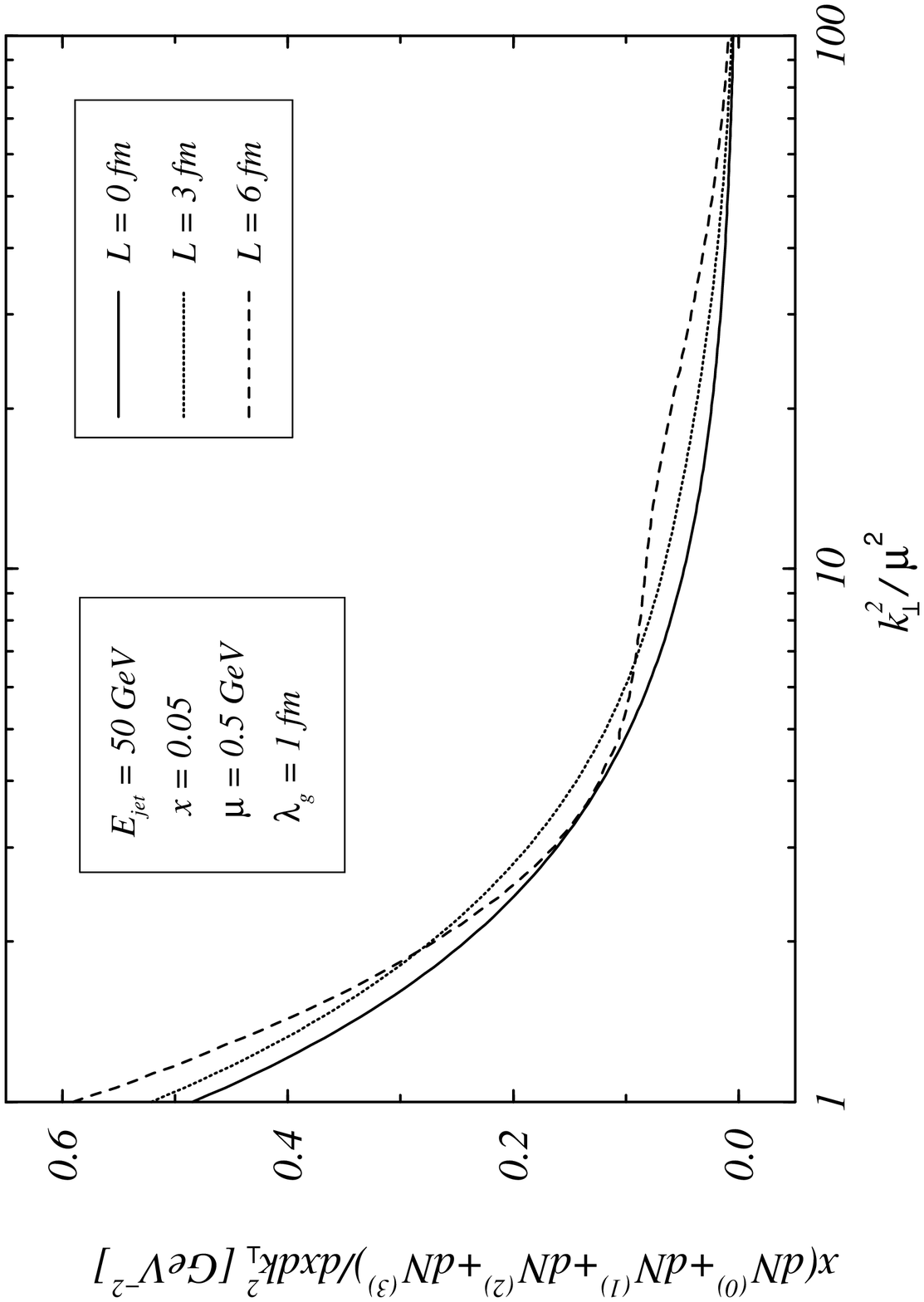}
\vskip - 35pt
\begin{minipage}[t]{15.0 cm}
{\small {FIG~4.}  (a)-(c) The  medium-induced double 
differential gluon distributions
are plotted vs. ${\bf k}^2 /\mu^2$ up to first 
$(dN^{(1)})$, second $(dN^{(1+2)})$ 
and third $(dN^{(1+2+3)})$ order in the opacity
$(L/\lambda_g)$ expansion, divided  by the medium independent hard
radiation $(dN^{(0)})\propto 1/k^2$. 
Curves are obtained numerically 
for exponential geometry with $L/\lambda_g=5$, $E_{jet}=50$ GeV and
$\mu=0.5$ GeV. The first three figures (a)-(c) are for  typical 
soft, semi-soft and hard gluons respectively, $x=0.05,\, 0.2,\, 0.5$.  
(d) The {\em full} gluon differential distribution
up to third order in opacity
$dN^{(0)}+ dN^{(1)}+dN^{(2)}+dN^{(3)}$  is shown
for $x=0.05$ for $L=0,3,6$ fm.}
\end{minipage}
\end{center}
\vskip 4truemm

We  consider  radiation 
outside a cone with  $|{\bf k}|>\mu$. 
and with the upper $|{\bf k}|$ bound  determined 
from  the  three body (jet+jet+glue)
kinematics. The gluon kinematic boundaries are therefore
\beq
{\bf k}^2_{ \, \min}=\mu^2\;,  
\quad {\bf k}^2_{ \, \max}=\min\, [4E^2x^2,4E^2x(1-x)] \;\;.
\eeq{klimits} 
The radiation intensity 
integrated 
over this range of  ${\bf k}$  is
\beq 
\frac{dI^{(0)}}{dx} = \frac{2 C_R \alpha_s}{\pi} 
\left( 1- x +\frac{x^2}{2} \right)  E \, \log 
\frac{|{\bf k}|_{\rm max}}{|{\bf k}|_{\rm min}}
 \;\; ,    
\eeq{di0} 
where $\left|{\bf k}\right|_{\rm max}$ and 
$|{\bf k}|_{\rm min}$ are given 
by Eq.~(\ref{klimits}). Note that the differential intensity
is roughly uniform with the exception of the kinematical edges
$x \rightarrow 0$ and $x \rightarrow 1$ as seen in Fig.~5a.
In the Leading Log of the Leading Pole Approximation (LLA) the
radiative total energy loss of a quark jet originating from a
hard vertex outside a cone defined by Eq.~(\ref{klimits}) is given by 
\beq
\Delta E^{(0)}=\frac{4C_R\alpha_s}{3\pi}\, E \,
\log \frac{E}{\mu} \;\; .
\eeq{de0}
While this overestimates the 
radiative  energy loss in the vacuum (self-quenching), 
it is important to note
that $\Delta E^{(0)}/E \sim 50\%$ is typically rather large.
As shown below, the  medium induced energy loss 
is small by comparison.
 For a gluon jet one has to replace 
 the quark splitting function 
$q\rightarrow qg$  by the gluon one $g\rightarrow gg$
in Eq.~(\ref{di0}) and 
use $C_A\equiv N_c\approx 2 C_F$.

The total energy loss should reduce to the medium 
independent one  in the limit of vanishing opacity
\beq
\lim_{L\rightarrow 0} \frac{\Delta E^{(tot)}}{\Delta E^{(0)} } 
\; = \; 1 
\eeq{limthin}
and for jets of asymptotically high energies due to
factorization 
\beq
\lim_{E \rightarrow \infty} \frac{\Delta E^{(tot)}}{\Delta E^{(0)}} 
\;=\; 1 \;\;.
\eeq{limfac}

\subsubsection{First Order in Opacity Correction}

The first order $(n=1)$ in opacity contribution, $dI^{(1)}/dx$,
to the induced radiation intensity 
  can be read of  from Eq.~(\ref{ndifdis}). 
The longitudinal coordinate 
average over the equivalent exponential target 
profile~(\ref{expgeom}) is done with   $L_e(2)=L/2$.
Including the 
$q\rightarrow qg$ splitting function as in~(\ref{hdist}) we have 
\beqar
\frac{dI^{(1)}}{dx}&=& 
\frac{C_R\alpha_s}{\pi}\left(1-x+\frac{x^2}{2}\right)
\frac{L}{\lambda_g}\, E \, 
\int_{{\bf k}^2_{ {\rm min}}}^{{\bf k}^2_{ {\rm max}}}
\frac{d {\bf k}^2_{}}{{\bf k}^2_{}}  
\nonumber \\[1.ex] &&
\qquad  \qquad \qquad \qquad \qquad
\int_0^{{\bf q}_{1 \max}^2} d^2{\bf q}_{1} \, 
\frac{ \mu_{eff}^2 }{\pi ({\bf q}_{1}^2 + \mu^2)^2 }
\, \frac{ 2\,{\bf k} \cdot {\bf q}_{1}
  ({\bf k} - {\bf q}_1)^2  L^2}
{16x^2E^2 +({\bf k} - {\bf q}_1)^4  L^2 } \;\;.          
\eeqar{dnx1}
To obtain a simple analytic result, we ignore
the kinematic boundaries and set
$|{\bf k}_{}|_{\rm min}=0$, $|{\bf k}_{}|_{\rm max}=\infty$ 
that is motivated by  Eq.~(\ref{limn})). We also set
$|{\bf q}_{1}|_{\rm max}=\infty$ (i.e. $\mu^2_{eff}=\mu^2$). 
This allows us to change variables  
${\bf q}^{\;\prime}\equiv {\bf k}-{\bf q}_{1}$ 
in Eq.~(\ref{dnx1}) 
\beqar
\frac{dI^{(1)}}{dx}&=& 
\frac{C_R\alpha_s}{\pi}\left(1-x+\frac{x^2}{2}\right)
\frac{\mu^2 L}{\lambda_g}\, E \, 
\int_0^\infty d{\bf q}^{\prime\, 2} 
\frac{ {\bf q}^{\prime \, 2}  L^2}
{16x^2E^2 +{\bf q}^{\prime \, 4}  L^2 }
\nonumber \\[1.ex] &&
\qquad  \qquad \qquad \qquad \qquad
\int_0^\infty \frac{d {\bf k}^2_{}}{{\bf k}^2_{}}  
\int_0^{2\pi} \frac{d\phi}{2\pi}
\frac{2 {\bf k} \cdot ({\bf k} +{\bf q}^\prime ) }
{({\bf q}^{\prime\, 2}+{\bf k}^2 
+ 2 |{\bf q}^{\prime}| \, 
|{\bf k}| \cos(\phi)  +\mu^2)^2}
\eeqar{dnx2}
and express the integrand in the azimuthal 
$\phi$ integral as a partial derivative with respect to 
${\bf k}^2$  
$$-2 {\bf k}^2\partial_{{\bf k}^2} 
\int_0^{2\pi} \frac{d \phi}{2 \pi} 
\frac{1}{({\bf q}^{\prime\, 2}+{\bf k}^2 
+ 2 |{\bf q}^{\prime}| \, 
|{\bf k}| \cos(\phi)  +\mu^2)}
=-2 {\bf k}^2 \partial_{{\bf k}^2}
\frac{1}{\sqrt{(({\bf k}^2+{\bf q}^{\prime \,2}+\mu^2)^2
-4{\bf k}^2 {\bf q}^{\prime\,2}) } }\;\; .$$ 
The remaining  
${\bf q}^{\;\prime}$
integral 
\beqar
\frac{dI^{(1)}}{dx}&=& 
\frac{C_R\alpha_s}{\pi}\left(1-x+\frac{x^2}{2}\right)
\frac{L}{\lambda_g}\, E \, 
\int_0^\infty d{\bf q}^{\prime\, 2} 
\frac{2\mu^2}{ {\bf q}^{\prime \, 2}  + \mu^2} \,
\frac{ {\bf q}^{\prime \, 2}  L^2}
{16x^2E^2 +{\bf q}^{\prime \, 4}  L^2 }\;\;. 
\eeqar{dnx3}
can be performed then analytically, resulting in
\begin{equation}
\frac{dI^{(1)}}{dx} =\frac{C_R\alpha_s}{\pi} \,
\left( 1 - x + \frac{x^2}{2} \right) \,E \, 
\frac{L}{\lambda_g}\, f(\gamma) \;, 
\qquad \gamma=\frac{L\mu^2}{4xE} \;\;,
\label{didx1}
\end{equation} 
where $\gamma$ is a measure of the formation probability.
The formation function $f(\gamma)$ is given by
\beq
f(\gamma) = 
\frac{\gamma \,\left(\pi
+2 \gamma \log \gamma \right)}
{(1+\gamma^2)} \approx \left\{ 
\begin{array}{ll}
\pi \gamma  &\qquad {\rm if} \quad \gamma \ll 1   \\
2 \log \gamma    & \qquad {\rm if} \quad \gamma \gg 1 \ 
\end{array}   \right.  \;\;.
\eeq{fapprox}
It is the $\gamma \ll 1$ limit of the formation function Eq.~(\ref{fapprox}) 
in which the the first order in opacity medium-induced intensity 
distribution reduces to a simple form with a characteristic quadratic
dependence on $L$:
\begin{equation}
\frac{dI^{(1)}}{dx} \approx \frac{C_R\alpha_s}{4} \,
\frac{ 1-x + \frac{x^2}{2} }{x} \, 
\frac{L^2\mu^2}{\lambda_g}\; .
\label{didx1app}
\end{equation} 
This formula  breaks down at both $x\rightarrow 0$ and
$x\rightarrow 1 $ because $|{\bf k}|_{\rm max}$ 
and $|{\bf q_{1}}|_{\rm max}$ cannot be 
approximated by $\infty$ and because the small $x$ approximations used above
break down as $x\rightarrow 1$.

The induced radiative energy loss to first order in opacity in the
framework of the above approximations is then given by
\beq
\Delta E^{(1)}=\frac{C_R\alpha_s}{4}\, 
\frac{L^2\mu^2}{\lambda_g} \,\log \frac{E}{\mu}  \;\;.
\eeq{de1}
This  equation is directly comparable  to Eq.(\ref{17}) of
Ref.~\cite{BDMPS3}.

\subsection{Induced radiation intensity to higher orders 
in opacity. Numerical results}

Whereas  Eq.~(\ref{de1}) displays analytically the
main qualitative features of non-abelian energy loss, in practice
at the finite energies available
kinematical bounds do affect quantitatively the
results. This naturally leads to a {\em reduction} relative
to the analytic
estimate.

In Fig.~5a the induced intensity distribution $dI^{(ind)}/dx$ 
(sets of dashed curves) is compared to the medium independent 
$dI^{(0)}$.
We consider the example of a $50$~GeV quark jet ($\mu=0.5$ GeV) 
at RHIC. The hard  radiation intensity result  is  roughly
constant with $x$, whereas 
for most of the $x$ region the induced radiation
intensity falls like $1/x^\alpha,\; \alpha \sim 1$.
We note that 
for relatively thin plasmas  $L/\lambda_g  \leq 3$   the medium induced
energy loss remains small compared to  $dI^{(0)}$. 

\newpage

\begin{center}
\vspace*{9.2cm}
\includegraphics{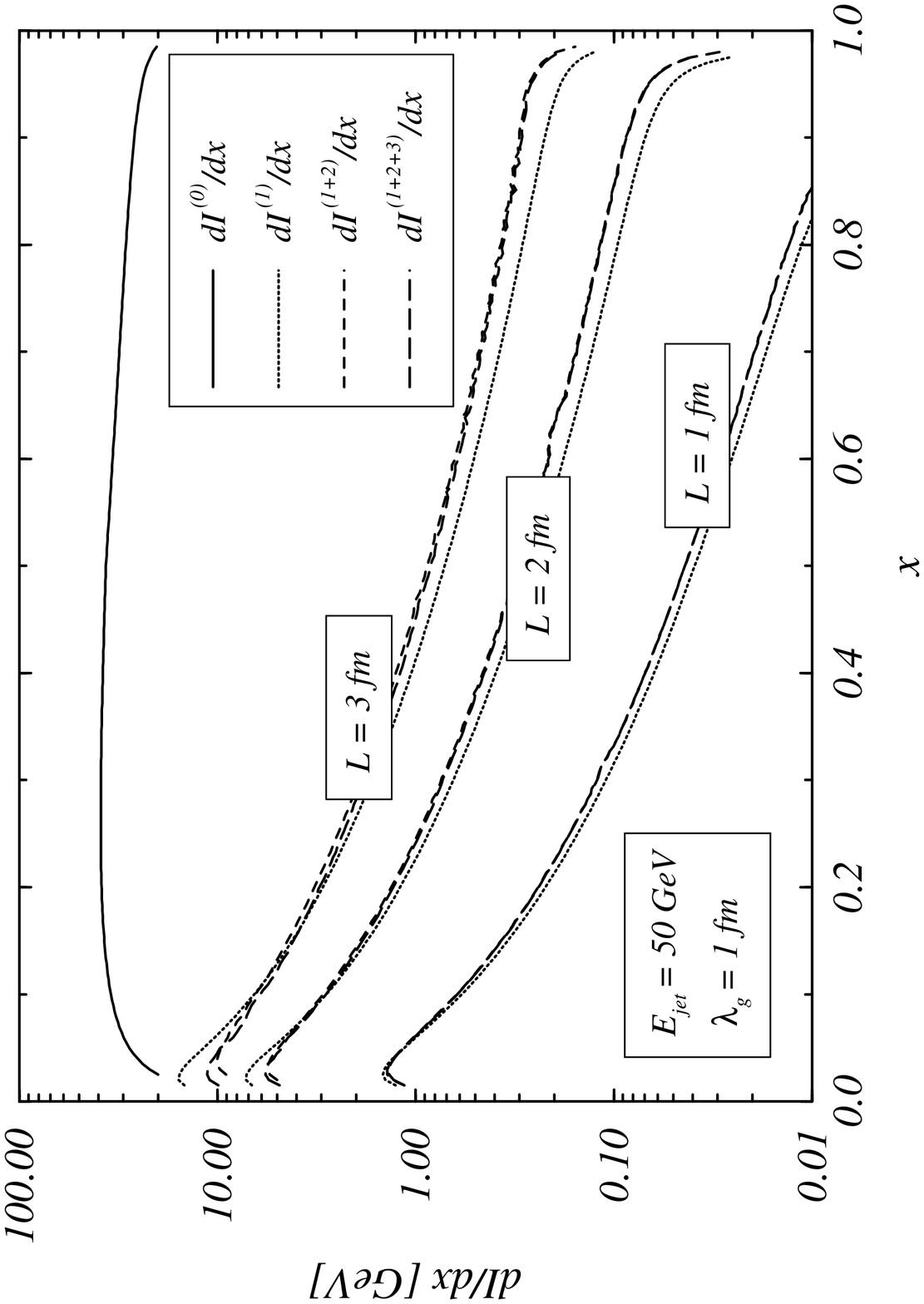}
\includegraphics{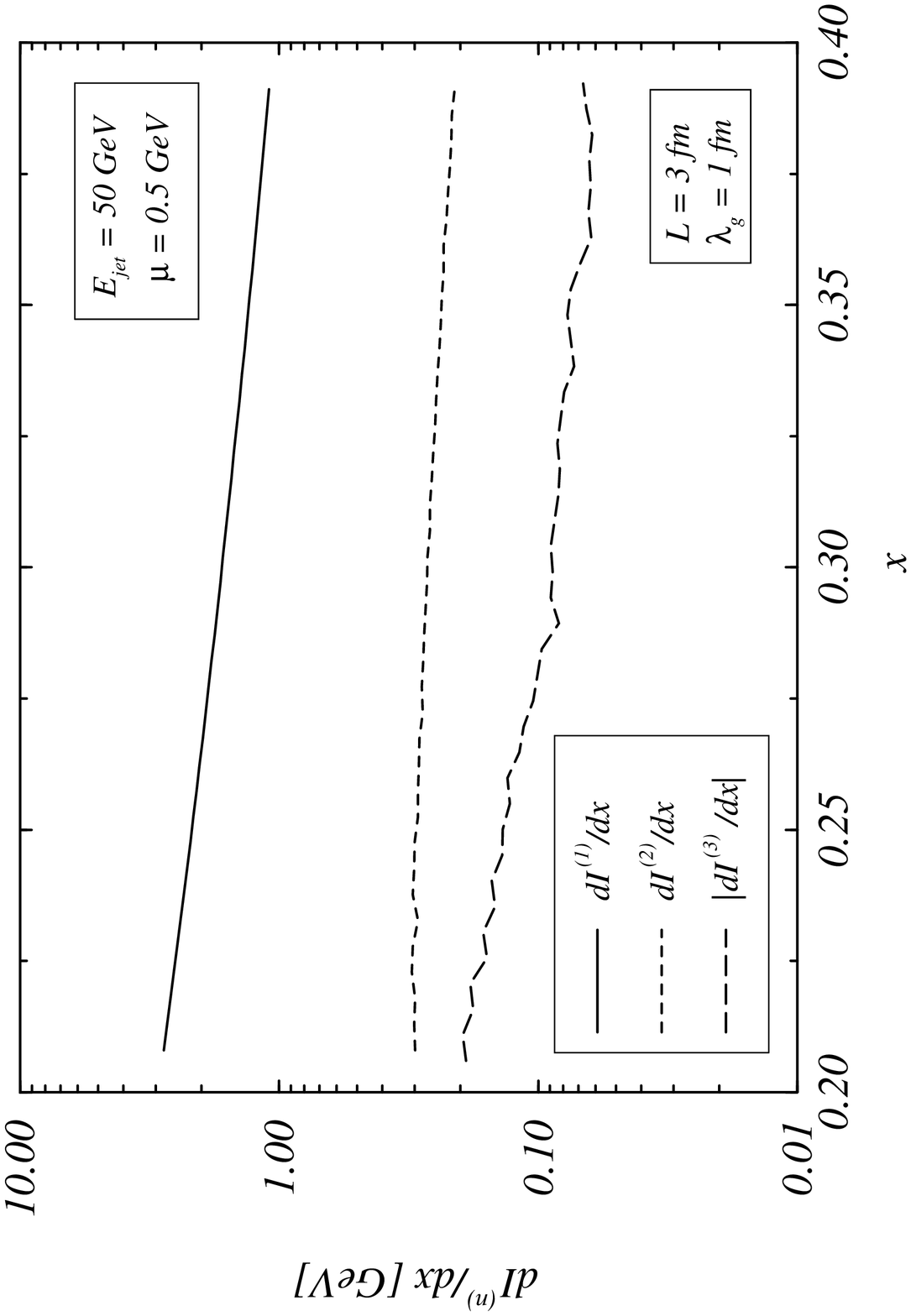}
\vskip - 20pt
\begin{minipage}[t]{15.0 cm}
{\small {FIG~5.}  (a) The radiation intensity distribution 
is plotted vs. the light-cone momentum fraction $x$ of the gluon. 
We consider a $50$~GeV quark jet in a plasma with screening scale
$\mu=0.5$~GeV and $\lambda_g=1$ fm. 
The solid curve shows  the dominant medium-independent radiation
intensity. The medium-induced gluon spectrum 
is plotted for up to third order in 
opacity ($dI^{(1)},dI^{(1+2)}$ and $dI^{(1+2+3)}$ )   
 for opacities $L/\lambda_g=1,2,3$.
(b) The absolute value of the orders in opacity $dI^{(1)},dI^{(2)}$
and $dI^{(3)}$ that contribute
in part (a)
 are plotted for the same energy and opacity $L/\lambda_g = 3$.}
\end{minipage}
\end{center}
\vskip 4truemm

It is surprising how much the first order result dominates
the induced intensity distribution\cite{GLVPRL}.
Fig.~5b illustrates how rapidly the
contributions from higher orders in opacity
decrease. While the differential angular distribution
continue to show some weak sensitivity to higher order contributions
in opacity,
the angle integrated intensity is much less affected  beyond first order.
The third order contribution
is almost buried in the  ``numerical noise''.
 For opacities  $L/\lambda_g \simeq 6$ some more pronounced
probability redistribution can be seen in 
the double differential level Fig.~4.  However, when
integrated over ${\bf k}^2$ and then
 over   $x_{\min}  < x < x_{\max}$ to obtain the 
induced non-abelian energy loss $\Delta E~{(ind)}$,
the first order result dominates very strongly
as seen in Fig.~6.
\begin{center}
\vspace*{10.2cm}
\includegraphics{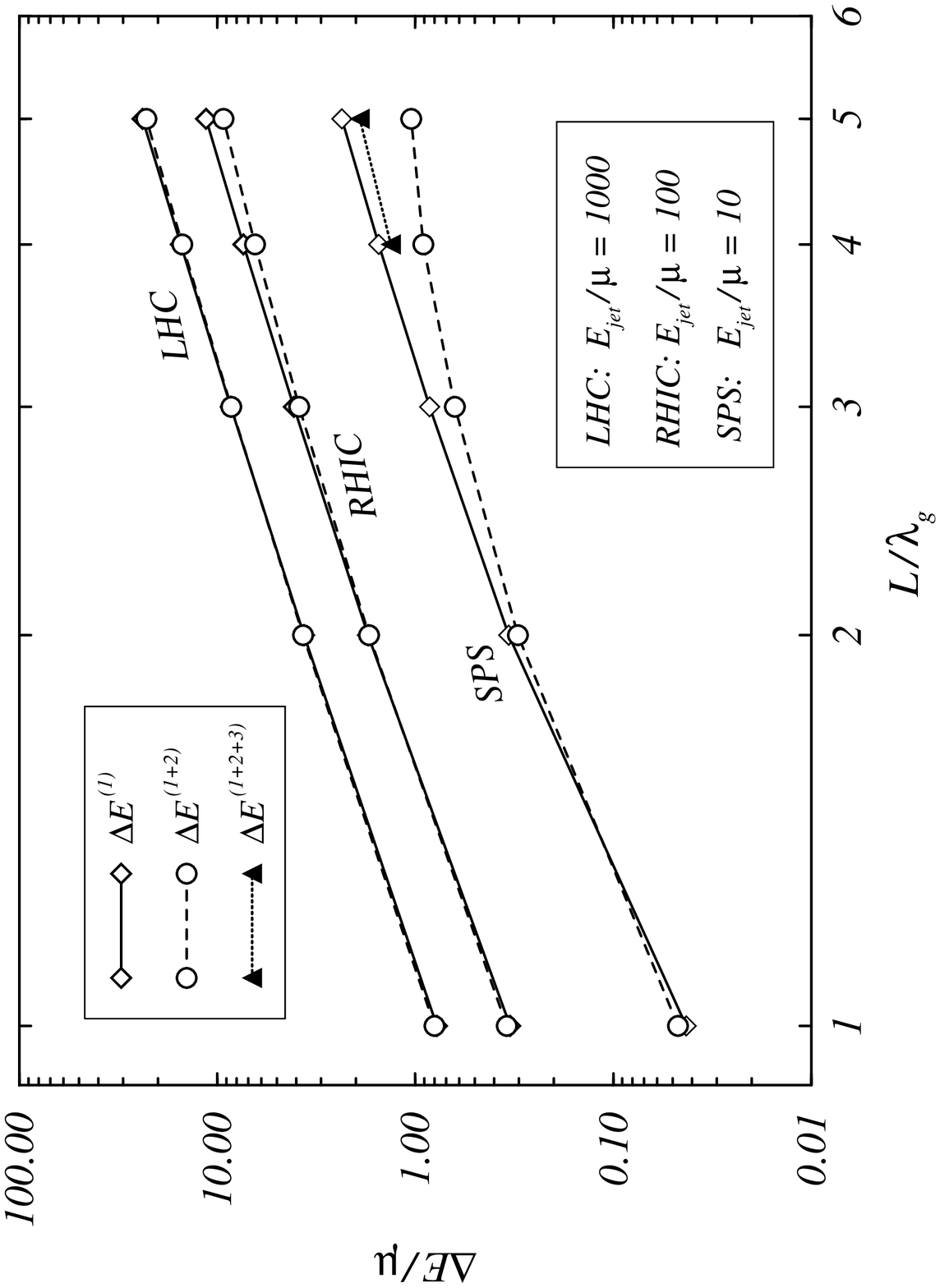}
\vskip - 20pt
\begin{minipage}[t]{15.0 cm}
{\small {FIG~6.}  The radiated energy loss of a quark jet
        with energy $E_{jet}=5,50,500$~GeV  (at SPS, RHIC, LHC) 
        is plotted as a function of 
        the opacity $L/\lambda_g$.  ($\lambda_g=1$~fm, $\mu=0.5$~GeV).
        Solid curves show the first order in opacity
results. The  dashed curves show
        results up to second order in opacity, and two third order
 results are shown by solid triangles for SPS energies. 
}
\end{minipage}
\end{center}
\vskip 4truemm    

The induced energy loss shown in Fig.~6 is
for quark jets with energies $E_{jet} = 5,\, 50,\, 500$~GeV typical
for SPS, RHIC and LHC. Higher orders in opacity $(L/\lambda_g)^n,\;\; 
n\geq 2$ have little effect except at very low SPS
energies. The kinematic bounds at SPS suppress very much
the energy loss in comparison to RHIC and LHC energies.
An analysis of the slopes as a function
of the opacity, $L/\lambda_g$,  shows that 
$\Delta E(ind)\propto L^{2\pm0.1}$
 at all energies even with finite kinematic
boundaries included. As a measure of the deviation
of the simple first order analytic estimate, 
we  generalize Eq.~(\ref{de1})  as follows:
\beq
\Delta E^{(ind)}=\frac{C_R\alpha_s}{N(E)}\, 
\frac{L^2\mu^2}{\lambda_g} \,\log \frac{E}{\mu}  \;\;,
\eeq{degen}
If the kinematic bounds are ignored as in Eq.~(\ref{didx1app}),
then $N(E)=4$.
Including  finite kinematic constraints
cause $N(E)$ to deviate considerably from this asymptotic value.
We find numerically that 
 $N(E)=7.3,\, 10.1,\, 24.4$  for $E=500,\, 50,\, 5$ GeV.
Together with the logarithmic
dependence of energy, these kinematic effects suppress greatly
the energy loss at lower (SPS) energies as seen in Fig.~6.
This is in contrast to the asymptotic,  energy independent result
quoted in Ref.~\cite{BDMPS2} 
where the finite kinematic bounds were neglected.
Our numerical results, however, agree with Ref.~\cite{BDMPS2} near LHC
energies as reported in \cite{GLVPRL}.

\vspace{0.5cm} 
\subsection*{Acknowledgments} 
\noindent 

We thank Al Mueller and Urs Wiedemann for numerous
clarifying discussions. 
This work was supported by the DOE Research Grant under Contract No.
De-FG-02-93ER-40764, partly by the US-Hungarian Joint Fund No.652
and OTKA No. T032796.

\vspace{1cm}

\begin{appendix}

\section{Diagrams $M_{1,0,0}=G_0X_{1,0}$,   $M_{1,1,0}=G_0G^{-1}X_{1,0}G_1$  
and   $M_{1,0,1}=G_0X_{1,1}$  }

To make contact with the results in 
Ref.~\cite{GLV1B,GLV1A}  we present explicit calculation
of  the  first nontrivial diagrams shown in Fig.~7. 
\begin{center}
\vspace*{4.9cm}
\includegraphics{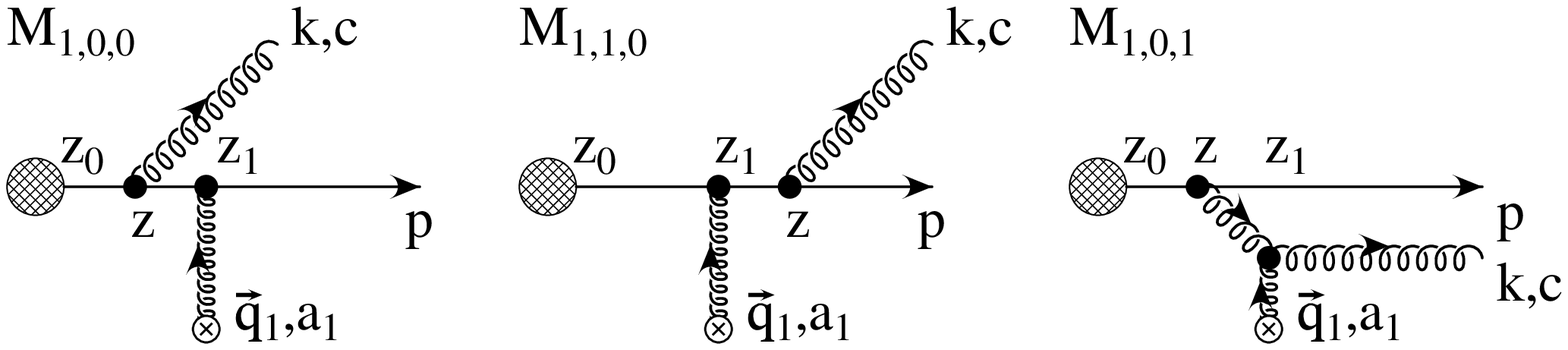}  
\vskip -20pt
\begin{minipage}[t]{15.0cm}
{\small {FIG~7.}
Three ``direct'' terms 
 $M_{1,0,0}=G_0X_{1,0}$,   $M_{1,1,0}=G_0G^{-1}X_{1,0}G_1$  
and   $M_{1,0,1}=G_0X_{1,1}$
contribute to the soft gluon radiation amplitude 
to first order in opacity $  L/\lambda \propto \sigma_{el}/A_\perp$.}
\end{minipage}
\end{center}
\vskip 4truemm
As a first application, consider the one rescattering
amplitude $M_{1,0,1}$, in the notation of Ref.~\cite{GLV1B}.  
\beqar
M_{1,0,1} &=& \,\int\frac{d^4 q_1}{(2\pi)^4} \; iJ(p+k-q_1)
e^{i(p+k-q_1)x_0}\,\Lambda_1(p,k,q_1)V(q_1)e^{iq_1x_1} 
\, \times \nonumber \\[1ex]
&\;& \times \; i\Delta(p+k-q_1)(-i)\Delta(k-q_1)  
\nonumber \\[1ex]
& \approx&  J(p+k)
e^{i(p+k)x_0} \;[c,a_1] T_{a_1} \,(-i)\int
 \frac{d^2 {\bf q}_{1}}{(2\pi)^2}
\,e^{-i{\bf q}_{1} \cdot ({\bf x}_{ 1}-{\bf x}_{ 0})}
\, 2g_s\,{ \bbox{\epsilon} 
\cdot({\bf k}-{\bf q}_{1})} \, \times
\nonumber \\[1ex]
&\;& \times \; E^+  \, \int \frac{dq_{1z}}{2\pi}v(q_{1z},{\bf q}_{1}) 
\Delta(p+k-q_1)\Delta(k-q_1)
\,e^{-iq_{1z}(z_1-z_0)}  \;\; .
\eeqar{101}
The longitudinal momentum transfer integral 
\beq
I_1(p,k,{\bf q}_{1},z_1-z_0) \equiv
 \, \int \frac{dq_{1z}}{2\pi}v(q_{1z},{\bf q}_{1}) 
\Delta(p+k-q_1)\Delta(k-q_1) \,e^{-iq_{1z}(z_1-z_0)}  \;\; 
\eeq{int1}
can be performed via
closing the contour below the real axis since $z_1 > z_0$.
In addition to the potential singularities at $\pm i\mu_{1}$ 
($\mu^2_i\equiv \mu^2_{i\perp}={\bf q}_i^2+\mu^2 $),
the two propagators have four poles in the $q_{1z}$ plane,
which up to leading correction are located at
\beq
\begin{array}{ll}
\bar{q}_{1}= E^+  +i\epsilon\;, \qquad   &
\bar{q}_{2}=-\omega_0-i\epsilon\;\;,    \\ 
\bar{q}_{3}=k^+ - \omega_1 +i\epsilon\;,\qquad   &
\bar{q}_{4}=-\omega_0 + \omega_1 -i\epsilon\;\;. 
\end{array}
\eeq{res20}
The residues give
\beq
{\rm Res}(\bar{q}_{2})\approx-v(-\omega_0,{\bf q}_{1})
\frac{e^{i\omega_0(z_1-z_0)}}{E^+k^+\omega_1}\;, \quad
{\rm Res}(\bar{q}_{4})\approx 
v(\omega_1-\omega_0,{\bf q}_{1})\frac{e^{i (\omega_0 -\omega_1 )
(z_1-z_0)}}{E^+k^+\omega_1}\;\;, 
\eeq{resid1}
while
\beq
{\rm Res}(-i\mu_{1})\approx\frac{4\pi \alpha_s \, 
e^{-\mu_{1}(z_1-z_0)}}
{(-2i\mu_{1}) E^+k^+(-i\mu_{1})^2} \;\;,
\eeq{resid2}
where we assumed that $k^+\gg\mu_{1}\gg \omega_i$.
In the well-separated case where $\mu(z_1-z_0) = \mu \lambda\gg 1$,
this potential residue is exponentially suppressed
and therefore 
\beqar
I_1(p,k,{\bf q}_{1},z_1-z_0)&\approx& i
\frac{e^{i\omega_0(z_1-z_0)}}{E^+k^+\omega_1}
\left(v(-\omega_0,{\bf q}_{1})-
 v(\omega_1-\omega_0,{\bf q}_{1}) e^{-i\omega_1 
(z_1-z_0)}\right)
\nonumber \\[1ex]
&\approx& i v(0,{\bf q}_{1})\frac{e^{i\omega_0(z_1-z_0)}}
{E^+k^+\omega_1}
\left(1-e^{-i\omega_1 (z_1-z_0)}\right) \;\;.
\eeqar{int1a}
We  thus  recover in this eikonal limit the time ordered 
perturbation result in~\cite{GLV1B,GLV1A}
\beqar
&&M_{1,0,1} = J(p)e^{ipx_0} 
 \, (-i) \int \frac{d^2 {\bf q}_{1}}{(2\pi)^2}  v(0,{\bf q}_{1})
\,e^{-i{\bf q}_{1} \cdot {\bf b}_1}
\; \times \nonumber \\[1ex]
&\;&  \qquad \qquad \times \; 2ig_s\,\frac{ { \bbox{\epsilon} 
\cdot({\bf k}-{\bf q}_{1})}}
{({\bf k}-{\bf q}_1 \,)^2} \;  
e^{i(\omega_0-\omega_1)z_1}(e^{i\omega_1 z_1}
-e^{i \omega_1 z_0}  ) \; [c,a_1] T_{a_1}\;\; ,
\eeqar{101b} 
where ${\bf b}_1={\bf x}_{1}-{\bf x}_{0}$.
Similarly, we recover the other two amplitudes for one center, e.g., 
\beqar
&&M_{1,1,0} = J(p)e^{ipx_0} 
\, (-i)\int \frac{d^2 {\bf q}_{1}}{(2\pi)^2}  v(0,{\bf q}_{1})
\,e^{-i{\bf q}_{1} \cdot {\bf b}_1} \; \times \nonumber \\[1ex]
&\;& \qquad \qquad
\times \; (-2ig_s)\,\frac{ { \bbox{\epsilon} \cdot {\bf k}}}
{{\bf k}^2} \;  e^{i\omega_0 z_1} \; c a_1 T_{a_1}\;\; .
\eeqar{110}

Note that while the direct $\langle M_{1,m,l} 
M^\dagger_{1,m^{\prime},l^{\prime}}\rangle$
ensemble averages lead to  non-vanishing modifications
to the initial hard spectrum, $|M_0|^2\propto |J(p)|^2 /{\bf k}^2$,
the interference between $M_0$ and $M_{1,l,m}$
vanishes because the ensemble average is over a color neutral
medium with $\tr\, T_a(i)=0$. 

\section{Diagram $M_{2,0,3}=G_0X_{1,1}X_{2,1}$}

Consider next the gluon two-scatterings  amplitude $M_{2,0,3}$ 
in~\cite{GLV1A}. 
Fig.~8 shows that for inclusive processes two interesting cases arise.
\begin{center}
\vspace*{5.0cm}
\includegraphics{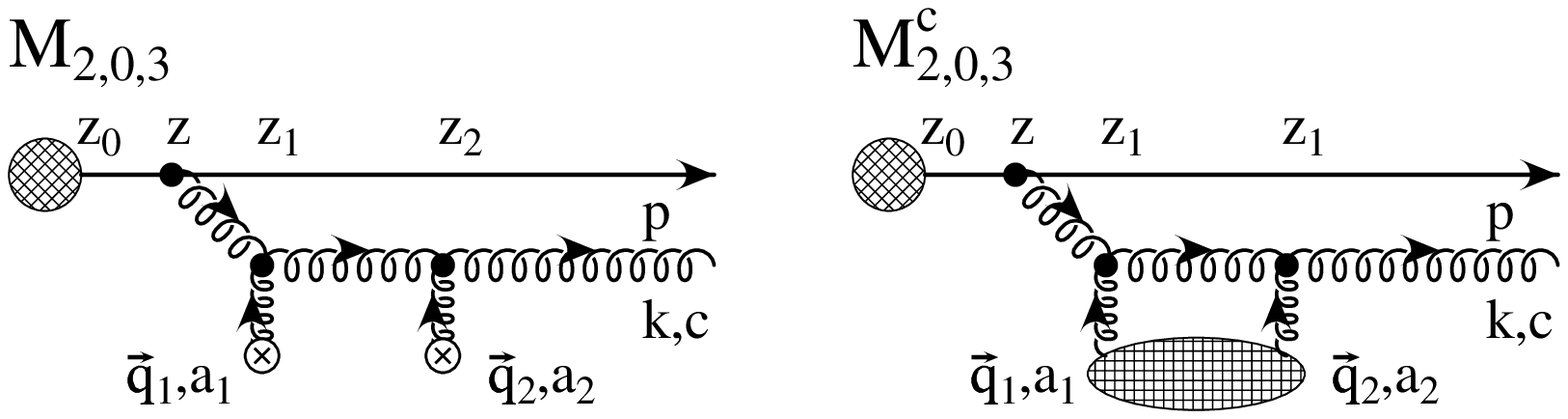}  
\vskip -15pt
\begin{minipage}[t]{15.0cm}
{\small { FIG~8.}
$M_{2,0,3}=G_0X_{1,1}X_{2,1}$  ``direct'' contributes to  
second order  in opacity 
$\propto (\sigma_{el}/A_\perp)^2$,
whereas $M^c_{2,0,3}=M_{2,0,3}(z_2=z_1) \equiv G_0O_{1,1}$ 
``contact-limit'' may  contribute to first order  in opacity 
$\propto (\sigma_{el}/A_\perp)^1$ as well.}
\end{minipage}
\end{center}
\vskip 4truemm
In the Feynman diagram approach
\beqar
M_{2,0,3}&=&\int\frac{d^4 q_1}{(2\pi)^4}\frac{d^4 q_2}{(2\pi)^4}
\; iJ(p+k-q_1-q_2)e^{i(p+k-q_1-q_2)x_0} 
V(q_1)e^{iq_1x_1} V(q_2)e^{iq_2x_2} \; \times \nonumber\\[1ex]
&\;& \times \; \Lambda_{12}(p,k,q_1,q_2) \; i\Delta(p+k-q_1-q_2) 
(-i)\Delta(k-q_1-q_2)(-i)\Delta(k-q_2) \; \times
\nonumber \\[1ex]
&\approx&  J(p+k)e^{i(p+k)x_0}\; [[c,a_2],a_1] (T_{a_2}(2)T_{a_1}(1))
\; \times \nonumber \\[1ex]
&\;& \times \; (-i)\int\frac{d^2 {\bf q}_{1}}{(2\pi)^2}
 (-i)\int \frac{d^2 {\bf q}_{2}}{(2\pi)^2}
\; 2ig_s {\bbox{\epsilon}
\cdot({\bf k}-{\bf q}_{1}-{\bf q}_{2})} 
e^{-i{\bf q}_{1}\cdot{\bf b}_{1}} 
e^{-i{\bf q}_{2}\cdot{\bf b}_{2}}
\; \times \nonumber \\[1ex]
&\;&\times \; \int\frac{d q_{1z}}{(2\pi)}\frac{d q_{2z}}{(2\pi)}
\frac{(E^+k^+)\, v(\vec{\bf q}_1)v(\vec{\bf q}_2)e^{-iq_{1z}(z_{1}-z_0)}
e^{-iq_{2z}(z_{2}-z_0)}}
{((p+k-q_1-q_2)^2+i\epsilon)((k-q_1-q_2)^2+i\epsilon)
((k-q_2)^2+i\epsilon)} \;\;,  
\eeqar{203b}
where ${\bf b}_i={\bf x}_{i}-{\bf x}_{0}$ are transverse impact parameters,
and we used the soft glue and rescattering kinematical simplifications
Eqs.~(\ref{softness},\ref{lam2}), e.g. $J(p+k-q_1-q_2)\approx J(p+k)
\approx J(p)$.
For the $q_{1z}$ integral, it is convenient to  rewrite the phase as
$$ e^{-iq_{1z}(z_{1}-z_0)}
e^{-iq_{2z}(z_{2}-z_0)}=e^{-i(q_{1z}+q_{2z})(z_{1}-z_0)}
e^{-iq_{2z}(z_{2}-z_1)}\;\; .$$ 
The first longitudinal integral
is closely related to Eq.~(\ref{int1})
\beq
I_2(p,k,{\bf q}_{1}, \vec{\bf q}_2,z_1-z_0) = \int\frac{d q_{1z}}{2\pi}
\frac{ v(q_{1z},{\bf q}_{1})e^{-i(q_{1z}+q_{2z})(z_{1}-z_0)}}
{((p+k-q_1-q_2)^2+i\epsilon)((k-q_1-q_2)^2+i\epsilon)} \;\;.
\eeq{i2}
Since $z_1-z_0\gg 1/\mu$,  we again close the contour  in the lower 
half $q_{1z}$ plane and neglect the pole at $- i\mu_{1}$.
The  remaining $q_{1z}$ poles are shifted by 
$-q_{2z}$ and ${\bf q}_{1}\rightarrow {\bf q}_{1}+{\bf q}_{2}$
relative to  Eq.~(\ref{res20}):
\beq
\begin{array}{ll}
\bar{q}_{1}= -q_{2z}+E^+  +i\epsilon\;, \qquad &
\bar{q}_{2}=-q_{2z}-\omega_0-i\epsilon\;,  \\
\bar{q}_{3}=-q_{2z}+k^+ -
\omega_{(12)} +i\epsilon\;,\qquad   &
\bar{q}_{4}=-q_{2z}-\omega_0 +
\omega_{(12)} -i\epsilon\;\;,
\end{array}
\eeq{res21}
where now $k^+ \omega_{(12)}=
({\bf k}-{\bf q}_{1}-{\bf q}_{2})^2.$
The residues at $\bar{q}_2,\, \bar{q}_4$ then give
\beq
I_2\approx i \frac{e^{i\omega_0(z_1-z_0)}}{E^+k^+\omega_{(12)}}
\left(v(-q_{2z}-\omega_0,{\bf q}_{1}) -
v(-q_{2z}-\omega_0+\omega_{(12)},{\bf q}_{1})
e^{-i\omega_{(12)}(z_1-z_0)}\right)
\;\; ,\eeq{i2a}
where we have neglected ${\cal O} ({\rm exp}(-\mu \lambda))$ contributions. 
This differs from Eq.~(\ref{int1a}) mainly in that the potential is evaluated
near $-q_{2z}$, which still remains to be integrated over,
and $\omega_1\rightarrow \omega_{(12)}$.

Next we need the following  {\em critical} $q_{2z}$ integral
\beqar
I_3(k,{\bf q}_{1},{\bf q}_{2},z_2-z_1) &\equiv& 
\int\frac{d q_{2z}}{2\pi}
\frac{ v(-q_{2z}+\delta\omega,
{\bf q}_{1})v(q_{2z},{\bf q}_{2})e^{-iq_{2z}(z_2-z_1)}}
{((k-q_2)^2+i\epsilon)} \;\;.
\eeqar{i3a}
Fortunately, we are interested in only two extreme limits: 
\begin{itemize}
\item{The limit of  well-separated scattering centers $z_2-z_1\gg 1/\mu$} ;
\item{The special ``contact'' $z_2=z_1$ limit to compute unitary 
contributions.}
\end{itemize}
In the first case, we can proceed ignoring the $q_{2z}=-i\mu_{1}$ 
and $-i\mu_{2}$ potential singularities. Then only the pole at 
$q_{2z}=\omega_2-\omega_0-i\epsilon$ contributes and yields
\beqar
I_3(k,{\bf q}_{1},{\bf q}_{2},z_2-z_1\gg1/\mu) &\approx&
-\frac{i}{k^+}v(0,{\bf q}_{1}) v(0,{\bf q}_{2})
e^{-i(\omega_2-\omega_0)(z_2-z_1)}
\;\; .
\eeqar{i3a1}
In this case we recover the result quoted in Ref.~\cite{GLV1B}
 for $M_{2,0,3}$
\beqar
M_{2,0,3}&\approx &  J(p)e^{ipx_0}
(-i)\int\frac{d^2 {\bf q}_{1}}{(2\pi)^2}
\,v(0,{\bf q}_{1}) \, e^{-i{\bf q}_{1}\cdot{\bf b}_{1}}
(-i)\int\frac{d^2 {\bf q}_{1}}{(2\pi)^2}
\, v(0,{\bf q}_{2}) \, e^{-i{\bf q}_{2}\cdot{\bf b}_{2}}
\; \times \nonumber \\[1ex] 
&\;& \times \; (2ig_s)\frac{{ \bbox{\epsilon}
\cdot({\bf k}-{\bf q}_{1}-{\bf q}_{2})}} 
{({\bf k}-{\bf q}_{1}-{\bf q}_{2})^2}
e^{i(\omega_0-\omega_2)z_2}e^{i(\omega_2-\omega_{(12)})z_1}
\left(e^{i\omega_{(12)}z_1}- e^{i\omega_{(12)}z_0} \right)
\; \times \nonumber \\[1ex]
&\;& \times \; [[c,a_2],a_1] (T_{a_2}T_{a_1})\;\;.
\eeqar{203c}

In the  general case  (including the special contact case with $z_2=z_1$)
both $q_{2z}=-i\mu_{2},\; -i\mu_{1}  $ singularities
in the Yukawa potential contribute together with the pole at 
$q_{2z}=\omega_2-\omega_0-i\epsilon$,
resulting in 
\beqar
I_3(k,{\bf q}_{1},{\bf q}_{2},z_2-z_1) &\approx&
\frac{-i}{k^+} \left( \, v(0,{\bf q}_{1}) v(0,{\bf q}_{2})
\;e^{-i(\omega_2-\omega_0)(z_2-z_1)}  \right. \nonumber \\[1ex]
 &\;& \; \left.\;\;  - 
\frac{(4\pi \alpha_s)^2}{2\,(\mu_{1}^2-\mu_{2}^2)} \left(
\frac{e^{-\mu_{2} (z_2-z_1)}}{\mu_{2}^2}
-\frac{e^{-\mu_{1} (z_2-z_1)} e^{-i\delta\omega (z_2-z_1)}}
{\mu_{1}^2} \right)\, \right)
\;\; .
\eeqar{i3a2}
For $z_2-z_1=\lambda\gg 1/\mu$ this 
reduces to Eq.~(\ref{i3a1}). For the special 
contact contribution $z_2-z_1=0$ it reduces to
\beqar
I_3(k,{\bf q}_{1},{\bf q}_{2},0) &\approx&
\frac{-i}{2\,k^+} v(0,{\bf q}_{1}) v(0,{\bf q}_{2})\;\; ,
\eeqar{i3a3}
i.e., exactly $\half$ of the strength in Eq.~(\ref{i3a1}).
The contact limit of this amplitude is therefore
\beqar
M^c_{2,0,3}&\approx &  J(p)e^{ipx_0}
(-i)\int \frac{d^2 {\bf q}_{1}}{(2\pi)^2}
\, v(0,{\bf q}_{1})\,
e^{-i{\bf q}_{1}\cdot {\bf b}_{1}}
(-i)\int\frac{d^2 {\bf q}_{2}}{(2\pi)^2}
\, v(0,{\bf q}_{2})\,
e^{-i{\bf q}_{2}\cdot {\bf b}_{2}}
\; \times \nonumber \\[1ex]
&\;&  \times 
\; \frac{1}{2}\;(2ig_s)\;\frac{{ \bbox{\epsilon}
\cdot({\bf k}-{\bf q}_{1}-{\bf q}_{2})}} 
{({\bf k}-{\bf q}_{1}-{\bf q}_{2})^2}
\;  e^{i\omega_0z_1}
\left(1- e^{-i\omega_{(12)}(z_1-z_0)} \right)
\; [[c,a_2],a_1] (T_{a_2}T_{a_1})\;\;.
\eeqar{203d}

After this explicit derivation of the factor $\half$ in the contact limit,
we can generalize it to  any functional form of the potential
as follows. First we  assume that a contact interactions only contributes 
for  $\vec{\bf q}_1=-\vec{\bf q}_2$, corresponding to no net transverse 
momentum exchange {\bf inside the potential function only}, 
i.e. $v(q_{1z},{\bf q}_{1})=v(-q_{2z},-{\bf q}_{2})$. However  the 
distinction has to be kept in the propagators and phases.
In this case the potential function $v(\vec{\bf q}_1), $ does not appear in the 
first integral $I_2$, Eq.~(\ref{i2}), which is modified to
\beqar
\bar{I}_2(p,k,{\bf q}_{1}, \vec{\bf q}_2,z_1-z_0) 
&=& \int\frac{d q_{1z}}{2\pi}
\frac{ e^{-i(q_{1z}+q_{2z})(z_{1}-z_0)}}
{((p+k-q_1-q_2)^2+i\epsilon)((k-q_1-q_2)^2+i\epsilon)}
\nonumber \\[1ex]
&\approx i&\frac{e^{i\omega_0(z_1-z_0)}}{E^+k^+\omega_{(12)}}
\left(1- e^{-i\omega_{(12)}(z_1-z_0)}\right) \;\; .
\eeqar{i2aa} 
This is independent of the functional form of the potential.
The second integral, $I_3$, then is modified to
\beqar
\bar{I}_3(k, {\bf q}_{1}={\bf q}_{2},z_2-z_1) 
&\equiv& \int\frac{d q_{2z}}{2\pi}
\frac{ |v(q_{2z},{\bf q}_{2})|^2 
e^{-iq_{2z}(z_2-z_1)}}{((k-q_2)^2+i\epsilon)}
\nonumber \\[1ex]
&\approx& \frac{1}{k^+} \int\frac{d q_{2z}}{2\pi}
 \frac{ |v(q_{2z},{\bf q}_{2})|^2 e^{-iq_{2z}(z_2-z_1)}}
{q_{2z}-\delta \omega +i\epsilon}
\;\; ,
\eeqar{i33}
where $z_2-z_1 \equiv \lambda$ and we have 
used  the finite range of $|v(q_z,q)|^2$, i.e. $v$ is limited 
within $(-\mu, +\mu)$,  to neglect the residue at $q_{2z} \approx k^+$.
 For large $z_2-z_1$ compared to the range of $v$, we close below
and obtain the dominant residue
\beq
 \bar{I}_3\approx -\frac{i}{k^+} |v(0,{\bf q}_{2})|^2
\; e^{-i(z_2-z_1) \delta\omega}
\eeq{i33a1}
independent approximately of the actual form of $v$
as long as
 all singularities of $v$ in the lower half plane have imaginary parts
$-i\mu_{i}$ with 
$ \mu_{i} \lambda  = \mu_{i}(z_2-z_1)  \gg 1$.

To extract the contact limit, $z_2-z_1=0$,
we can integrate
along the real $q_{2z}$ axis and use the Dirac representation
of a pole approaching the real axis:
$$\frac{1}{q_{2z}-\delta\omega+i\epsilon}=\frac{q_{2z}-\delta\omega}
{ (q_{2z}-\delta\omega)^2+\epsilon^2} - i\pi \delta(q_{2z}-\delta\omega)
\;\; ..$$
This gives
\beqar
k^+e^{i\delta\omega (z_2-z_1)} \bar{I_3}&\approx& 
-\frac{i}{2}|v(\delta\omega,{\bf q}_{2})|^2 +
\int_{-\infty}^\infty
\frac{d\tilde{q}_z}{2\pi}  |v(\tilde{q}_z+\delta\omega,{\bf q}_{2})|^2 
\frac{\tilde{q}_z\cos(\tilde{q}_z\lambda)-i\tilde{q}_z\sin(\tilde{q}_z\lambda)}
{\tilde{q}_z^2+\epsilon^2} \;\;,
\eeqar{kib3}
where we have introduced $\tilde{q}_z=q_{2z}-\delta\omega$.
For our high energy eikonal applications $\delta\omega \ll \mu$,
and therefore we can now expand $|v|^2$ around $\delta\omega=0$. 
The correction of ${\cal O}(\delta\omega/\mu)$
can be neglected. This leads to a vanishing real part since that
part of the integrand is odd. On the other hand, the imaginary part
has a finite $\epsilon=0$ limit
\beq
-i \int_{-\infty}^\infty
\frac{d \tilde{q}_z}{2\pi}  |v(\tilde{q}_z,{\bf q}_{2})|^2 
\frac{\sin(\tilde{q}_z\lambda)}{\tilde{q}_z}
 \;\;=\;\;  \left\{
\begin{array}{ll}
0  & \quad {\rm if } \;\; \mu\lambda =\mu(z_2-z_1)  \rightarrow 0 
\\[.5ex]
-\frac{i}{2} |v(0,{\bf q}_{2})|^2 &  \quad{\rm if } \;\; 
\mu\lambda =\mu(z_2-z_1)   \rightarrow \infty
\end{array} \right. \;\;,
\eeq{lims}
where $\mu$ is the range of the potential $v$ as indicated above.
We see that while the detailed interpolation between
the asymptotic limits depends on the actual functional form
of $v$, the limits Eqs.~(\ref{i3a1},\ref{i3a3}) and the factor of $\half$
reduction in the contact $(\lambda=0)$ limit is very general.

We thus confirm  that $O_{i,1}$ diagrams where the two legs are 
attached to the gluon line can be obtained from the higher order 
direct diagram  by setting $z_{i+1}=z_i$ and reducing the strength
twice (i.e. times $\half$). 

\section{Diagrams $M_{2,0,0}=G_0X_{1,0}X_{2,0}$ and 
$M_{2,2,0}=G_0X_{1,0}G^{-1}X_{2,0}G_2$}

In those graphs it is the jet rather than the gluon
that suffers two sequential scatterings as seen from Fig.~9.
\begin{center}
\vspace*{11.0cm}
\includegraphics{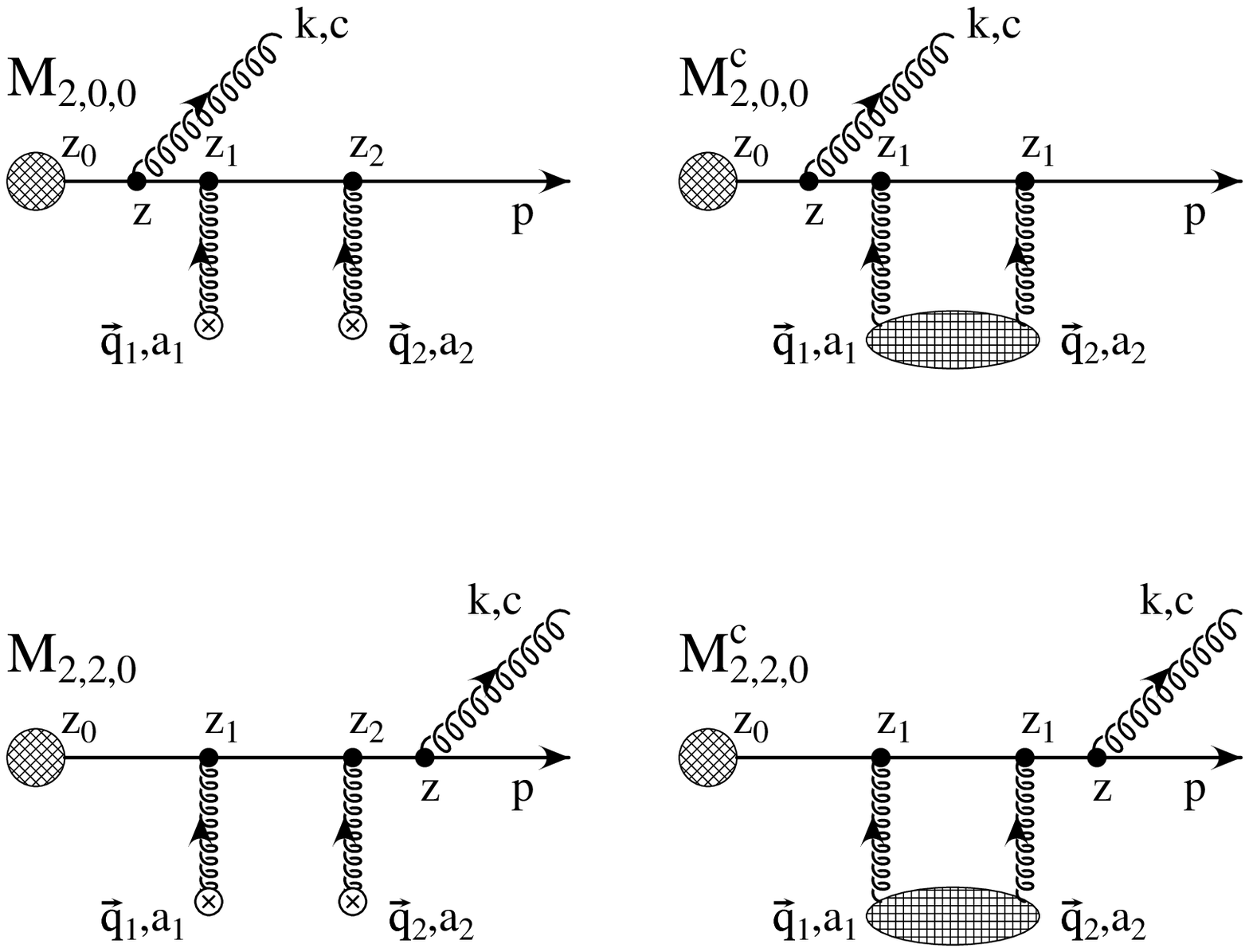}  
\vskip -30pt
\begin{minipage}[t]{15.0cm}
{\small { FIG~9.}
 $M_{2,0,0}=G_0X_{1,0}X_{2,0}$ and 
$M_{2,2,0}=G_0X_{1,0}G^{-1}X_{2,0}G_2$  graphs in the well-separated case 
together with their $z_2=z_1$ limits  
$M^c_{2,0,0}\equiv G_0O_{1,0} $, $M^c_{2,2,0}\equiv 
G_0G^{-1}O_{1,0}G_1  $.}
\end{minipage}
\end{center}
\vskip 4truemm
It is straightforward to write the expression for the amplitude
\beqar
M_{2,0,0}&=&\int\frac{d^4 q_1}{(2\pi)^4}\frac{d^4 q_2}{(2\pi)^4}
\; iJ(p+k-q_1-q_2)e^{i(p+k-q_1-q_2)x_0} 
V(q_1)e^{iq_1x_1} V(q_2)e^{iq_2x_2} \; \times \nonumber\\[1ex]
&\;& \times \; (-iE^+)^2ig_s(2p+k)_\mu \epsilon^\mu \; i\Delta(p+k-q_1-q_2) 
\,i\Delta(p-q_1-q_2)\,i\Delta(p-q_2) \; \times
\nonumber \\[1ex]
&\;& \times \; a_2 a_1 c(T_{a_2}T_{a_1}) \nonumber \\[1ex]
&\approx&  J(p)e^{ipx_0}(-i)\int\frac{d^2 {\bf q}_{1}}{(2\pi)^2}
e^{-i{\bf q}_{1}\cdot{\bf b}_{1}} 
(-i)\int \frac{d^2 {\bf q}_{2}}{(2\pi)^2}
e^{-i{\bf q}_{2}\cdot{\bf b}_{2}}  \; \times \nonumber  \\[1ex]
&\;& \times \frac{2ig_s(\bbox{\epsilon}\cdot{\bf k})}{x} 
\; e^{i \omega_0 z_0} \; a_2a_1 c (T_{a_2}T_{a_1})\;(E^+)^2\
\; \times \nonumber \\[1ex]
&\;&\times \; \int\frac{d q_{1z}}{(2\pi)}\frac{d q_{2z}}{(2\pi)}
\frac{ v(\vec{\bf q}_1)v(\vec{\bf q}_2)e^{-iq_{1z}(z_{1}-z_0)}
e^{-iq_{2z}(z_{2}-z_0)}}
{((p+k-q_1-q_2)^2+i\epsilon)((p-q_1-q_2)^2+i\epsilon)
((p-q_2)^2+i\epsilon)} \;\;. 
\eeqar{200}
In this case we define
\beq
I_2(p,k,{\bf q}_{1},\vec{\bf q}_2,z_1-z_0) = \int\frac{d q_{1z}}{2\pi}
\frac{ v(q_{1z},{\bf q}_{1})e^{-i(q_{1z}+q_{2z})(z_{1}-z_0)}}
{((p+k-q_1-q_2)^2+i\epsilon)((p-q_1-q_2)^2+i\epsilon)} \;\;..
\eeq{i2prim}
Since $z_1-z_0\gg 1/\mu$, we  neglect the pole at $- i\mu_{1}$.
The  remaining $q_{1z}$ poles are 
\beq
\begin{array}{ll}
\bar{q}_{1}= -q_{2z}+E^+ +k^+ +i\epsilon\;, \qquad   &
\bar{q}_{2}=-q_{2z}-\omega_0-i\epsilon\;,        \\
\bar{q}_{3}=-q_{2z}+E^+ +i\epsilon\;,\qquad   &
\bar{q}_{4}=-q_{2z}  -i\epsilon\;\;,
\end{array}
\eeq{res2002}
where we discarded $({\bf p}+{\bf k} - 
{\bf q}_{1}-{\bf q}_{2})^2/E^+$
relative to $\omega_0$.  The $\bar{q}_2,\, \bar{q}_4$ residues then give
\beq
I_2\approx -i \frac{v(-q_{2z},{\bf q}_{1})}{(E^+)^2\omega_{0}}
\left(e^{i\omega_0(z_1-z_0)}-1\right) \;\;.
\;\; \eeq{i2ab}
Not that $\omega_0$ has been neglected in the potential relative to
$\mu_{1}$. The second integral, $I_3$, is then modified to
\beqar
\bar{I}_3(p,{\bf q}_{1},{\bf q}_{2},z_2-z_1) 
&\equiv& \int\frac{d q_{2z}}{(2\pi)}
\frac{ v(-q_{2z},{\bf q}_{1}) \, v(q_{2z},
{\bf q}_{2}) \, e^{-iq_{2z}(z_2-z_1)}}{((p-q_2)^2+i\epsilon)}
\nonumber \\[1ex]
&\approx & \frac{i}{E^+}v(0,{\bf q}_{1})v(0,{\bf q}_{2}) 
\, \times \, \left\{\begin{array}{ll}
1 \quad  &{\rm if } \;\; \mu\lambda = \mu(z_2-z_1) \rightarrow \infty \\[.5ex]
\frac{1}{2} \quad  & {\rm if } \;\; \mu\lambda = \mu(z_2-z_1) \rightarrow 0
\end{array} 
\right. \;.
\eeqar{i3200}
Note that $E^+\omega_0={\bf k}^2/x$.
With the help of Eqs.~(\ref{i2ab},\ref{i3200})  in the case of well-separated
scattering centers we obtain
\beqar
M_{2,0,0}&=& J(p)e^{ipx_0}(-i)\int\frac{d^2 {\bf q}_{1}}{(2\pi)^2}
v(0,{\bf q}_{1})e^{-i{\bf q}_{1}\cdot{\bf b}_{1}} 
(-i)\int \frac{d^2 {\bf q}_{2}}{(2\pi)^2} v(0,{\bf q}_{2})
e^{-i{\bf q}_{2}\cdot{\bf b}_{2}}  \; \times \nonumber  \\[1ex]
&\;& \times \; \frac{2ig_s(\bbox{\epsilon}
\cdot{\bf k})}{{\bf k}^{\;2}} 
\; (e^{i\omega_0 z_1} - e^{i\omega_0 z_0}) \; a_2a_1 c (T_{a_2}T_{a_1}) \;\;.
\eeqar{200fin}
We thus recover the result from~\cite{GLV1B}.
In the contact limit there is extra factor of $\half$ relative to the
naive expectation and Eq.~(\ref{200fin}). Analogous calculation leads to
the expected result for $M_{2,2,0}$. 

We thus arrive at a conclusion similar to the one in Appendix~B, i.e. 
in the case when the two legs of the virtual contribution are attached to the
quark line,  $O_{i,0}$, there is an additional 
factor of $\half$ relative to just setting $z_{i+1}=z_i$ in the 
higher order well-separated case amplitudes.

\section{Diagrams $M_{2,0,1}=G_0X_{1,1}X_{2,0}$ and $M_{2,0,2} 
                = G_0X_{1,0}X_{2,1}$}

In the case when one of the hits is on the parent parton and the other hit is
on the radiated gluon we encounter a different situation. Explicit 
calculation shows this on the example of $M^c_{2,0,1} 
= M_{2,0,1}(z_2=z_1) $ in Fig.~10. 
\begin{center}
\vspace*{5.0cm}
\includegraphics{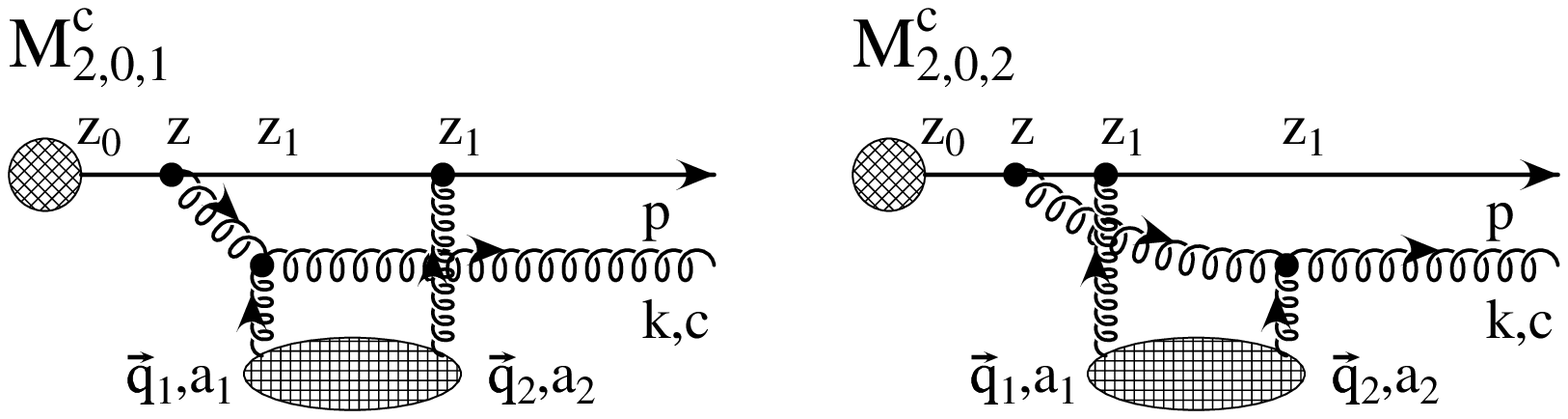}  
\vskip -20pt
\begin{minipage}[t]{15.0cm}
{\small { FIG~10.}
$M^c_{2,0,1}=G_0O_{1,2}$  and $M^c_{2,0,2}= G_0O_{1,2}$ topologically 
indistinct   contact diagrams. 
There are no additional factors of $\half$ arising
from the integration  in taking the $z_2= z_1$ limit
in  $M_{2,0,1}=G_0X_{1,1}X_{2,0}$ and $M_{2,0,2} 
= G_0X_{1,0}X_{2,1}$     .}
\end{minipage}
\end{center}
\vskip 4truemm

\beqar
M_{2,0,1}&=&\int\frac{d^4 q_1}{(2\pi)^4}\frac{d^4 q_2}{(2\pi)^4}
\; iJ(p+k-q_1-q_2)e^{i(p+k-q_1-q_2)x_0} 
V(q_1)e^{iq_1x_1} V(q_2)e^{iq_2x_2} \; \times \nonumber\\[1ex]
&\;& \times \; (-iE^+) \Lambda_1 \; i\Delta(p+k-q_1-q_2) 
\,(-i)\Delta(k-q_1)\,i\Delta(p-q_2) \; \times
\nonumber \\[1ex]
&\approx&  J(p)e^{ipx_0}(-i)\int\frac{d^2 {\bf q}_{1}}{(2\pi)^2}
e^{-i{\bf q}_{1}\cdot{\bf b}_{1}} 
(-i)\int \frac{d^2 {\bf q}_{2}}{(2\pi)^2}
e^{-i{\bf q}_{2}\cdot{\bf b}_{2}}  \; \times \nonumber  \\[1ex]
&\;& \times \;2ig_s(\bbox{\epsilon}\cdot({\bf k}-{\bf q}_{1})) 
\; e^{i \omega_0 z_0} \; a_2 [c,a_1] (T_{a_2}T_{a_1})\;(E^+)^2\
\; \times \nonumber \\[1ex]
&\;&\times \; \int\frac{d q_{1z}}{(2\pi)}\frac{d q_{2z}}{(2\pi)}
\frac{ v(\vec{\bf q}_1)v(\vec{\bf q}_2)e^{-iq_{1z}(z_{1}-z_0)}
e^{-iq_{2z}(z_{2}-z_0)}}
{((p+k-q_1-q_2)^2+i\epsilon)((k-q_1)^2+i\epsilon)
((p-q_2)^2+i\epsilon)} \;\;.  \nonumber \\[1ex]
\eeqar{201}
We perform the $q_{1z}$ integral first
\beq
I_2(p,k, {\bf q}_{1},  \vec{\bf q}_2,z_1-z_0) = \int\frac{d q_{1z}}{2\pi}
\frac{ v(q_{1z},{\bf q}_{1})e^{-i(q_{1z}+q_{2z})(z_{1}-z_0)}}
{((p+k-q_1-q_2)^2+i\epsilon)((k-q_1)^2+i\epsilon)} \;\;.
\eeq{201i2}
The pole at $-i\mu_{1}$ is again exponentially suppressed. 
The poles of interest in the lower half plane are 
 $q_{1z} = -q_{2z}-\omega_0 - i\epsilon$ and
$q_{1z} = -\omega_0 + \omega_1 - i\epsilon$.
Taking the residues leaves us with
\beq
I_2 = i\; \frac{e^{i\omega_0(z_{1}-z_0)}}
{E^+ k^+ (q_{2z}+\omega_1)} (v(-q_{2z}-\omega_0, {\bf q}_{1} )
 -  v(\omega_1-\omega_0, {\bf q}_{1} )
e^{-i (q_{2z} +\omega_1) (z_{1}-z_0)} ) \;\;.  
\eeq{i2res}
It is important to  notice that there is  no pole
at $q_{2z} = -\omega_1$ in Eq.~(\ref{i2res}). The
remaining integral over $q_{2z}$ is
\beqar
I_3(p,k,{\bf q}_{1},{\bf q}_{2},z_1-z_0,z_2-z_1 ) &=& 
\int\frac{d q_{2z}}{2\pi} \frac{1}{q_{2z}+\omega_1}
\; \left( \frac{e^{-iq_{2z}(z_{2}-z_1)}}{(p-q_2)^2+i\epsilon}  
\; v(-q_{2z}-\omega_0, {\bf q}_{1} )v(q_{2z}, {\bf q}_{2} )
\right. \nonumber \\[1ex]
&\;& \left.  \quad - \;\frac{e^{-i(q_{2z}(z_{2}-z_0)+\omega_1(z_{1}-z_0)  )}}
{(p-q_2)^2+i\epsilon} \; v(\omega_1 -\omega_0, {\bf q}_{1} )
v(q_{2z}, {\bf q}_{2} )  \right) \;\;.
\eeqar{201i3}
Unlike in previous examples,
we now show that no factor of $\half$ arises in the contact limit
$z_2=z_1>z_0$.
The poles in the lower half plane are $q_{2z}=-i\epsilon$,
$q_{2z}=-i\mu_{2}$, and  $q_{2z}=-\omega_0-i\mu_{2}$ and the 
corresponding residues contribute 
\beqar
{\rm Res}(-i\epsilon) & \approx & 
\frac{v(0,{\bf q}_{1})v(0,{\bf q}_{2})}
{E^+ \omega_1} ( 1 - e^{-i\omega_1(z_1-z_0)} )\;\;, \nonumber \\[1ex]
{\rm Res}(-i\mu_{2}) & \approx  & 
\frac{ (4\pi \alpha_s)^2 e^{-\mu_{2}(z_2-z_1)}}
{E^+\mu^2_{2}(2i\mu_{2})( \mu_{1}^2 -  
\mu_{2}^2 -2i\omega_0 \mu_{2}   )} \;\;,
\nonumber \\[1ex]
{\rm Res}(-i\mu_{1} -\omega_0) & \approx  & 
\frac{(4\pi \alpha_s)^2 e^{-\mu_{1}(z_2-z_1)}e^{+i\omega_0 (z_2-z_1) }}
{E^+\mu^2_{1}(2i\mu_{1})( \mu_{2}^2 -  
\mu_{1}^2 +2i\omega_0 \mu_{1}   )} \;\;.
\eeqar{201rescont}
The exponentially suppressed contributions  
$\propto \exp[-\mu_{2}(z_1-z_0)]$ 
in the second and third residues have been neglected, 
and we  made use of the
approximation $\omega_i \ll \mu_{j}$. In the limit of well separated 
scattering centers, the second and third residues are also exponentially
suppressed. In the contact limit, on the other 
hand, both the second and third residues in Eq.~(\ref{201rescont}) 
contribute. However, impact parameter averaging via Eq.~(\ref{bave})
sets $\mu_{1} =  \mu_{2}$,  and the those two contributions
cancel exactly. The result is then
\beqar
M^c_{2,0,1}&=& J(p)e^{ipx_0}(-i)\int\frac{d^2 {\bf q}_{1}}{(2\pi)^2}
e^{-i{\bf q}_{1}\cdot{\bf b}_{1}}v(0,{\bf q}_{1})
(-i)\int \frac{d^2 {\bf q}_{2}}{(2\pi)^2}
e^{-i{\bf q}_{2}\cdot{\bf b}_{2}} v(0,{\bf q}_{2}) 
\; \times \nonumber  \\[1ex]
&\;& \times \;2ig_s\, \frac{\bbox{\epsilon}
\cdot({\bf k}-{\bf q}_{1})}
{({\bf k} - {\bf q}_{1})^2} 
\; e^{i (\omega_0 -\omega_1) z_1}( e^{i \omega_1 z_1} 
- e^{i \omega_1 z_0} ) \; a_2 [c,a_1] (T_{a_2}T_{a_1})\;\;.
\eeqar{201recov}
We emphasize that in this case, there is {\bf no} factor of $\half$ emerging
in the $z_2=z_1$ contact limit  $O_{i,2}$ when one of the legs is
attached to the quark line and the other on to the gluon line. 

Similarly, the diagram $M_{2,0,2}$ leads to the result quoted in 
Ref.~\cite{GLV1B}. In the contact limit, it reduces to 
$M_{2,0,2}(z_2=z_1) = M_{2,0,1}(z_2=z_1)$. However, it
is important to point out that in the contact limit only one
of the diagrams must be taken into account in order to avoid over counting. 
This can be directly  seen from the expansion in the powers of the 
interaction Lagrangian. Alternatively, we get the correct contact
answer by taking the limit of both cross contact diagrams
and multiplying each by $\half$.

\section{Zero measure contact limit of \\  $M_{2,1,0}=X_{1,0}G_1X_{2,0}$
               and $M_{2,1,1}=X_{1,0}G_1X_{2,1}$}

In calculating the different contributions coming from two interactions with 
the same potential centered around $\vec{\bf x}_1$ we have to  take into
account the two graphs given in Fig.~11, where one of the hits occurs before
the gluon emission vertex and the other one after. 

\begin{center}
\vspace*{5.5cm}
\includegraphics{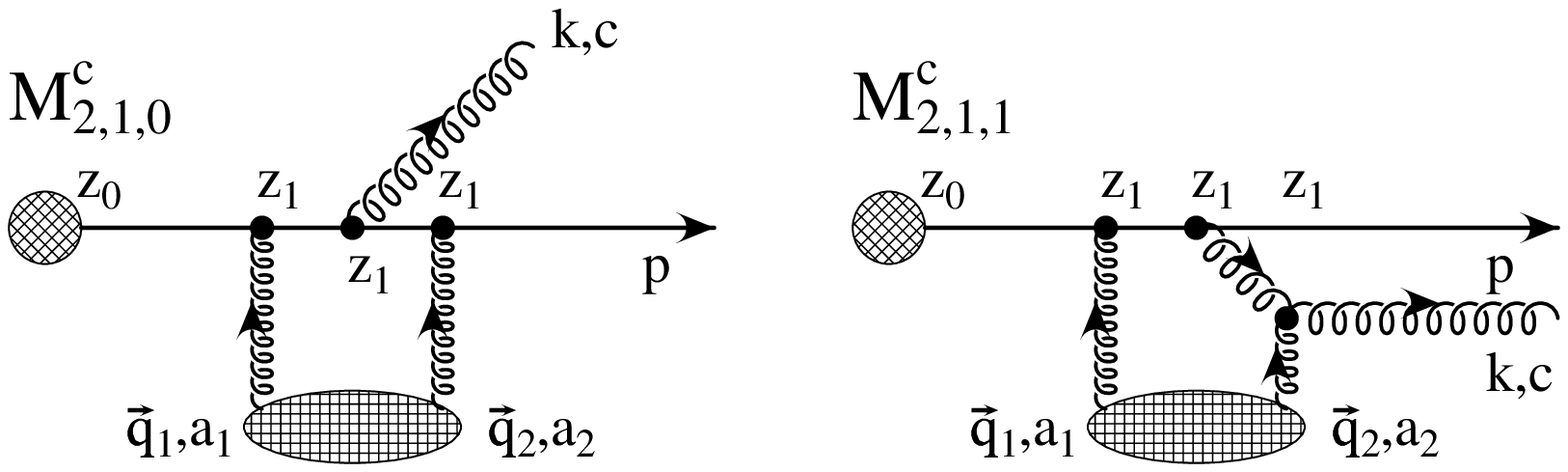}  
\vskip -20pt
\begin{minipage}[t]{15.0cm}
{\small { FIG~11.}
Diagrams  $M^c_{2,1,0}=X_{1,0}G_1X_{2,0}(z_2=z_1)$  and 
$M^c_{2,1,1}X_{1,0}G_1X_{2,1}(z_2=z_1) $ of ${\cal O}(0)$ 
according to the time-ordered perturbation theory.} 
\end{minipage}
\end{center}
\vskip 4truemm

In the framework of
time-ordered perturbation theory from~\cite{GLV1B,GLV1A} the 
graphs are identically zero
in the contact limit because $\int_{t_1}^{t_1} dt \cdots \equiv 0$. We 
here present a more detailed study of the validity of this argument.
In the contact limit $M_{2,1,0} \rightarrow M^c_{2,1,0}=
X_{1,0}G_1 X_{1,0}$.  We will show that  this  contribution vanishes
in the small $x=k^+/E^+$ limit
\beqar
M^c_{2,1,0} &= &\int\frac{d^4 q_1}{(2\pi)^4}\frac{d^4 q_2}{(2\pi)^4}
\; iJ(p+k-q_1-q_2)e^{i(p+k-q_1-q_2)x_0} 
V(q_1)e^{iq_1x_1} V(q_2)e^{iq_2x_1} \; \times \nonumber\\[1ex]
&\;& \times \; (-iE^+)^2ig_s(2(p-q_2)+k)_\mu \epsilon^\mu 
\; i\Delta(p+k-q_1-q_2) \,i\Delta(p+k-q_2)\,i\Delta(p-q_2) \; \times
\nonumber \\[1ex]
&\;& \times \; a_2 c a_1 (T_{a_2}T_{a_1}) \nonumber \\[1ex]
&\approx&  J(p)e^{ipx_0}(-i)\int\frac{d^2 {\bf q}_{1}}{(2\pi)^2}
e^{-i{\bf q}_{1}\cdot{\bf b}_{1}} 
(-i)\int \frac{d^2 {\bf q}_{2}}{(2\pi)^2}
e^{-i{\bf q}_{2}\cdot{\bf b}_{1}}  \; \times \nonumber  \\[1ex]
&\;& \times \; 2ig_s \,
\frac{\bbox{\epsilon}\cdot{\bf k}}{{\bf k}^{\;2}} 
\; e^{i \omega_0 z_0} \; a_2 c a_1  (T_{a_2}T_{a_1})\;(E^+)^2
\; \int\frac{d q_{1z}}{(2\pi)}\frac{d q_{2z}}{(2\pi)} 
\;v(\vec{\bf q}_1)v(\vec{\bf q}_2)
\; \times \nonumber \\[1ex] 
&\;& \;  \times \; \frac{e^{-i(q_{1z}+q_{2z})(z_{1}-z_0)}}
{(p+k-q_1-q_2)^2+i\epsilon} \left( \frac{1}{(p-q_2)^2+i\epsilon}
\;- \; \frac{1}{(p+k-q_2)^2+i\epsilon} \right) \;\;.
\eeqar{210}
In writing this we have made use  the simplifying 
soft gluon and weak interaction  approximations in Eq.~(\ref{softness}).

We can now perform the $q_{1z}$ integral, which brings a factor of
$1/E^+$ and sets the overall phase to 0. We have to now perform  
\beq
I_3 = \int \frac{q_{2z}}{2\pi} \; v(-q_{2z},{\bf q}_{1})
v(q_{2z},{\bf q}_{2}) \;
 \left( \frac{1}{(p-q_2)^2+i\epsilon}
\;- \; \frac{1}{(p+k-q_2)^2+i\epsilon} \right) \;\;.
\eeq{210i3}
We close in the upper half plane where there are four poles $q_{2z}=
i\mu_{1},\; q_{2z}=i\mu_{2},\; q_{2z}=E^+ + i\epsilon, \;
q_{2z}=E^+ + k^+ + i\epsilon$. The residues at the poles involving 
$E^+$ are suppressed by a factor $(1/E^+)^4$ and have negligible 
contribution in the
high energy approximation. The remaining two residues give
\beqar 
{\rm Res}(i\mu_{1}) &\approx& \frac{(4\pi \alpha_s)^2}
{E^+ (2\mu_{1}^2)(\mu_{2}^2 - \mu_{1}^2)} \; 
\left(- \frac{k^+}{E^+} \right) \;\;, 
\nonumber \\[1ex] 
{\rm Res}(i\mu_{2}) &\approx& \frac{(4\pi \alpha_s)^2}
{E^+ (2\mu_{2}^2)(\mu_{1}^2 - \mu_{2}^2)} \; 
\left(- \frac{k^+}{E^+} \right) \;\;. 
\eeqar{resids}
We have discarded $\omega_0$ relative to $\mu_{i}$ and carried
the $k^+/E^+$ expansion to first order.
It is important to notice the sum of the residues is not singular for
the special case of interest when the contact limit contribution is 
averaged over the transverse position of the scatterer, i.e.
$\mu_{1}=\mu_{2}$. 
The above contribution is suppressed by an ${\cal O}(k^+/E^+)$ 
factor relative to
the other graphs. In the high energy $E^+\rightarrow \infty$
limit, where the time-ordered perturbation theory works, 
$M_{2,1,0}(z_2=z_1) \approx 0$.
Similar calculation for $M_{2,1,1}\rightarrow    
M^c_{2,1,1} X_{1,0}G_1 X_{1,1}$ shows 
suppression of the order ${\cal O} (\mu /k^+ )$ and is approximately 
zero   according to the $k^+\gg \mu$
regime~(\ref{eorder}) in which the problem is set up.

This leads us to the  conclusion that the  naive 
``zero measure'' argument for such graphs
is approximately valid for our kinematics.

\section{Amplitude iteration technique to second order}

To illustrate the general iteration procedure
and check the general iteration results using the
reaction operator approach, we construct the classes for
rank 1 and 2 explicitly.

For the two rank 1 classes, we apply 
 the operators $\htD_1$ and  $\htV_1$ on the hard vertex amplitude
Eq.~(\ref{hvert}) once  to obtain
\beqar
G_0D_1 &=& -{\bf H}\, e^{i\omega_0 z_0}\, a_1c - 
{\bf C}_1\, e^{i(\omega_0 z_1- \omega_1z_{10})}\, [c,a_1] 
-{\bf B}_1\, e^{i\omega_0 z_1} \, [c,a_1]  \;\;, \label{1} \\[1ex] 
 G_0V_1 &=& \frac{C_R+C_A}{2}\, {\bf H}\, e^{i\omega_0 z_0}\, c 
-\frac{C_A}{2}\,{\bf C}_1\, e^{i(\omega_0 z_1-\omega_1z_{10})}\, c 
- \frac{C_A}{2}\,{\bf B}_1\,e^{i\omega_0 z_1}\, c  \label{2}  \;\;.
\eeqar{1apl} 
We recall that $z_{ij} \equiv z_i-z_j,\; \Delta z_i \equiv z_i-z_{i-1}$.
These are all amplitudes needed for the first order in opacity
$L/\lambda$ calculation. 

Some of the rank two classes  are obtained from the rank 1
classes Eq.~(\ref{1apl}) through relabeling, i.e.
$G_0D_2\equiv G_0D_1(1\rightarrow 2),\; G_0V_2\equiv
 G_0V_1(1\rightarrow 2)$.
The rest are readily derived from  Eqs.~(\ref{1},\ref{2}) through
our iteration scheme Eqs.~(\ref{didamit},\ref{vidamit})  
\beqar
G_0D_1D_2 &=& -{\bf H}\, e^{i\omega_0 z_0}\, a_2a_1c - 
{\bf C}_1\, e^{i(\omega_0 z_1- \omega_1z_{10})}\, a_2 [c,a_1] 
- {\bf C}_2\, e^{i(\omega_0 z_2- \omega_2z_{20})}\, a_1 [c,a_2] 
\nonumber \\[1.5ex] 
&&-{\bf B}_1\, e^{i\omega_0 z_1} \, a_2[c,a_1]  
-{\bf B}_2\, e^{i\omega_0 z_2} \, [c,a_2] a_1  
\nonumber \\[1.5ex] 
&&-{\bf C}_{(12)}\, e^{i(\omega_0 z_2 -\omega_2 z_{21}-\omega_{(12)} z_{10})} 
\, [[c,a_2], a_1] 
-{\bf B}_{2(12)}\, e^{i(\omega_0 z_2 -\omega_2 z_{21} )} 
\, [[c,a_2], a_1]  \;\;, 
\label{3} \\[1ex] 
G_0D_1V_2 &=& \frac{C_R+C_A}{2}\, {\bf H}\, 
e^{i\omega_0 z_0}\, a_1c 
\nonumber \\[1.5ex] 
&&+ \frac{C_R+C_A}{2}\, {\bf C}_1\, 
e^{i(\omega_0 z_1- \omega_1z_{10})}\,  [c,a_1] 
+{\bf C}_2\, e^{i(\omega_0 z_2- \omega_2z_{20})}\, a_2 a_1 [c,a_2] 
\nonumber \\[1.5ex] 
&&+ \frac{C_R+C_A}{2}\,  {\bf B}_1\, e^{i\omega_0 z_1} \, [c,a_1]  
-  \frac{C_A}{2}\,  {\bf B}_2\, e^{i\omega_0 z_2} \, c a_1  
\nonumber \\[1.5ex] 
&&+ {\bf C}_{(12)}\, e^{i(\omega_0 z_2 -\omega_2 z_{21}-\omega_{(12)} z_{10})} 
\, a_2 [[c,a_2], a_1] 
+{\bf B}_{2(12)}\, e^{i(\omega_0 z_2 -\omega_2 z_{21} )} 
\,a_2 [[c,a_2], a_1]  \;\;, 
\label{4} \\[1ex] 
G_0V_1D_2 &=& \frac{C_R+C_A}{2}\, {\bf H}\, 
e^{i\omega_0 z_0}\, a_2c 
\nonumber \\[1.5ex] 
&&- \frac{C_A}{2}\, {\bf C}_1\, 
e^{i(\omega_0 z_1- \omega_1z_{10})}\,  a_2 c 
+  \frac{C_R+C_A}{2}\,  {\bf C}_2\, 
e^{i(\omega_0 z_2- \omega_2z_{20})}\, [c,a_2] 
\nonumber \\[1.5ex] 
&&- \frac{C_A}{2}\,  {\bf B}_1\, e^{i\omega_0 z_1} \, a_2 c  
+ \frac{C_R}{2}\,  {\bf B}_2\, e^{i\omega_0 z_2} \, [c, a_2]  
\nonumber \\[1.5ex] 
&&-\frac{C_A}{2}\,  {\bf C}_{(12)}\, 
e^{i(\omega_0 z_2 -\omega_2 z_{21}-\omega_{(12)} z_{10})} 
\, [c,a_2] 
-\frac{C_A}{2}\, {\bf B}_{2(12)}\, e^{i(\omega_0 z_2 -\omega_2 z_{21} )} 
\, [c,a_2]  \;\;, 
\label{5} \\[1ex] 
G_0V_1V_2 &=& - \frac{(C_R+C_A)^2}{4}\, {\bf H}\, 
e^{i\omega_0 z_0}\, c 
\nonumber \\[1.5ex] 
&&+ \frac{C_A(C_A+C_R)}{4}\, {\bf C}_1\, 
e^{i(\omega_0 z_1- \omega_1z_{10})}\,  c 
+  \frac{C_A(C_R+C_A)}{4}\,  {\bf C}_2\, 
e^{i(\omega_0 z_2- \omega_2z_{20})}\, c 
\nonumber \\[1.5ex] 
&&+\frac{C_A(C_R+C_A)}{4}\, {\bf B}_1\, e^{i\omega_0 z_1} \, c  
+ \frac{C_RC_A}{4}\,  {\bf B}_2\, e^{i\omega_0 z_2} \, c  
\nonumber \\[1.5ex] 
&&-\frac{C^2_A}{4}\,  {\bf C}_{(12)}\, 
e^{i(\omega_0 z_2 -\omega_2 z_{21}-\omega_{(12)} z_{10})} 
\, c 
-\frac{C^2_A}{4}\, {\bf B}_{2(12)}\, e^{i(\omega_0 z_2 -\omega_2 z_{21} )} 
\, c  \;\;. 
\label{6} 
\eeqar{2apl}  
With this explicit construction of the
relevant classes,  we can compute the differential gluon 
probability up to second order in the opacity expansion. 
Similar calculations can
be performed for the Gunion-Bertsch case as well. 

To first order in opacity  $L/\lambda$  the probability
to radiate a
gluon  from either  {\em quark}  or {\em gluon} jets is 
proportional to
\beqar
&&  
 \frac{1}{d_R} \, 
\left \langle \;{\rm Tr} \left[ G_0D_1D_1^\dagger G_0^\dagger + 
  2\, {\rm Re} ( G_0 V_1 G_0^\dagger ) \right] 
\; \right \rangle_v  
= C_R C_A  \, \left \langle \; \left[ \,
( - 2\,{\bf C}_1 \cdot {\bf B}_1 ) \,
\,\left( 1- \cos ( \omega_1 \Delta z_1 )\right) \, \right] 
\;  \right \rangle_v  
\nonumber \\[1.ex]
&& \qquad  \quad = C_R C_A \,   \int d{\bf q}_{1}\, 
(\bar{v}^2({\bf q}_{1}) - \delta^2({\bf q}_{1}) )
\left[ \, ( - 2\,{\bf C}_1 \cdot {\bf B}_1 ) \,
\,\left( 1- \cos ( \omega_1 \Delta z_1 )\right) \, \right] 
\;\;,
\label{dn1amps}
\eeqar{dn1amps }
where the $17$ terms in Eqs.~(\ref{h1_1},\ref{h1_2}) collapse to
a single term with  color trivial factor $C_R C_A d_R$ interpretable
as gluon final state interaction only. We note that if the 
normalized potential  $\bar{v}^2({\bf q}_{1})$ is replaced by 
 $\bar{v}^2({\bf q}_{1}) - \delta^2({\bf q}_{1})$ 
the result of Eq.~(\ref{dn1amps}) remains unchanged.   

To second order in opacity we find that
\beqar
&& \frac{1}{d_R} \,  \left \langle \;  {\rm Tr} 
\left[ \, G_0D_1D_2D_2^\dagger D_1^\dagger G_0^\dagger +  
( G_0 V_1D_2D_2^\dagger G_0^\dagger + {\rm h.c.})  
+( G_0D_1V_2 D_1^\dagger G_0^\dagger + {\rm h.c.})  
\right. \right. 
\nonumber \\[1.ex]
&& \left. \left. \qquad  \qquad 
+  ( G_0 V_1 V_2 G_0^\dagger + {\rm h.c.} ) 
+  ( G_0 V_1 V^\dagger_2 G_0^\dagger + {\rm h.c.} )  \,  \right] 
\;  \right \rangle_v  
\nonumber \\[1ex]
&& = C_R C_A^2  \, 
\left \langle \;
\left[ \, 2\,{\bf C}_1 \cdot {\bf B}_1 \,
\left( 1- \cos ( \omega_1 \Delta z_1 )\right) 
+\, 2\, {\bf C}_2 \cdot {\bf B}_2 \,
\left( \cos ( \omega_2 \Delta z_2 ) - 
\cos ( \omega_2 (\Delta z_1 + \Delta z_2 )\right) 
\right.  \right.
\nonumber \\[1.5ex]
&&\qquad \qquad \qquad \;
 -   \,2\, {\bf C}_{(12)} \cdot {\bf B}_{2(12)} \,
\left( 1- \cos ( \omega_{(12)} \Delta z_1 )\right)
\nonumber \\[1.5ex]
&&\qquad \qquad \qquad  \left. \left.
- \,2\, {\bf C}_{(12)} \cdot {\bf B}_2 \,
\left( \cos ( \omega_2 \Delta z_2 )
 - \cos ( \omega_{(12)} \Delta z_1 + \omega_2 \Delta z_2 )\right) 
\, \right]\; \right \rangle_v \;\; 
\nonumber \\[1.ex]
&&= C_R C_A^2\,    \prod_{i=1,2} \int d{\bf q}_{i}\, 
(\bar{v}^2({\bf q}_{i}) - \delta^2({\bf q}_{i}) )   
\left[ -   \,2\, {\bf C}_{(12)} \cdot {\bf B}_{2(12)} \,
\left( 1- \cos ( \omega_{(12)} \Delta z_1 )\right) \right.
\nonumber \\[1.ex]
&&\left.  \qquad \qquad \qquad \qquad \qquad  \qquad \qquad \quad \;\;
- \,2\, {\bf C}_{(12)} \cdot {\bf B}_2 \,
\left( \cos ( \omega_2 \Delta z_2 )
 - \cos ( \omega_{(12)} \Delta z_1 + \omega_2 \Delta z_2 )\right) 
\; \right] \;\;.
\eeqar{secord} 

Eqs.~(\ref{dn1amps},\ref{secord}) have simple physical interpretation.
We showed that for {\em both}  quark and gluon
jets a simple color trace $\propto C_A^n$ survives. 
In general this can be seen from 
Eqs.~(\ref{didamit},\ref{vidamit},\ref{dvid}).
Thus the distribution of the radiated gluons is interpretable 
as the interference
of the Cascade amplitude  that has the knowledge of all final state 
interactions with the effective color currents, i.e. the 
Gunion-Bertsch amplitudes,  
originating at the scattering centers and also having all possible 
subsequent interactions with the gluon line. The formation physics effects
arise from  phase differences that store the information on the
cumulative formation lengths 
 before the effective color current emission.  
The averaging over the momentum  
transfers is seen to be effectively over a modified potential  
that vanishes on integration.

\end{appendix}

\end{document}